\documentclass[10pt,twocolumn,journal]{IEEEtran}
\usepackage{setspace}
\usepackage{clrscode}
\usepackage{amsthm}
\usepackage{pict2e}
\usepackage{multirow}
\usepackage{enumitem}

\usepackage{amsmath}
\usepackage{amssymb}
\usepackage[dvips]{graphicx}
\usepackage{epsfig}
\usepackage{hyperref}
\usepackage{algorithm}
\usepackage{algorithmic}
\usepackage{caption}
\usepackage{subcaption}
\usepackage{capt-of}

\newcommand*{\Resize}[2]{\resizebox{#1}{!}{$#2$}}%

\newcommand{\rme}{{\mathrm{e}}}

\newcommand{\E}{\mathbb{E}}

\newcommand{\mH}{\mathcal{H}}
\newcommand{\hH}{\hat{\mathcal{H}}}

\newcommand{\tg}{\tilde{g}}

\makeatletter

\newcommand{\Rmnum}[1]{\expandafter\@slowromancap\romannumeral #1@}
\makeatother

\newcommand{\argmax}{\operatornamewithlimits{arg\,max}}

\newcommand{\avg}{\text{avg}}
\newcommand{\bP}{\mathbf{P}}
\newcommand{\nid}{\rm{d}}
\newcommand{\f}{\rm{f}}
\newcommand{\pk}{\rm{pk}}
\newcommand{\re}{\rm{re}}
\newcommand{\upper}{\rm{upper}}

\newcommand{\PSNR}{\text{PSNR}}

\newtheorem{Prop}{Proposition}

\begin{document}

\title{Multimedia Transmission over Cognitive Radio Channels under Sensing Uncertainty}

\author{\vspace{0.2cm}
Chuang Ye, Gozde Ozcan, M. Cenk Gursoy, and Senem Velipasalar
\thanks{The authors are with the Department of Electrical
Engineering and Computer Science, Syracuse University, Syracuse, NY, 13244
(e-mail: chye@syr.edu, gozcan@syr.edu, mcgursoy@syr.edu, svelipas@syr.edu).}
\thanks{This work has been funded in part by National Science Foundation CAREER grant CNS-1206291 and National Science Foundation grant CNS-1302559.}
}

\maketitle

\begin{abstract}
This paper studies the performance of hierarchical modulation-based multimedia transmission in cognitive radio (CR) systems with imperfect channel sensing results under constraints on both transmit and interference power levels. Unequal error protection (UEP) of data transmission using hierarchical quadrature amplitude modulation (HQAM) is considered in which high priority (HP) data is protected more than low priority (LP) data. In this setting, closed-form bit error rate (BER) expressions for HP data and LP data are derived in Nakagami-$m$ fading channels in the presence of sensing errors. Subsequently, the optimal power control that minimizes weighted sum of average BERs of HP bits and LP bits or its upper bound subject to peak/average transmit power and average interference power constraints is derived and a low-complexity power control algorithm is proposed. Power levels are determined in three different scenarios, depending on the availability of perfect channel side information (CSI) of the transmission and interference links, statistical CSI of both links, or perfect CSI of the transmission link and imperfect CSI of the interference link. The impact of imperfect channel sensing decisions on the error rate performance of cognitive transmissions is also evaluated. In addition, tradeoffs between the number of retransmissions, the severity of fading, and peak signal-to-noise ratio (PSNR) quality are analyzed numerically. Moreover, performance comparisons of multimedia transmission with conventional quadrature amplitude modulation (QAM) and HQAM, and the proposed power control strategies are carried out in terms of the received data quality and number of retransmissions.
\end{abstract}

\thispagestyle{empty}


\section{Introduction}

Recent overwhelming growth in the volume of multimedia content, multimedia traffic and wireless multimedia applications is drastically increasing the demand for more bandwidth. With this and the fact that prime portion of the spectrum has already been allocated, bandwidth scarcity has become one of the major bottlenecks in wireless services. At the same time, according to the report from the Spectrum-Policy Task Force of the Federal Communications Commission (FCC) \cite{fcc}, many portions of the allocated spectrum are mostly unused or inefficiently used. Recently, cognitive radio (CR) has been proposed to realize dynamic spectrum access (DSA) in order to overcome the spectrum underutilization problem by allowing the unlicensed users (i.e., cognitive or secondary users) to access the licensed spectrum without causing harmful interference to the licensed users (i.e., primary users) \cite{mitola}, \cite{haykin}. DSA strategies can be mainly categorized into three models, namely dynamic exclusive use model, open radio spectrum sharing, and hierarchical radio spectrum access model \cite{zhao}. Dynamic exclusive use model provides dynamic spectrum allocation and spectrum rights, which allow license holders to sell and trade the spectrum. Therefore, spectrum auction and market based policies for resource allocation lead to a profitable way of utilizing the spectrum \cite{niyato} -- \cite{mwangoka}. While users can access the spectrum on a non-priority basis in the open sharing model, there is a hierarchy between the access rights of the primary and cognitive users in the hierarchical spectrum access model. In particular, the primary users have priority in accessing the spectrum, and cognitive users can either coexist with the primary users by varying their transmission power according to primary user activity and interference constraints, or transmit only when there is no active primary user in the channel. Therefore, spectrum sensing is an essential functionality of CR systems in order to detect the temporarily unused frequency bands \cite{wang}. Along with this, efficient design of medium access control protocol has an important role for exploiting the spectrum opportunities \cite{mishra}.

\vspace{-0.37cm}
\subsection{Literature Overview}
Existing literature mainly focuses on the performance of spectrum sensing methods and the throughput of CR systems. There have been relatively limited number of studies on multimedia transmission in CR networks. The work in \cite{hu} mainly focused on the optimization of the overall received quality of MPEG-4 fine grained scalable video multicast by considering proportional fairness and also primary user protection from harmful interference in CR networks. In \cite{hu2}, the optimal channel and path selection strategy for streaming multiple videos over a multi-hop CR network was proposed in the presence of imperfect sensing decisions and a constraint on the collision probability. The authors in \cite{huang1} proposed an optimal packet loading strategy for multimedia transmissions of secondary users by considering each channel with different primary user activity. The authors in \cite{luo} jointly optimized the quantization step size of source coding, modulation type and channel coding parameters in order to minimize the expected video distortion over CR networks subject to a packet delay constraint. In \cite{bocus}, an optimal subcarrier and antenna selection scheme that maximizes the aggregate visual quality of the received video in downlink CR networks was proposed. In \cite{tigang}, a channel allocation scheme was introduced to meet the different quality of experience (QoE) requirements of the secondary users. The recent work in \cite{huang2} proposed a cross-layer scheduling scheme for OFDM-based CR systems in which optimal subcarrier assignment, power and modulation allocation were performed for each incoming multimedia packet. The authors in \cite{he} investigated the optimal assignment of cognitive users to idle-sensed channels to maximize the visual quality of downlink multiuser video streaming. Also, the work in \cite{yao} mainly focused on improving  the quality of H.264/SVC video at the secondary receiver in multi-channel CR networks. Moreover, the authors in \cite{jiang} studied joint adaptation of scalable video coding (SVC) and transmission rate to minimize the average energy consumption of cognitive users subject to quality of service (QoS) requirements.

\subsection{Main Contributions}
In this paper, we analyze the performance of multimedia transmission based on hierarchical quadrature amplitude modulation (HQAM) with power control in CR systems. The main contributions of this paper can be summarized as follows:
\begin{itemize}
\item Unlike the aforementioned works in \cite{hu} -- \cite{tigang}, we have considered an error-resilient method called unequal error protection (UEP), which provides different levels of protection to different parts of the multimedia data in order to increase the robustness of transmission against wireless channel impairments, e.g., noise, interference from other users and fading. HQAM is an efficient UEP technique in which high priority (HP) data bits are mapped to the first two most significant bits (MSBs) of each constellation point whereas low priority (LP) data bits are mapped to the rest of the bits. We identify the optimal maximum a posteriori probability (MAP) decision rule for HQAM and new expressions for computing the bit error rates (BERs) of HP data bits and LP data bits in the presence of sensing errors for any given fading distribution. We further derive closed-form expressions for BERs of HP bits and LP bits for 16-HQAM averaged over Nakagami-$m$ fading, which is chosen due to its ability of representing a wider range of fading severities.
\item HQAM based multimedia transmission without power control in non-cognitive context has been analyzed recently \cite{barmada} -- \cite{morimoto}. Different from these works, we obtain optimal power adaptation schemes to minimize the weighted sum of average BERs of HP bits and LP bits in sensing-based spectrum sharing CR systems subject to peak/average transmit power constraints along with average interference power constraint under imperfect sensing decisions. In sensing-based spectrum sharing CR systems, cognitive users sense the channel to determine the primary user activity and then adapt their transmission power levels according to the channel sensing decisions. It is assumed that either instantaneous channel side information (CSI) or statistical CSI is available to determine optimal power levels. We note that our results are also different from the work in \cite{milli}, where the authors derived optimal power control schemes by assuming that the primary user always exists in the channel, and therefore secondary users do not perform any channel sensing.
\item A low-complexity optimal power control algorithm under peak/average transmit power and average interference power constraints is proposed. Also, we analyze and approximate the optimal power control schemes at high SNR levels, and obtain closed-form power expressions in terms of the Lambert-W function, which is easy to evaluate.
\item We analyze the transmission of H.264/MPEG-4 coded video and JPEG200 coded image using conventional QAM and HQAM in terms of peak signal-to-noise ratio (PSNR) quality and number of retransmissions in a CR system. We further investigate the relations between sensing errors, optimal transmission powers, number of retransmissions and the received data quality.
\end{itemize}

The remainder of this paper is organized as follows: Section \ref{sec:system_model} introduces the system model including channel sensing and cognitive channel model. In Section \ref{sec:multimedia_transmission}, HQAM-based multimedia transmission in CR systems is described. In Section \ref{sec:BER_formulation}, closed-form BER expressions for HP data and LP data averaged over Nakagami-$m$ fading channel with 16-HQAM signaling are derived. In Section \ref{sec:opt_power}, optimal power control policies that minimize weighted sum of average BERs of HP bits and LP bits or its upper bound in the presence of imperfect sensing decisions subject to both transmit power and interference constraints are determined and the optimal power control algorithm is provided. Numerical and simulation results are presented and discussed in Section \ref{sec:num_results}. Finally, we conclude the paper in Section \ref{sec:conclusion}. Several proofs are relegated to the Appendix.

\section{System Model} \label{sec:system_model}

\subsection{Channel Sensing} \label{subsec:sensing}

\begin{figure}[h]
\centering
\includegraphics[width=0.3\textwidth]{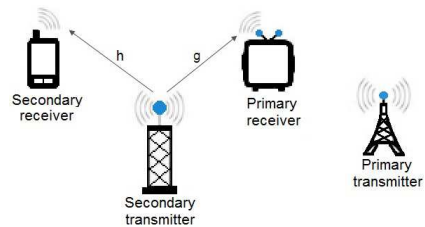}
\caption{Cognitive radio channel model}\label{fig:system_model}
\end{figure}

We consider a CR system in which a secondary transmitter sends multimedia data i.e., image and/or video to a secondary receiver by utilizing the spectrum licensed to the primary users as illustrated in Fig. \ref{fig:system_model}. To peacefully coexist with the primary users, secondary users should initially learn the primary users' activity through channel sensing. Channel sensing can be formulated as a binary hypothesis testing problem in which hypotheses $\mH_0$ and $\mH_1$ denote that the primary users are inactive and active in the channel, respectively.
\begin{figure*}
\centering
{\includegraphics[width = 0.7\linewidth]{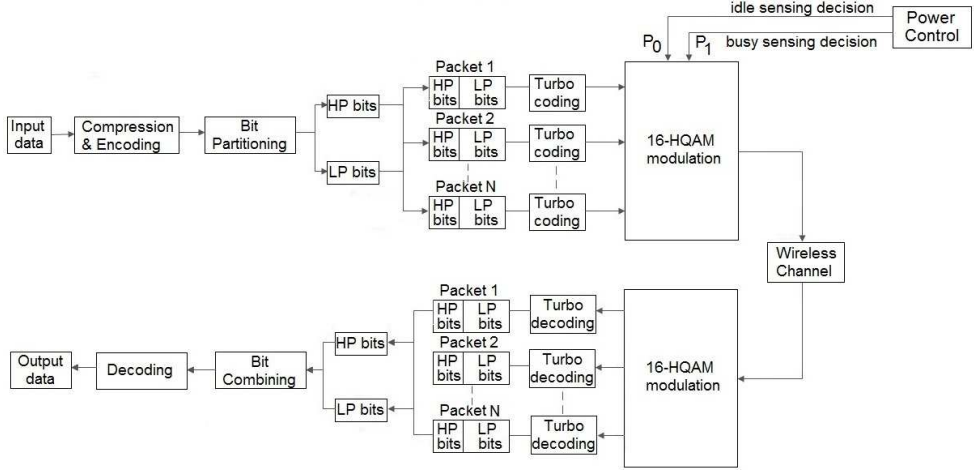}}
\caption{Block diagram of the multimedia transmission and reception system. }
\label{fig:transmission_model} 
\end{figure*}
Several spectrum sensing methods including matched filter detection, energy detection, and cyclostationary feature detection, have been developed in the literature \cite{axell} and the corresponding sensing performance is characterized by two parameters, namely the probabilities of detection and false alarm, which are defined as
\begin{align}
P_{\nid}= \Pr\{\hH_1|\mH_1\}, \hspace{0.3cm}P_{\f}=\Pr\{\hH_1|\mH_0\},
\end{align}
where $\hH_0$ and $\hH_1$ correspond to the events that the channel is detected as idle and busy, respectively. In a missed detection event, secondary users fail to detect active primary users and hence secondary users can collide with the primary users' transmission while in a false alarm event, secondary users detect the channel as busy while in fact there is no active primary user, resulting in the underutilization of the channel.

\subsection{Cognitive Channel Model} \label{subsec:channel_model}
After performing channel sensing, the secondary transmitter starts sending multimedia data to a secondary receiver over a flat-fading channel. It is assumed that the secondary users are allowed to transmit under both idle and busy sensing decisions. Under this assumption, the channel input-output relation is given by
\begin{align}
\small
\begin{split}
y = \begin{cases} hs+ n &\text{in the absence of primary user activity}\\
hs+ n + w &\text{in the presence of primary user activity}
\end{cases}.
\label{eq:received_signal}
\end{split}
\normalsize
\end{align}
Above, $s$ and $y$ are the complex-valued transmitted and received signals, respectively and $w$ denotes the primary users' received faded signal distributed according to a circularly symmetric complex Gaussian distribution with zero mean and variance $\sigma_w^2$. Also, $n$ represents the circularly symmetric complex Gaussian noise with zero mean and variance $\sigma_{n}^2$. In addition, $h$ is the channel fading coefficient of the transmission link between the secondary transmitter and the secondary receiver as shown in Fig. \ref{fig:system_model}.

\subsection{Multimedia Transmission System}\label{sec:multimedia_transmission}
The block diagram of the multimedia transmission system is depicted in Fig. \ref{fig:transmission_model}. Input image or video is first compressed before transmission. JPEG2000 image coder is chosen as the compression technique for image transmission. In the case of video transmission, H.264/MPEG-4 codec is employed to compress the video content \cite{puri}.

\begin{figure}
\centering
{\includegraphics[width = 0.7\linewidth]{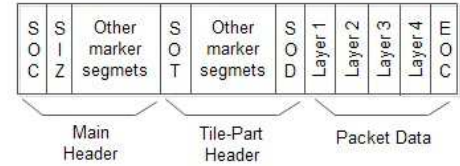}}
\caption{JPEG2000 codestream structure}
\label{fig:JPEG2000} 
\vspace{-.3cm}
\end{figure}

Following compression, data partitioning is applied. In particular, the compressed data is divided into two priority levels, namely HP and LP. The structure of JPEG2000 codestream is shown in Fig \ref{fig:JPEG2000}, which consists of a sequence of marker segments and layers with unequal importance \cite{natu}. Main header and tile-part header have a sequence of marker segments which contain important coding parameters and the layers in the packet data have different sensitivity to the corruption of the data. Therefore, for the images, the codestream header (i.e., main header and tile-part header) and lower layers are classified as HP data whereas the rest of the codestream is assigned as LP data. In the case of videos, there are three types of frames, namely I, P and B frames. I frame is the key frame in the coded video sequence. It can be encoded independently from other frames by using only its own information. Therefore, this frame is used as a reference frame for coding inter-coded frames such as P frames and B frames, and it is also employed for indexing and prevention of error propagation \cite{puri}. Any loss of I frames has more devastating impact on video quality than loss in other frames. Therefore, I frame is regarded as HP data while the rest of the frames are assigned as LP data.

After identifying HP data and LP data, the compressed data sequence is divided into $N$ packets of equal size. Each packet contains both HP data and LP data in such a way that the ratio of HP bits and LP bits is the same. Subsequently, channel coding based on Turbo codes \cite{Valenti} is employed in order to enhance the resilience of the compressed data to wireless channel impairments, e.g., noise, interference from other users, and fading. Finally, HP bits and LP bits within packets are modulated using 16-HQAM and transmission power is determined based on the sensing decision, as further discussed in the following sections, and then each packet is transmitted over the wireless channel. At the receiver, ARQ mechanism is employed. More specifically, if the received power of the packet is less than a certain threshold, the secondary receiver requests the retransmission of the packet. On the other hand, if the received power of the packet is greater than the threshold, the output data is obtained by performing the inverse operations i.e., demodulation, turbo decoding, bit combining, and source decoding as shown in Fig. \ref{fig:transmission_model}.

\section{HQAM modulation and Bit error rate analysis}\label{sec:BER_formulation}
In this section, we first present the signal constellation of 16-HQAM. Then, we provide the optimal detection rule for the CR system in the presence of channel sensing errors. Subsequently, the BER performance of Gray-encoded 16-HQAM  associated with this optimal detector over Nakagami-$m$ fading is analyzed .

\subsection{16-HQAM Constellation} \label{subsec:optimaldecision}
\begin{figure}
\centering
\includegraphics[width=0.261\textwidth]{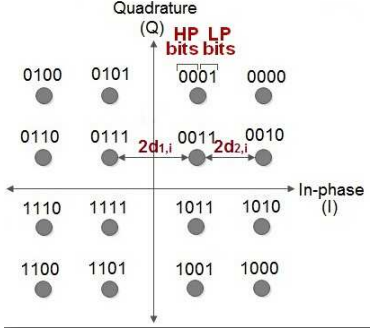}
\caption{Signal constellation diagram of Gray coded 16-HQAM}\label{fig:signal_constellation}
\end{figure}

Secondary users are assumed to employ 16-HQAM,  which provides two priority layers, HP and LP. In particular, HP data bits occupy the two most significant bits of each symbol point while LP data bits occupy the rest of the bits of the symbol. On the other hand, the conventional 16-QAM is non-hierarchical with each layer having the same reliability. Fig. \ref{fig:signal_constellation} shows the constellation diagram of Gray-encoded 16-HQAM, in which neighboring signal points differ only by one bit and the signal points in the same quadrant have the same HP bits.
In the figure, $2d_{1,i}$ and $2d_{2,i}$ represent the minimum distance between each quadrant and the minimum distance between the signal constellation points within each quadrant, respectively. Let us define the ratio $\alpha_i=d_{1,i}/d_{2,i}$. By changing the value of $\alpha_i$, we can control the protection level for HP and LP bits. More specifically, for a given average signal power, increasing the value of $\alpha_i$ increases the distance between quadrants, which leads to diminished BER for HP bits. On the other hand, the distance between the constellation points within the quadrant decreases, and hence BER for LP bits increases. As a result, HP data is protected more against errors than LP data.

The minimum distance between the quadrants and the minimum distance between the signal constellation points within the quadrants under the sensing decision $\hH_i$ can be written respectively as
\begin{equation}
\small
\begin{split}
d_{1,i} =\sqrt{\frac{\alpha_i^2P_i}{2(\alpha_i+1)^2+2}}, \hspace{0.3cm}d_{2,i} =\sqrt{\frac{P_i}{2(\alpha_i+1)^2+2}}
\end{split}
\normalsize
\end{equation}
where $P_i$ denotes the average transmission power under the sensing decision $\hH_i$ for $i \in \{0,1\}$. In particular, the average power level is $P_0$ if the channel is detected as idle whereas the secondary user transmits at average power level $P_1$ if the channel is detected as busy.
\subsection{Bit Error Rate Analysis}
It is assumed that the sensing decisions and the perfect knowledge of the fading realizations are available at the secondary receiver. Thus, any phase shift due to fading can be removed by multiplying the received signal with the phase of the fading coefficient $h$. Under these assumptions, the optimal MAP decision rule for any arbitrary $M$-ary digital modulation under sensing decision $\hH_i$ is given as follows:
\begin{align}
\label{eq:MAP_decision_rule}
\hat{s} &= \argmax_{0\le k\le M-1}\, \Pr\{s_k|y,h,\hH_i\}
\\
&= \argmax_{0\le k\le M-1}\, p_kf(y | s_k, h, \hH_i) \label{eq:MAP_decision_rule-2}
\\
&= \argmax_{0\le k\le M-1}\, \sum_{j=0}^1p_k\Pr\{\mH_j|\hH_{i}\}f(y | s_k, h, \hH_{i}, \mH_j), \label{eq:cond-prob-yr}
\end{align}
where $\hat{s}$ is the MAP detector output, $p_k$ is the prior probability of the signal constellation point $s_k$. Above, (\ref{eq:MAP_decision_rule-2}) is obtained by Bayes' rule, and can further be expanded by conditioning the density function $f(y | s_k, h, \hH_i)$ on the hypotheses $\mH_0$ and $\mH_1$ as in (\ref{eq:cond-prob-yr}). Also, $f(y|s_k, h, \hH_i, \mH_j)$ in (\ref{eq:cond-prob-yr}) is the conditional distribution of the received real signal $y$ given the transmitted signal $s_k$, channel fading coefficient $h$, channel sensing decision $\hH_i$, and true state of the channel $\mH_j$, and can be expressed as
\begin{align}
f(y|s_k, h, \hH_i, \mH_j) = \begin{cases} \frac{1}{\pi\sigma_n^2}\rme^{-\frac{|y - s_kh|^2}{\sigma_n^2}}, &j=0\\
 \frac{1}{\pi(\sigma_n^2 + \sigma_w^2)}\rme^{-\frac{|y - s_kh|^2}{\sigma_n^2 + \sigma_w^2}}, &j=1
\end{cases}.
\label{eq:distrubition_yr}
\end{align}
Note that the sensing decision $\hH_i$ has an impact on the density function through $P_i$, the power of the transmitted signal $s_k$. Additionally, the conditional probabilities in (\ref{eq:cond-prob-yr}) can be written as
\begin{align}
\small
\begin{split}
\Pr\{\mathcal{H}_j | \hat{\mathcal{H}}_i\} \!=\! \frac{\Pr\{\mathcal{H}_j\} \Pr\{\hat{\mathcal{H}}_i | \mathcal{H}_j\}}{ \Pr\{\mathcal{H}_0\} \Pr\{\hat{\mathcal{H}}_i | \mathcal{H}_0\}\!+\!\Pr\{\mathcal{H}_1\} \Pr\{\hat{\mathcal{H}}_i | \mathcal{H}_1\}}  \,\, i \! \in\! \{0,1\}. \nonumber
\end{split}
\normalsize
\end{align}
Above, $\Pr\{\mH_0\}$ and $\Pr\{\mH_1\}$ denote the prior probabilities of primary users being inactive and active in the channel, respectively.

The average bit error probability for the MAP decision rule in (\ref{eq:MAP_decision_rule}) can be computed as
\begin{equation}
\small
\begin{split}
&\text{BER} \!=\! 1 \!-\!  \frac{1}{\log_2M}\!\!\sum_{m=0}^{M-1}\sum_{v=1}^{\log_2M}\!\sum_{i,j=0}^{1}\!p_m\!\Pr\{\mH_j,\hH_i\} \!\Pr\{b_v|s_m,\hH_i,\mH_j\}.
\end{split}
\normalsize
\end{equation}
where $b_v$ is the $v$-th bit for the symbol and $\Pr\{b_v|s_m,\hH_i,\mH_j\}$ denotes the probability of correctly detecting the bit $b_v$ given the symbol $s_m$, sensing decision $\hH_i$ and true channel state $\mH_j$.

It was shown in \cite{ozcan} that the midpoints between the signal constellation points are optimal thresholds for rectangular QAM signaling in the presence of channel sensing errors. Since HQAM is a modification of conventional QAM primarily through the new bit assignment scheme, the optimal detector structure in HQAM is the same as in QAM signaling.

Next, we analyze the BER performance of HP and LP bits in 16-HQAM. The signals are assumed to be equally likely. Since HP data is mapped to two most significant bits in the signal constellation, the corresponding BER can be found by analyzing the change of in-phase bits. Hence, BER of HP bits for a given fading coefficient can be expressed as
\begin{equation}
\small
\begin{split}
P_{\text{HP}}(\bP,h) = \frac{1}{32}\sum_{k=0}^{15}\sum_{i,j=0}^{1}\Pr\{\mH_j,\hH_i\}\bigg(&P_e(b_{1}|s_k,h,\hH_i,\mH_j)\\+&P_e(b_{2}|s_k,h,\hH_i,\mH_j)\bigg)
\end{split}
\normalsize
\end{equation}
where $\bP=[P_0,P_1]$ and $P_e(.)$ denotes the probability of an error in a single bit. As seen in Fig. \ref{fig:signal_constellation}, the most significant bit $b_1$ does not change in the in-phase direction, and only changes in the quadrature direction in the form of $0-0-1-1$. Similarly, the second bit $b_2$ just changes in the in-phase direction in the same form of $0-0-1-1$. Hence, BER expression can be calculated as
\begin{equation}
\begin{split}\label{eq:BER_HP}
\hspace{-0.25cm}P_{\text{HP}}(\bP,h) = \frac{1}{2}\sum_{j=0}^{1}\sum_{i=0}^{1}\sum_{l=0}^{1}\Pr\{\mH_j,\hH_i\}&Q\Bigg(\sqrt{\frac{c_{l,i} P_i|h|^2}{\sigma_j^2}}\Bigg),
\end{split}
\end{equation}
where $c_{0,i}=\frac{(\alpha_i+2)^2}{(\alpha_i+1)^2+1}$ and $c_{1,i} =\frac{\alpha_i^2}{(\alpha_i+1)^2+1}$. Also, $Q(x) = \int_{x}^{\infty}\frac{1}{\sqrt{2\pi}}e^{-t^2/2}dt$ is the Gaussian $Q$-function and $\sigma_j^2$ is defined as
\begin{align}
\small
\begin{split}
\sigma_j^2 = \begin{cases} \sigma_n^2, &j=0\\
\sigma_n^2+\sigma_w^2,&j=1
\end{cases}.
\label{eq:distrubition_yr}
\end{split}
\normalsize
\end{align}
Subsequently, LP bits correspond to the two least significant bits in the signal constellation. Thus, BER of LP bits can be calculated by considering the change of quadrature bits as follows:
\begin{equation}
\small
\begin{split}
\hspace{-0.15cm}P_{\text{LP}}(\bP,h) = \frac{1}{32}\sum_{k=0}^{15}\sum_{i,j=0}^{1}\Pr\{\mH_j,\hH_i\}\bigg(&P_e(b_{3}|s_k,h,\hH_i,\mH_j)\\+&P_e(b_{4}|s_k,h,\hH_i,\mH_j)\bigg).
\end{split}
\normalsize
\end{equation}
As observed from Fig. \ref{fig:signal_constellation}, the third bit, $b_{3}$, changes according to the pattern $0-1-1-0$ in the quadrature direction while it does not change in the in-phase direction. The last bit, $b_{4}$, has similar changes but in the other direction. As a result, BER expression is given by (\ref{eq:BER_LP}) shown at the top of the next page.
\begin{figure*}
\small
\begin{equation}
\begin{split} \label{eq:BER_LP}
P_{\text{LP}}(\bP,h) = \frac{1}{2}\sum_{j=0}^{1}\sum_{i=0}^{1}&\Pr\{\mH_j,\hH_i\}\Bigg\{2Q\Bigg(\sqrt{\frac{\beta_{0,i}P_i|h|^2}{\sigma_j^2}}\Bigg)+Q\Bigg(\sqrt{\frac{\beta_{1,i}P_i|h|^2}{\sigma_j^2}}\Bigg)-Q\Bigg(\sqrt{\frac{\beta_{2,i}P_i|h|^2}{\sigma_j^2}}\Bigg)\Bigg\}\\
&\hspace{-2.8cm}\text{where } \beta_{0,i} =\frac{1}{(\alpha_i+1)^2+1} \hspace{0.3cm}\beta_{1,i}=\frac{(2\alpha_i+1)^2}{(\alpha_i+1)^2+1} \hspace{0.3cm} \beta_{2,i}=\frac{(2\alpha_i+3)^2}{(\alpha_i+1)^2+1}.
\end{split}
\end{equation}
\hrule
\end{figure*}
Note that the above BER expressions are for a given instantaneous realization of the fading coefficient, $h$. The averaged BER of HP bits and LP bits over Nakagami-$m$ fading distribution are given in (\ref{eq:BER_HP_averaged_fading}) and (\ref{eq:BER_LP_averaged_fading}), respectively, at the top of the next page, where $_2F_1(.,.;.;.)$ denotes the Gauss hypergeometric function \cite[eq. 9.10]{gradshteyn}. The derivation steps of these expressions are given in Appendix \ref{app:derivation-eqs}.
\begin{figure*}
\begin{equation}
\begin{split} \label{eq:BER_HP_averaged_fading}
P_{\text{HP}}(\bP) =\frac{1}{4\sqrt{\pi}}\frac{\Gamma(m+\frac{1}{2})}{\Gamma(m+1)}&\sum_{j=0}^{1}\sum_{i=0}^{1}\sum_{l=0}^{1}\Bigg\{\frac{\Pr\{\mH_j,\hH_i\}\sqrt{\frac{c_{l,i}P_i\Omega}{2m\sigma_j^2}}}{\Big(\frac{c_{l,i}P_i\Omega}{2m\sigma_j^2}+1\Big)^{m+\frac{1}{2}}} {}_2F_1\Bigg(1,m+1/2;m+1;\frac{2m\sigma_j^2}{c_{l,i}P_i\Omega+2m\sigma_j^2}\Bigg)\Bigg\}
\end{split}
\end{equation}
\hrule
\end{figure*}
\begin{figure*}
\small
\begin{equation}
\begin{split}\label{eq:BER_LP_averaged_fading}
&P_{\text{LP}}(\bP) =\frac{1}{4\sqrt{\pi}}\frac{\Gamma(m+\frac{1}{2})}{\Gamma(m+1)}\sum_{j=0}^{1}\sum_{i=0}^{1}\Bigg\{\frac{2\Pr\{\mH_j,\hH_i\}\sqrt{\frac{\beta_{0,i}P_i\Omega}{2m\sigma_j^2}}}{\Big(\frac{\beta_{0,i}P_i\Omega}{2m\sigma_j^2}+1\Big)^{m+\frac{1}{2}}} {}_2F_1\Bigg(1,m+1/2;m+1;\frac{2m\sigma_j^2}{\beta_{0,i}P_i\Omega+2m\sigma_j^2}\Bigg)\\&+\frac{\Pr\{\mH_j,\hH_i\}\sqrt{\frac{\beta_{1,i}P_i\Omega}{2m\sigma_j^2}}}{\Big(\frac{\beta_{1,i}P_i\Omega}{2m\sigma_j^2}+1\Big)^{m+\frac{1}{2}}} {}_2F_1\Bigg(1,m+1/2;m+1;\frac{2m\sigma_j^2}{\beta_{1,i}P_i\Omega+2m\sigma_j^2}\Bigg)-\frac{\Pr\{\mH_j,\hH_i\}\sqrt{\frac{\beta_{2,i}P_i\Omega}{2m\sigma_j^2}}}{\Big(\frac{\beta_{2,i}P_i\Omega}{2m\sigma_j^2}+1\Big)^{m+\frac{1}{2}}} {}_2F_1\Bigg(1,m+1/2;m+1;\frac{2m\sigma_j^2}{\beta_{2,i}P_i\Omega+2m\sigma_j^2}\Bigg)\Bigg\}.
\end{split}
\end{equation}
\hrule
\end{figure*}
For the special case where $m$ is an integer in the BER expression of HP bits given in (\ref{eq:BER_HP_averaged_fading}), using the property for Gauss hypergeometric function with integer argument \cite[Appendix A]{eng}, we can simplify the corresponding BER expression as
\begin{equation}
\small
\begin{split} \label{eq:BER_HP_averaged_fading_int}
P_{\text{HP}}(\bP) =\frac{1}{2}&\sum_{j=0}^{1}\sum_{i=0}^{1}\sum_{l=0}^{1}\Pr\{\mH_j,\hH_i\}\Bigg[H\Bigg(\frac{c_{l,i}P_i\Omega}{2m\sigma_j^2}\Bigg)\Bigg]^m\\ &\times\sum_{r=0}^{m-1}\binom{m-1+r}{r}\Bigg[1-H\Bigg(\frac{c_{l,i}P_i\Omega}{2m\sigma_j^2}\Bigg)\Bigg]^r
\end{split}
\normalsize
\end{equation}
where
\begin{equation}
\small
\begin{split}
H(x)=\frac{1}{2}\Bigg(1-\sqrt{\frac{x}{1+x}}\Bigg) \hspace{1cm}x \geq 0.
\end{split}
\normalsize
\end{equation}
In a similar fashion, the BER of LP bits for integer values of $m$ is given in (\ref{eq:BER_LP_averaged_fading_int}) on the next page.
\begin{figure*}
\small
\begin{equation}
\begin{split} \label{eq:BER_LP_averaged_fading_int}
\hspace{-0.2cm}P_{\text{LP}}(\bP) \!=\!\frac{1}{2}\!\!\sum_{i,j=0}^{1}\!\!\Pr\{\mH_j,\hH_i\}\Bigg\{2\Bigg[\!H\Bigg(\frac{\beta_{1,i}P_i\Omega}{2m\sigma_j^2}\Bigg)\Bigg]^m\sum_{r=0}^{m-1}\!\!\binom{m\!-\!1\!+\!r}{r}\Bigg[1\!-\!H\Bigg(\frac{\beta_{1,i}P_i\Omega}{2m\sigma_j^2}\Bigg)\!\Bigg]^r\!\!\!+&\!\Bigg[H\Bigg(\frac{\beta_{2,i}P_i\Omega}{2m\sigma_j^2}\Bigg)\Bigg]^m\!\sum_{r=0}^{m-1}\!\!\binom{m\!-\!1\!+\!r}{r}\Bigg[1\!-\!H\Bigg(\frac{\beta_{2,i}P_i\Omega}{2m\sigma_j^2}\Bigg)\Bigg]^r\\&\hspace{-1.4cm}-\Bigg[H\Bigg(\frac{\beta_3P_i\Omega}{2m\sigma_j^2}\Bigg)\Bigg]^m\sum_{r=0}^{m-1}\binom{m-1+r}{r}\Bigg[1-H\Bigg(\frac{\beta_3P_i\Omega}{2m\sigma_j^2}\Bigg)\Bigg]^r\Bigg\}
\end{split}
\end{equation}
\hrule
\end{figure*}

\section{Optimal Power Control} \label{sec:opt_power}
In this section, we characterize the optimal power control policies that minimize the weighted sum of BERs of HP bits and LP bits or its upper bound subject to peak/average transmit power and average interference power constraints, assuming the availability of either the instantaneous or statistical CSI of the transmission link and interference link at the secondary transmitter.
\subsection{Peak transmit and average interference power constraints}
In this subsection, we consider peak transmit and average interference power constraints being imposed on secondary transmissions.
\subsubsection{Perfect CSI of both transmission and interference links}
Here, we assume that the instantaneous values of the fading coefficients of the transmission link, $h$, and interference link, $g$, are perfectly known by the secondary transmitter. In this case, the optimal power control problem is given by
\begin{align}
\label{eq:Opt_Qavg_Ppeak_inst}
&\hspace{0.5cm}\min_{
\substack{P_0(h,g),P_1(h,g) }} \E\big\{\lambda P_{\text{HP}}(\bP,h) + (1-\lambda)P_{\text{LP}}(\bP,h)\big\} \\ \nonumber &\hspace{0.1cm}\text{subject to} \\ &\label{eq:peak_P0_constraint}P_0(h,g) \le P_{\pk} \\ &\label{eq:peak_P1_constraint}P_1(h,g) \le P_{\pk} \\ \label{eq:interference_power_constraint_inst}
&\E\{(1-P_{\nid})\,P_0(h,g) \,|g|^2 + P_{\nid} \,P_1(h,g) \, |g|^2\} \le Q_{\avg}
\end{align}
where $P_{\text{HP}}(\bP,h)$ and $P_{\text{LP}}(\bP,h)$ are instantaneous BER expressions for given fading coefficients $h$ and $g$, and $\lambda \in [0,1]$. Above, when $\lambda=1$ or $0$, the optimal power levels are chosen to minimize only the BER of HP bits or LP bits, respectively. In the case of $\lambda=1/2$, BER of HP bits and LP bits are equally weighed in the objective function to determine the optimal transmission power levels. Hence, the value of $\lambda$ can be adjusted to reflect the importance of the HP and LP bits. In (\ref{eq:peak_P0_constraint}) and (\ref{eq:peak_P1_constraint}), $P_{\pk}$ denotes the peak transmit power limit of the secondary transmitter due to hardware$/$battery constraints and in (\ref{eq:interference_power_constraint_inst}), $Q_{\avg}$ represents average interference power limit at the primary receiver, which is imposed to satisfy the long-term QoS requirements of the primary users. In addition, since instantaneous CSI is available at the secondary transmitter, the power levels $P^{(0)}(h,g)$ and $P^{(1)}(h,g)$ are functions of both $h$ and $g$.

Note that the objective function in (\ref{eq:Opt_Qavg_Ppeak_inst}), or in particular $P_{\text{LP}}(\bP,h)$, consists of a sum of Gaussian $Q$ functions with positive and negative weights. Therefore, the Hessian of the objective function is not necessarily positive semidefinite due to the sum of exponential functions with different positive and negative weights. On the other hand, by removing the negative-weighted $Q$ functions in (\ref{eq:BER_LP}), we can obtain an upper bound on the BER expression in the objective function. Now, being composed of only positive weighted sum of $Q$ functions that are convex for positive arguments, this upper bound is convex. Therefore, the minimization problem becomes convex with affine constraints in (\ref{eq:peak_P0_constraint}), (\ref{eq:peak_P1_constraint}) and (\ref{eq:interference_power_constraint_inst}). In the following result, we identify the optimal power control scheme that minimizes this upper bound.
\begin{Prop}\label{prop_1} The optimal power control policy that minimizes the BER upper bound under the constraints in (\ref{eq:peak_P0_constraint}), (\ref{eq:peak_P1_constraint}) and (\ref{eq:interference_power_constraint_inst}) is given by
\begin{align} \label{eq:P0_opt_QavgPpeak_ins}
P^{(0)}_{\text{opt}}(h,g)=\min\big(P_{\pk},P_0^*\big)\\  \label{eq:P1_opt_QavgPpeak_ins}
P^{(1)}_{\text{opt}}(h,g)=\min\big(P_{\pk},P_1^*\big)
\end{align}
where $P_0^*$ is solution to
\begin{equation} \label{eq:P0_sol}
\small
\begin{split}
\hspace{-0.2cm}\sum_{j,l=0}^{1}&\frac{P(\mH_j,\!\hH_0)}{4\sqrt{2\pi}}\!\Bigg\{\lambda\frac{\rme^{\frac{-c_{l,0}P_0^*|h|^2}{2\sigma_j^2}}}{\sqrt{\frac{\sigma_j^2 P_0^*}{c_{l,0}|h|^2}}}\!+\!(1\!-\!\lambda)\rho_l \frac{\rme^{\frac{-\beta_{l,0}P_0^*|h|^2}{2\sigma_j^2}}}{\sqrt{\frac{ \sigma_j^2P_0^*}{\beta_{l,0}|h|^2}}}\!\Bigg\}\!\!=\!\mu_1(1\!-\!P_{\nid})|g|^2
\end{split}
\normalsize
\end{equation}
and $P_1^*$ is solution to
\begin{equation}\label{eq:P1_sol}
\small
\begin{split}
\hspace{-0.1cm}\sum_{j,l=0}^{1}&\frac{P(\mH_j,\!\hH_1)}{4\sqrt{2\pi}}\Bigg\{\lambda\frac{\rme^{\frac{-c_{l,1}P_1^*|h|^2}{2\sigma_j^2}}}{\sqrt{\frac{\sigma_j^2 P_1^*}{c_{l,1}|h|^2}}}\!+\!(1\!-\!\lambda)\rho_l \frac{\rme^{\frac{-\beta_{l,1}P_1^*|h|^2}{2\sigma_j^2}}}{\sqrt{\frac{\sigma_j^2P_1^*}{\beta_{l,1}|h|^2}}}\Bigg\}\!=\!\mu_1P_{\nid} |g|^2.
\end{split}
\normalsize
\end{equation}
Above, $\rho_0=2$, $\rho_1=1$, and $\mu_1$ is the Lagrange multiplier, which can be determined by satisfying the average interference constraint in (\ref{eq:interference_power_constraint_inst}) with equality.
\end{Prop}
\emph{Proof:} See Appendix \ref{app:proof-prop1}.

The above expressions are strictly monotonically decreasing functions of $P_0^*$ and $P_1^*$, respectively. By taking the first derivate of the above expressions and analyzing the limits as $P_0^*$ and $P_1^*$ approach $0$ and $\infty$, respectively, it can be easily shown that there always exists unique solutions for $P_0^*$ and $P_1^*$ due to the strict monotonicity. The optimal power control algorithm for this scenario is given in Algorithm \ref{algorithm1}.

In the following result, we identify closed-form approximations for the power levels in a specific scenario.
\begin{Prop}\label{prop_2} At high SNRs, the optimal power control policy that minimizes the BER of HP bits, (i.e., when $\lambda=1$) under perfect sensing decision (i.e., when $P_{\nid}=1$ and $P_{\f}=0$) subject to the constraints (\ref{eq:peak_P0_constraint}), (\ref{eq:peak_P1_constraint}) and (\ref{eq:interference_power_constraint_inst}) can be approximated in closed-form as
\begin{align}
&\hspace{-0.1cm}P^{(0)}_{\text{opt}}(h,g)=P_{\pk}\\
\label{eq:P1_opt_perfect}
&\hspace{-0.1cm}P^{(1)}_{\text{opt}}(h,g)=\min\Bigg(P_{\pk},\frac{W_{0}\Big(\frac{(c_{1,1}|h|^2P(\mH_1))^2}{32 \pi ((\sigma_n^2+\sigma_w^2)\mu_1|g|^2)^2}\Big)}{\frac{c_{1,1}|h|^2}{\sigma_n^2+\sigma_w^2}}\Bigg)
\end{align}
where $W_{0}(.)$ represents the primary branch of the Lambert function \cite{corless}.
\end{Prop}
\emph{Proof:} See Appendix \ref{app:proof-prop2}.

\vspace{-0.2cm}
\begin{algorithm}[H]
    \caption{The optimal power control algorithm under the peak transmit power and average interference power constraints}
    \begin{algorithmic}[1]\label{algorithm1}
  \STATE Initialize $\epsilon > 0$, $t > 0$, $\mu_1^{(0)}=\mu_{1,\text{init}}$, $n=0$
      \REPEAT
       \STATE Solve $P_0^*$ and $P_1^*$ in (\ref{eq:P0_sol}) and (\ref{eq:P1_sol}), respectively by bisection search and then determine $P^{(0)}_{\text{opt}}(h,g)$ in (\ref{eq:P0_opt_QavgPpeak_ins}) and $P^{(1)}_{\text{opt}}(h,g)$ in (\ref{eq:P1_opt_QavgPpeak_ins}).
	\STATE Update $\mu_1$ using the projected subgradient method as follows
       	\STATE $\Resize{8cm}{\mu_1^{(n+1)}=\big(\mu_1^{(n)}+t\big(\E\{(1-P_{\nid})\,P_0(h,g) \,|g|^2 + P_{\nid} \,P_1(h,g) \, |g|^2\} -Q_{\avg}\big)\big)^+}$ where $(.)^{+}=\max(.,0)$
	\STATE $n \leftarrow n+1$
       \UNTIL{$\Resize{7cm}{|\mu_1^{(n)}(\E\{(1-P_{\nid})\,P_0(h,g) \,|g|^2 + P_{\nid} \,P_1(h,g) \, |g|^2\} -Q_{\avg})| \le \epsilon}$}
    \end{algorithmic}
  \end{algorithm}

\subsubsection{Perfect CSI of transmission link and imperfect CSI of interference link}
In this case, we assume the transmitter has imperfect CSI of the interference link fading coefficient $g$, which is expressed as $g = \hat{g} + \tg,$ where $\hat{g}$ is the estimate of the interference link and $\tg$ is the error in the estimate. It is assumed that $\hat{g}$ and $\tg$ are independent, circularly symmetric complex Gaussian distributed with mean zero and variances $\sigma_g^2-\sigma_e^2$ and $\sigma_e^2$, respectively. Thus, the average interference constraint can be written as
\begin{equation}
\small
\begin{split} \label{eq:interference_power_constraint_inst_Imper}
Q_{\avg} & \geq \E\{[(1-P_{\nid})\,P_0(h,\hat{g})+P_{\nid} \,P_1(h,\hat{g})] \,|g|^2\} \\
& = \E\{[(1-P_{\nid})\,P_0(h,\hat{g})+P_{\nid} \,P_1(h,\hat{g})] \,(|\hat{g}|^2+|\tg|^2)\} \\
& = \E\{[(1-P_{\nid})\,P_0(h,\hat{g})+P_{\nid} \,P_1(h,\hat{g})] \,(|\hat{g}|^2+\sigma_e^2)\}.
\end{split}
\normalsize
\end{equation}
Hence, the optimal power control problem is expressed as
\small
\begin{align}
\label{eq:Opt_Qavg_Ppeak_inst_Imper}
&\hspace{-0.4 cm}\min_{
\substack{P_0(h,\hat{g}),P_1(h,\hat{g}) }} \E\big\{\lambda P_{\text{HP}}(\bP,h,\hat{g}) + (1-\lambda)P_{\text{LP}}^{u}(\bP,h,\hat{g})\big\} \\ &\label{eq:peak_P0_constraint_Imper} \text{subject to} \hspace{0.2cm}P_0(h,\hat{g}) \le P_{\pk},~ P_1(h,\hat{g}) \le P_{\pk} \\ \label{eq:interference_power_constraint_inst1_Imper}
&\E\{[(1\!-\!P_{\nid})\,P_0(h,\hat{g})\!+\!P_{\nid} \,P_1(h,\hat{g})] \,(|\hat{g}|^2+\sigma_e^2)\} \le Q_{\avg}
\end{align}
\normalsize
where $P_{\text{HP}}(\bP,h,\hat{g})$ and $P_{\text{LP}}^{u}(\bP,h,\hat{g})$ are the instantaneous BER expressions for given fading coefficients $h$ and $\hat{g}$.
In this setting, the optimal power control scheme is determined as follows:
\begin{Prop}\label{prop_3} The optimal power control scheme subject to the constraints in (\ref{eq:peak_P0_constraint_Imper}) and  (\ref{eq:interference_power_constraint_inst1_Imper}) under imperfect CSI of the interference link is given by
\small
\begin{align} \label{eq:opt_power_P0_PpeakQavg_Imper}
P^{(0)}_{\text{opt}}(h,\hat{g})&=\min(P_0^*(h,\hat{g}),P_{\pk}),\\ \label{eq:opt_power_P1_PpeakQavg_Imper}P^{(1)}_{\text{opt}}(h,\hat{g})&=\min(P_1^*(h,\hat{g}),P_{\pk})
\end{align}
\normalsize
where $P_0^*$ and $P_1^*$ are solutions to the following equations, respectively:
\small
\begin{align} \nonumber
&\hspace{-0.3cm}\sum_{j,l=0}^{1}\!\!\frac{P(\mH_j,\!\hH_0)}{4\sqrt{2\pi}}\!\Bigg\{\lambda\frac{\rme^{\frac{-c_{l,0}P_0^*|h|^2}{2\sigma_j^2}}}{\sqrt{\frac{\sigma_j^2 P_0^*}{c_{l,0}|h|^2}}}\!+\!(1\!-\!\lambda)\rho_l \frac{\rme^{\frac{-\beta_{l,0}P_0^*|h|^2}{2\sigma_j^2}}}{\sqrt{\frac{\sigma_j^2 P_0^*}{\beta_{l,0}|h|^2}}}\!\Bigg\}\\ \label{eq:P0_sol_avg_Imper}&\hspace{2cm}\!\!=\!\mu_1(1\!-\!P_{\nid})(|\hat{g}|^2+\sigma_e^2), \\ \nonumber
&\hspace{-0.3cm}\sum_{j,l=0}^{1}\!\!\frac{P(\mH_j,\!\hH_1)}{4\sqrt{2\pi}}\Bigg\{\lambda\frac{\rme^{\frac{-c_{l,1}P_1^*|h|^2}{2\sigma_j^2}}}{\sqrt{\frac{\sigma_j^2 P_1^*}{c_{l,1}|h|^2}}}\!+\!(1\!-\!\lambda)\rho_l \frac{\rme^{\frac{-\beta_{l,1}P_1^*|h|^2}{2\sigma_j^2}}}{\sqrt{\frac{\sigma_j^2 P_1^*}{\beta_{l,1} |h|^2}}}\Bigg\}\\ \label{eq:P1_sol_avg_Imper} &\hspace{2cm}\!=\!\mu_1P_{\nid}(|\hat{g}|^2+\sigma_e^2),
\normalsize
\end{align}
\normalsize
where $\mu_1$ is the Lagrange multiplier associated with the average interference power constraints in (\ref{eq:interference_power_constraint_inst1_Imper}).
\end{Prop}
The proof of Proposition \ref{prop_3} is similar to that of Proposition \ref{prop_1}, and hence it is omitted for brevity.

\subsubsection{Statistical CSI of both transmission and interference links}\label{subsec:Qavg_Ppeak_cons} Different from the previous subsections where the knowledge (or the estimate) of the instantaneous values of the fading coefficients is available at the secondary transmitter, the secondary transmitter in this case is assumed to know only the statistics of the transmission and interference links, (i.e., only the distributions of the fading coefficients are known). Hence, the optimal power levels are no longer functions of $h$ and $g$ (or $\hat{g}$). Under this assumption, we can formulate the optimization problem as follows:
\begin{align}
\small
\label{eq:Opt_Qavg_Ppeak_cons}
&\hspace{0.3cm}\min_{
\substack{P_0, P_1}} \lambda P_{\text{HP}}(\bP) + (1-\lambda)P_{\text{LP}}(\bP)\\ \nonumber &\hspace{-1cm}\text{subject to} \\ &\hspace{-1cm}\label{eq:peak_P0_constraint_statistical}P_0 \le P_{\pk}, \hspace{0.1cm} P_1\le P_{\pk}\\
&\hspace{-1cm}\label{eq:interference_power_constraint_inst1}(1-P_{\nid})\,P_0 \,\E\{|g|^2\} + P_{\nid} \,P_1 \, \E\{|g|^2\} \le Q_{\avg}
\end{align}
%
where $P_{\text{HP}}(\bP)$ and $P_{\text{LP}}(\bP)$ are closed-form expressions of the average BER over Nakagami-$m$ fading, given in (\ref{eq:BER_HP_averaged_fading}) and (\ref{eq:BER_LP_averaged_fading}), respectively. We solve (\ref{eq:Opt_Qavg_Ppeak_cons}) exactly by performing an exhaustive search, which has low complexity due to being performed over a one-dimensional bounded line which defines the boundary of the region of feasible power pairs $(P_0, P_1)$ satisfying (\ref{eq:peak_P0_constraint_statistical}) and (\ref{eq:interference_power_constraint_inst1}). Additionally, as we describe in the previous subsection, if a convex upper bound on error rates is obtained using a similar approach, convex optimization tools can be employed to find the optimal power levels, $P^{(0)}_{\text{opt}}$ and $P^{(1)}_{\text{opt}}$, that minimize this upper bound.

\subsection{Average transmit and average interference power constraints}
Now, we consider the presence of average transmit and average interference power constraints. We again address the cases of instantaneous and statistical CSI.
\subsubsection{Perfect CSI of both transmission and interference links}
In this case, the optimization problem subject to average transmit power and average interference power constraints is formulated as follows:
\begin{align}
\label{eq:Opt_Qavg_Ppavg_inst}
&\min_{
\substack{P_0(h,g),P_1(h,g)}} \E\big\{\lambda P_{\text{HP}}(\bP,h) + (1-\lambda)P_{\text{LP}}(\bP,h)\big\} \\ \nonumber &\hspace{0.1cm}\text{subject to}\\ &\hspace{0.1cm}\label{eq:Pavg_constraint1}\E\{P(\hH_0)\,P_0(h,g)  + P(\hH_1) P_1(h,g)\} \le P_{\avg}\\ \label{eq:interference_power_constraint_inst2}
&\hspace{0.1cm}\E\{(1-P_{\nid})\,P_0(h,g) \,|g|^2 + P_{\nid} \,P_1(h,g) \, |g|^2\} \le Q_{\avg}
\end{align}
where $P_{\avg}$ denotes the average transmit power limit at the secondary transmitter. Similarly as in the previous subsection, we again consider an upper bound on the BER in the objective function. Under these constraints, the optimal power control scheme is determined as follows:
\begin{Prop}\label{prop_4} The optimal power control policy that minimizes the BER upper bound under the constraints in (\ref{eq:Pavg_constraint1}) and (\ref{eq:interference_power_constraint_inst2}) is obtained as
\begin{align}
P^{(0)}_{\text{opt}}=P_0^*, \hspace{0.1cm} P^{(1)}_{\text{opt}}=P_1^*
\end{align}
where $P_0^*$ and $P_1^*$ are solutions to the following equations, respectively:

\small
\begin{align} \nonumber
&\hspace{-0.3cm}\sum_{j,l=0}^{1}\!\!\frac{P(\mH_j,\!\hH_0)}{4\sqrt{2\pi}}\!\Bigg\{\lambda\frac{\rme^{\frac{-c_{l,0}P_0^*|h|^2}{2\sigma_j^2}}}{\sqrt{\frac{\sigma_j^2 P_0^*}{c_{l,0}|h|^2}}}\!+\!(1\!-\!\lambda)\rho_l \frac{\rme^{\frac{-\beta_{l,0}P_0^*|h|^2}{2\sigma_j^2}}}{\sqrt{\frac{ \sigma_j^2P_0^*}{\beta_{l,0}|h|^2}}}\!\Bigg\}\!\!=\!\mu_1(1\!-\!P_{\nid})|g|^2 \\ \label{eq:P0_sol_avg}&\hspace{7.2cm}+\mu_2 P(\hH_0)
\end{align}
\begin{align} \nonumber
&\hspace{-0.3cm}\sum_{j,l=0}^{1}\!\!\frac{P(\mH_j,\!\hH_1)}{4\sqrt{2\pi}}\Bigg\{\lambda\frac{\rme^{\frac{-c_{l,1}P_1^*|h|^2}{2\sigma_j^2}}}{\sqrt{\frac{\sigma_j^2 P_1^*}{c_{l,1}|h|^2}}}\!+\!(1\!-\!\lambda)\rho_l \frac{\rme^{\frac{-\beta_{l,1}P_1^*|h|^2}{2\sigma_j^2}}}{\sqrt{\frac{ \sigma_j^2P_1^*}{\beta_{l,1}|h|^2}}}\Bigg\}\!=\!\mu_1P_{\nid} |g|^2\\ \label{eq:P1_sol_avg} &\hspace{7.2cm}+\mu_2 P(\hH_1)
\end{align}
\normalsize
where $\mu_1$ and $\mu_2$ are the Lagrange multipliers associated with the average transmit power and average interference power constraints in (\ref{eq:Pavg_constraint1}) and (\ref{eq:interference_power_constraint_inst2}), respectively.
\end{Prop}
Proposition \ref{prop_4} is proved similarly as Proposition \ref{prop_1}, and hence we omit the proof for brevity. Below, we provide Algorithm \ref{algorithm2} for obtaining the optimal power levels.

\begin{algorithm}[H]
    \caption{The optimal power control algorithm under average transmit power and average interference power constraints}
    \begin{algorithmic}[1]\label{algorithm2}
  \STATE Initialize $\epsilon, t_1,t_2 > 0$, $\mu_1^{(0)}=\mu_{1,\text{init}}$, $\mu_2^{(0)}=\mu_{2,\text{init}}$, $n=0$
      \REPEAT
       \STATE Solve $P_0^*$ and $P_1^*$ in (\ref{eq:P0_sol_avg}) and (\ref{eq:P1_sol_avg}), respectively by bisection search.
	\STATE Update $\mu_1$ and $\mu_2$ using the projected subgradient method as follows
       	\STATE $\Resize{8cm}{\mu_1^{(n+1)}=\big(\mu_1^{(n)}+t_1\big(\E\{(1-P_{\nid})\,P_0(h,g) \,|g|^2 + P_{\nid} \,P_1(h,g) \, |g|^2\} -Q_{\avg}\big)\big)^+}$
        \STATE $\Resize{8cm}{\mu_2^{(n+1)}=\big(\mu_2^{(n)}+t_2\big(\E\{P(\mH_0)\,P_0(h,g)  + P(\mH_1) P_1(h,g)\}- P_{\avg}\big)\big)^+}$
	\STATE $n \leftarrow n+1$
       \UNTIL{$\Resize{7cm}{|\mu_1^{(n)}(\E\{(1-P_{\nid})\,P_0(h,g) \,|g|^2 + P_{\nid} \,P_1(h,g) \, |g|^2\} -Q_{\avg})| \le \epsilon}$ and $\Resize{7cm}{|\mu_2^{(n)}(\E\{P(\mH_0)\,P_0(h,g)  + P(\mH_1) P_1(h,g)\}- P_{\avg})| \le \epsilon}$}
    \end{algorithmic}
  \end{algorithm}

With slight change in Algorithm \ref{algorithm2}, we can incorporate a retransmission mechanism into the power control scheme. In particular, we can assume that the transmitter is silent and therefore does not send a packet if the channel fading coefficient is less than a certain threshold, e.g., during deep fading, which lowers the energy consumption. Hence, the power is set to zero if the channel fading coefficient is below this threshold in Algorithm \ref{algorithm2} and the corresponding Lagrange multipliers satisfying the constraints are found. In that case, more power is allocated for favorable channel conditions since the transmitter does not consume power when the channel undergoes deep fading.

Next, we discuss a special case for which we again have closed-form approximations for the optimal power levels.

\begin{Prop}\label{prop_5} At high SNRs, the optimal power control policy minimizing the BER of HP bits, (i.e., when $\lambda=1$) in the presence of perfect sensing results under the average transmit power constraint in (\ref{eq:Pavg_constraint1}) and average interference power constraint in (\ref{eq:interference_power_constraint_inst2}) can be approximated in closed-form as
\begin{align}
\small
&\hspace{-0.1cm}P^{(0)}_{\text{opt}}(h,g)=\frac{\sigma_n^2}{c_{1,0}\,|h|^2}W_{0}\Bigg(\frac{(c_{1,0}|h|^2P(\mH_0))^2}{32 \pi (\sigma_n^2\,\mu_2P(\mH_0))^2}\Bigg)\\
&\hspace{-0.1cm}P^{(1)}_{\text{opt}}(h,g)=\frac{W_{0}\Bigg(\frac{(c_{1,1}\,|h|^2P(\mH_1))^2}{32 \pi \big((\sigma_n^2+\sigma_w^2)(\mu_1|g|^2+\mu_2P(\mH_1))\big)^2}\Bigg)}{\frac{c_{1,1}|h|^2}{\sigma_n^2+\sigma_w^2}}.
\end{align}
\normalsize
\end{Prop}
Since the proof of Proposition \ref{prop_5} is similar to that of Proposition \ref{prop_2}, it is omitted for brevity.

\subsubsection{Perfect CSI of transmission link and imperfect CSI of interference link}
In this case, the optimal power control problem is expressed as
\small
\begin{align}
\label{eq:Opt_Qavg_Ppavg_inst_Imper}
&\hspace{-0.4 cm}\min_{
\substack{P_0(h,\hat{g}),P_1(h,\hat{g}) }} \E\big\{\lambda P_{\text{HP}}(\bP,h,\hat{g}) + (1-\lambda)P_{\text{LP}}^{u}(\bP,h,\hat{g})\big\} \\ \nonumber &\hspace{-0.5cm}\label{eq:Pavg_constraint1_Imper} \text{subject to} \\
&\hspace{-0.5cm}\E\{P(\hH_0)\,P_0(h,\hat{g})  + P(\hH_1) P_1(h,\hat{g})\} \le P_{\avg}\\ \label{eq:interference_power_constraint_inst2_Imper}
&\hspace{-0.5cm}\E\{[(1\!-\!P_{\nid})\,P_0(h,\hat{g})\!+\!P_{\nid} \,P_1(h,\hat{g})] \,(|\hat{g}|^2+\sigma_e^2)\} \le Q_{\avg}.
\end{align}
\normalsize
Under the above constraints, the optimal power control scheme is determined in the following:
\begin{Prop}\label{prop_6} The optimal power control scheme subject to average transmit power constraint in (\ref{eq:Pavg_constraint1_Imper}) and average interference power constraint in (\ref{eq:interference_power_constraint_inst2_Imper}) is given by
\small
\begin{align}
P^{(0)}_{\text{opt}}(h,\hat{g})=P_0^*, \hspace{0.1cm} P^{(1)}_{\text{opt}}(h,\hat{g})=P_1^*
\end{align}
\normalsize
where $P_0^*$ and $P_1^*$ are solutions to the following equations, respectively:
\small
\begin{align} \nonumber
&\hspace{-0.3cm}\sum_{j,l=0}^{1}\!\!\frac{P(\mH_j,\!\hH_0)}{4\sqrt{2\pi}}\!\Bigg\{\lambda\frac{\rme^{\frac{-c_{l,0}P_0^*|h|^2}{2\sigma_j^2}}}{\sqrt{\frac{\sigma_j^2 P_0^*}{c_{l,0}|h|^2}}}\!+\!(1\!-\!\lambda)\rho_l \frac{\rme^{\frac{-\beta_{l,0}P_0^*|h|^2}{2\sigma_j^2}}}{\sqrt{\frac{\sigma_j^2 P_0^*}{\beta_{l,0}|h|^2}}}\!\Bigg\}\\ \label{eq:P0_sol_avg_Imper}&\hspace{2cm}\!\!=\!\mu_1(1\!-\!P_{\nid})(|\hat{g}|^2+\sigma_e^2) +\mu_2 P(\hH_0), \\ \nonumber
&\hspace{-0.3cm}\sum_{j,l=0}^{1}\!\!\frac{P(\mH_j,\!\hH_1)}{4\sqrt{2\pi}}\Bigg\{\lambda\frac{\rme^{\frac{-c_{l,1}P_1^*|h|^2}{2\sigma_j^2}}}{\sqrt{\frac{\sigma_j^2 P_1^*}{c_{l,1}|h|^2}}}\!+\!(1\!-\!\lambda)\rho_l \frac{\rme^{\frac{-\beta_{l,1}P_1^*|h|^2}{2\sigma_j^2}}}{\sqrt{\frac{\sigma_j^2 P_1^*}{\beta_{l,1} |h|^2}}}\Bigg\}\\ \label{eq:P1_sol_avg_Imper} &\hspace{2cm}\!=\!\mu_1P_{\nid}(|\hat{g}|^2+\sigma_e^2)+\mu_2 P(\hH_1),
\normalsize
\end{align}
\normalsize
where $\mu_1$ and $\mu_2$ are the Lagrange multipliers associated with the average transmit power and average interference power constraints in (\ref{eq:Pavg_constraint1_Imper}) and (\ref{eq:interference_power_constraint_inst2_Imper}), respectively.
\end{Prop}
The proof of Proposition \ref{prop_6} is similar to that of Proposition \ref{prop_1}, and therefore, we omitted the proof for brevity.

\subsubsection{Statistical CSI of both transmission and interference links} In this case, the optimal power control problem is given by
\begin{align}
\label{eq:Opt_Qavg_Ppavg_cons}
&\hspace{0.3cm}\min_{
\substack{P_0, P_1}} \lambda P_{\text{HP}}(\bP) + (1-\lambda)P_{\text{LP}}(\bP) \\ \nonumber &\hspace{-1cm}\text{subject to}\\ &\hspace{-1cm}\label{eq:Pavg_constraint_cons} P(\hH_0)\,P_0  + P(\hH_1) P_1 \le P_{\avg} \\ \label{eq:interference_power_constraint_cons1}
&\hspace{-1cm}(1-P_{\nid})\,P_0 \,\E\{|g|^2\} + P_{\nid} \,P_1 \, \E\{|g|^2\} \le Q_{\avg}.
\end{align}
Similarly as in Section \ref{subsec:Qavg_Ppeak_cons}, transmission power levels, $P^{(0)}_{\text{opt}}$ and $P^{(1)}_{\text{opt}}$ can be obtained numerically by either exhaustive search or by employing convex optimization tools if upper bounds on error rates are considered as the objective function.

\begin{figure*}
\centering
\begin{subfigure}[b]{0.32\textwidth}
\centering
\includegraphics[width=\textwidth]{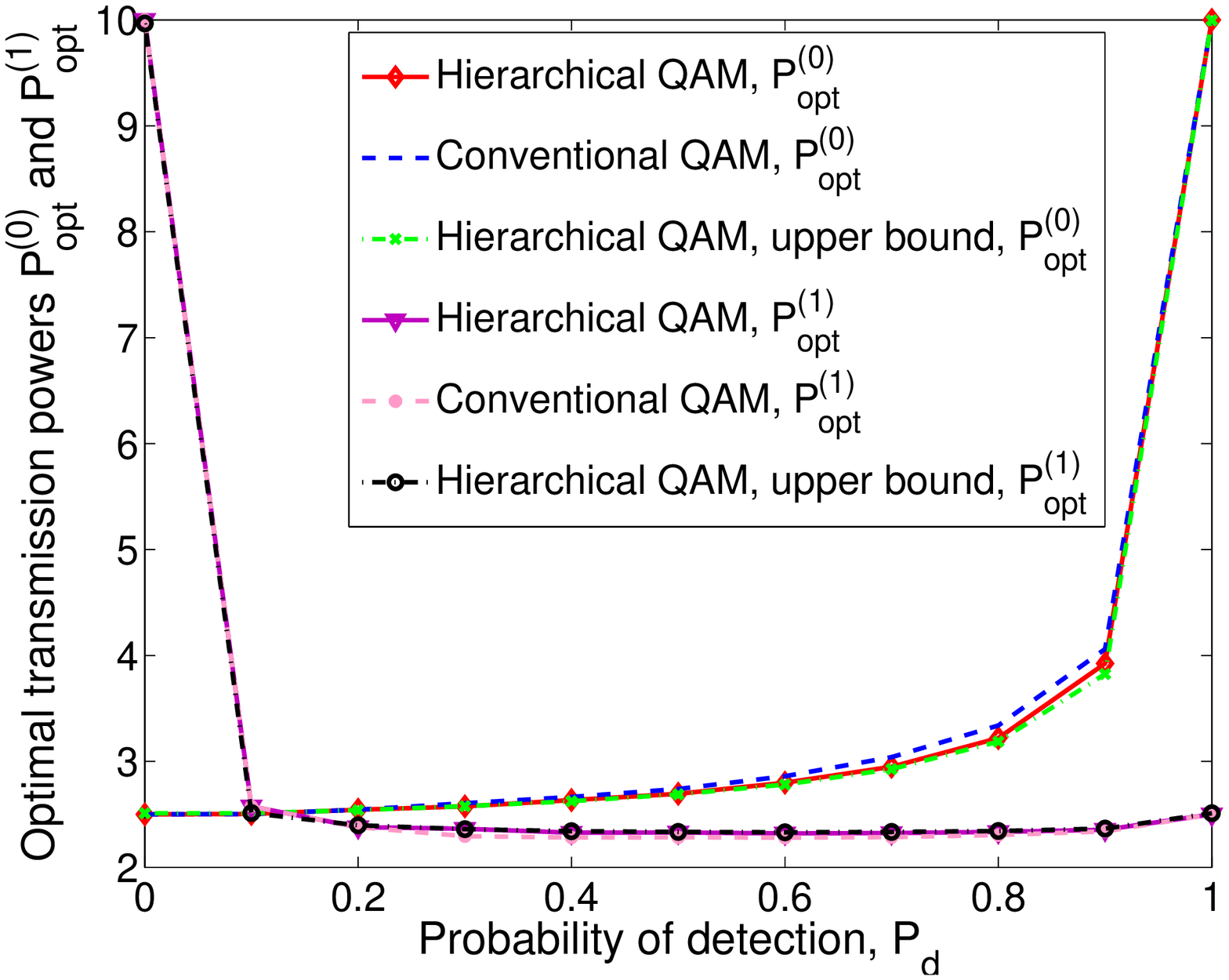}
\caption{$P_0$ and $P_1$ vs. $P_{\nid}$}
\end{subfigure}
\begin{subfigure}[b]{0.32\textwidth}
\centering
\includegraphics[width=\textwidth]{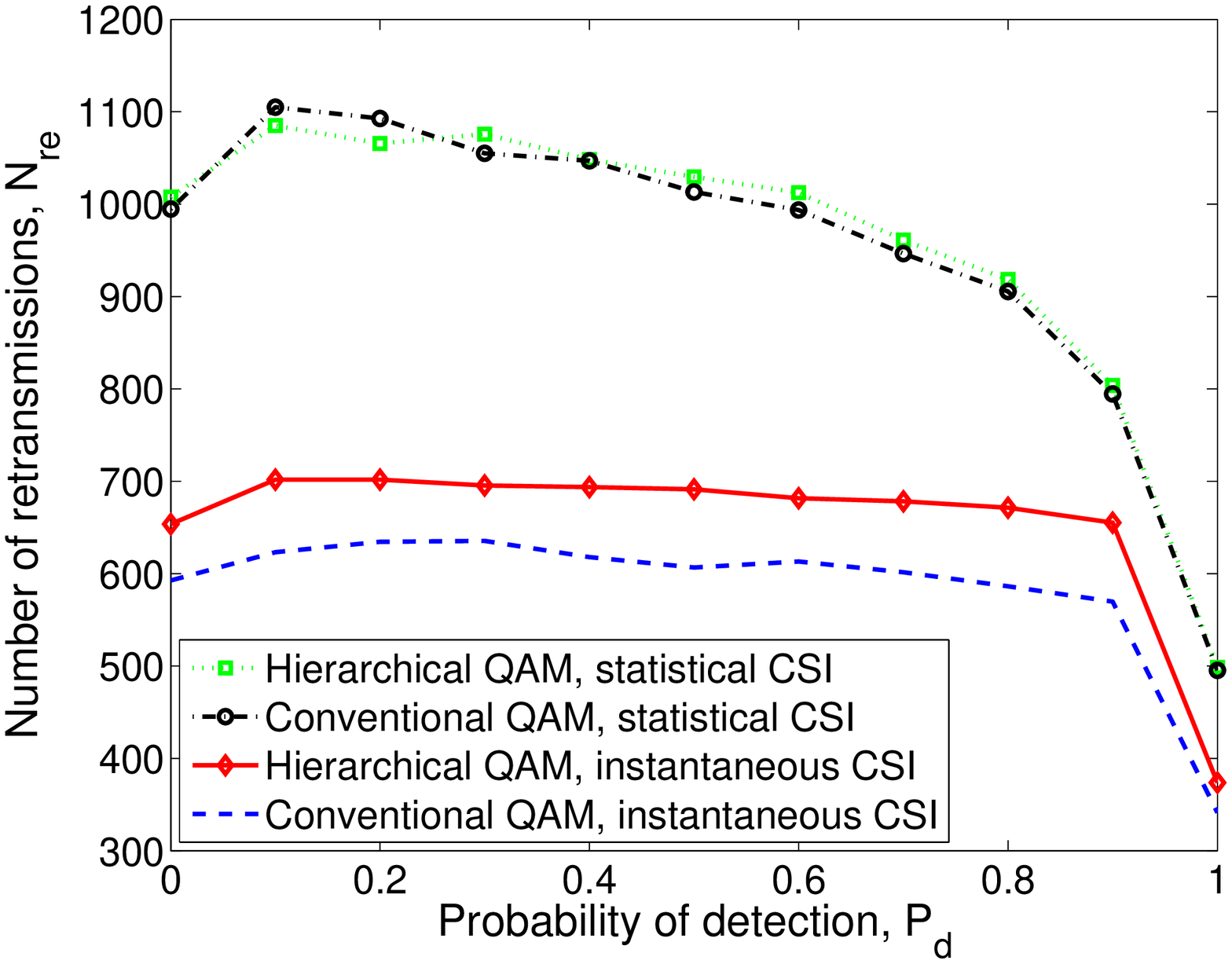}
\caption{$N_{re}$ vs. $P_{\nid}$}
\end{subfigure}
\begin{subfigure}[b]{0.32\textwidth}
\centering
\includegraphics[width=\textwidth]{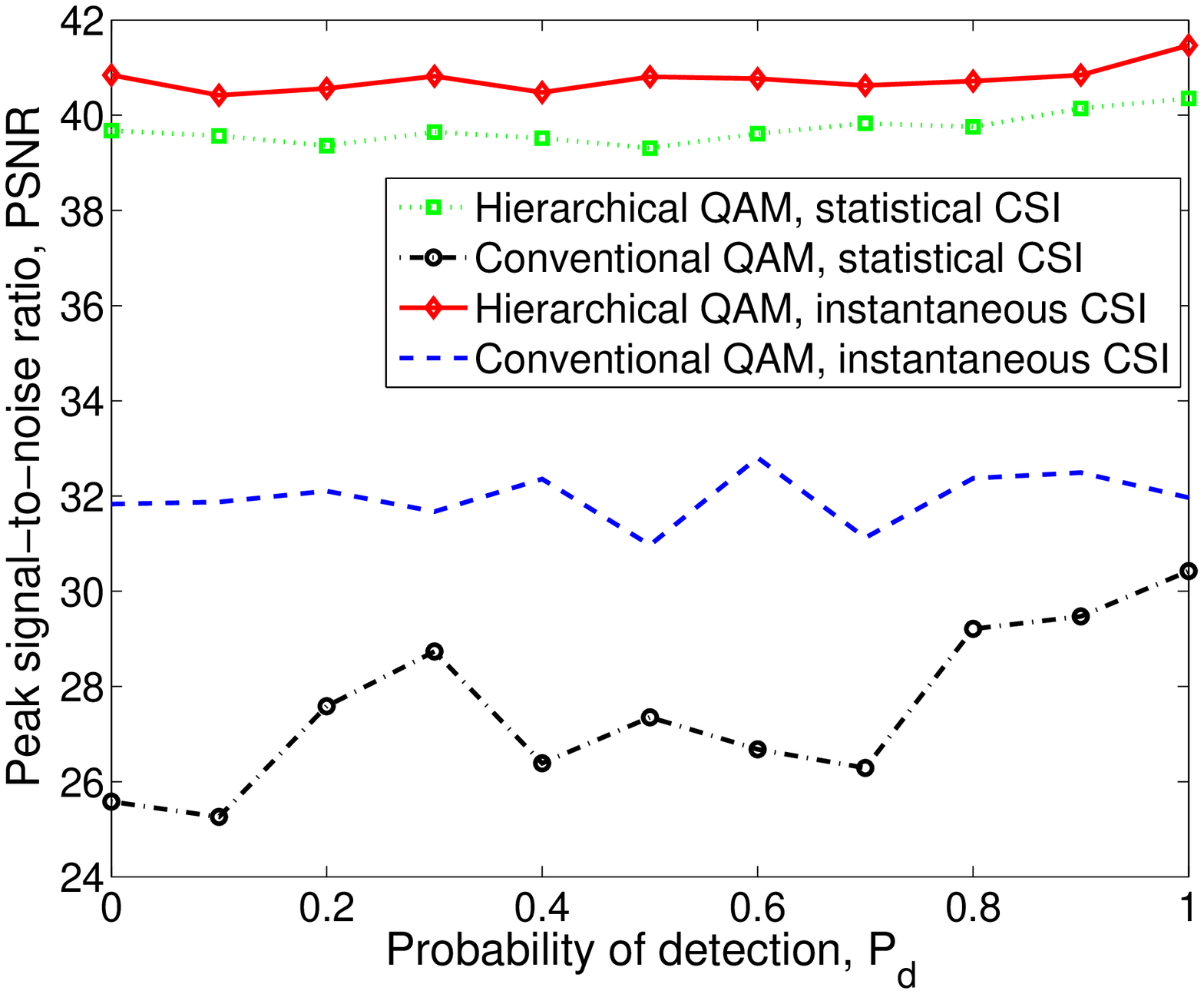}
\caption{PSNR vs. $P_{\nid}$}
\end{subfigure}
\caption{\small{(a) Optimal transmission powers $P_0$ and $P_1$ vs. $P_{\nid}$; (b) number of retransmissions, $N_{re}$ vs. $P_{\nid}$; (c) Peak signal-to-noise ratio, PSNR vs. $P_{\nid}$.}}\label{fig:P0P1_Nre_PSNR_vsPd}
\end{figure*}

\begin{figure*}
\centering
\begin{subfigure}[b]{0.32\textwidth}
\centering
\includegraphics[width=\textwidth]{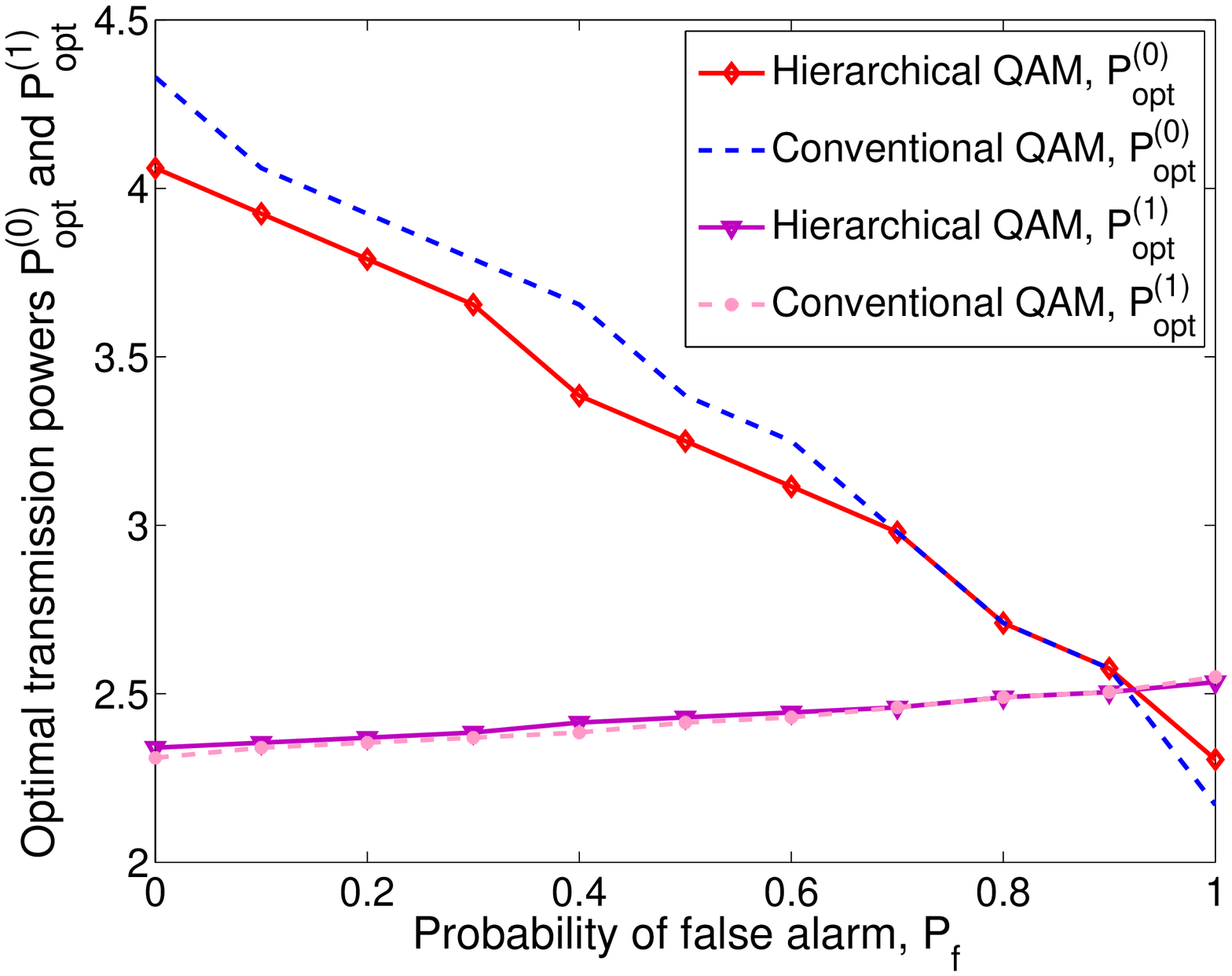}
\caption{$P_0$ and $P_1$ vs. $P_{\f}$}
\end{subfigure}
\begin{subfigure}[b]{0.32\textwidth}
\centering
\includegraphics[width=\textwidth]{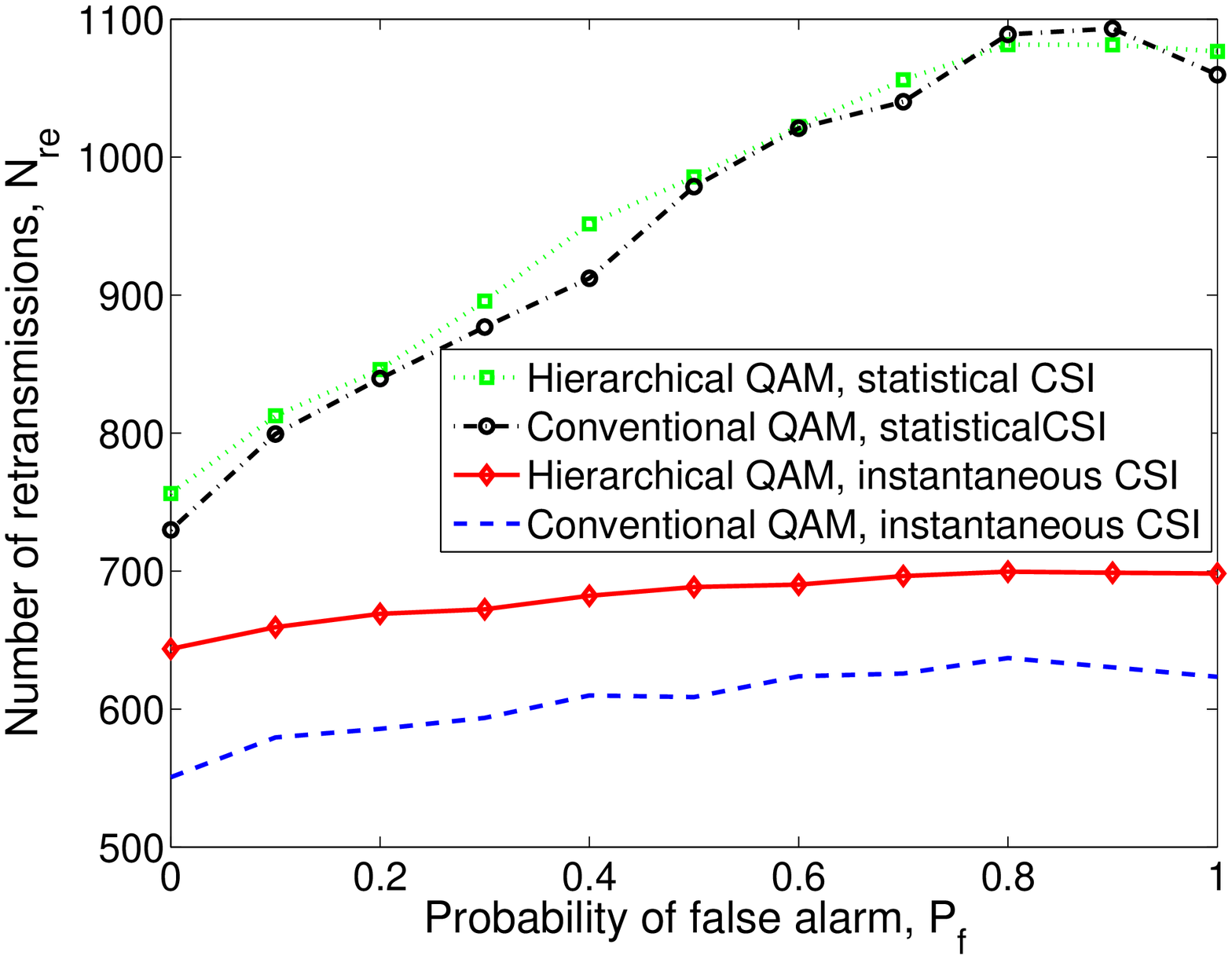}
\caption{$N_{re}$ vs. $P_{\f}$}
\end{subfigure}
\begin{subfigure}[b]{0.32\textwidth}
\centering
\includegraphics[width=\textwidth]{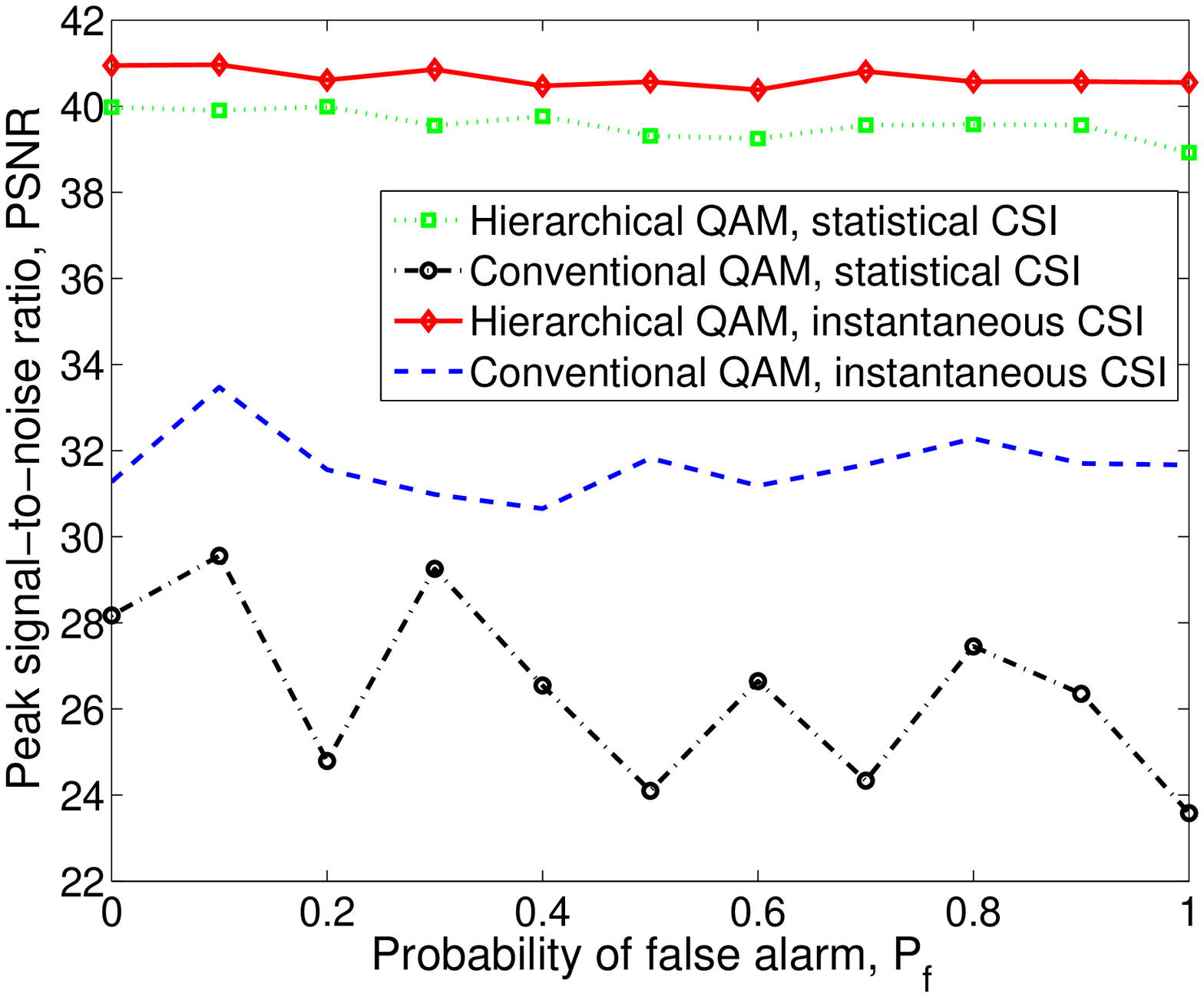}
\caption{PSNR vs. $P_{\f}$}
\end{subfigure}
\caption{\small{(a) Optimal transmission powers $P_0$ and $P_1$ vs. $P_{\f}$; (b) number of retransmissions, $N_{re}$ vs. $P_{\f}$; (c) Peak signal-to-noise ratio, PSNR vs. $P_{\f}$.}}\label{fig:P0P1_Nre_PSNR_vsPf}
\end{figure*}

\section{Numerical and Simulation Results}\label{sec:num_results}
In this section, we perform comprehensive numerical computations and simulations to evaluate the performance of multimedia transmissions of cognitive users with optimal power control and only imperfect sensing results under different severity levels of fading.

\subsection{Simulation Settings}
In the case of image transmission, test image is chosen as the gray-scale ``Lena'' and "Boat" images with size 512$\times$512 pixels. For video transmission, standard test video sequence ``Bus'' is used in the simulations. It is assumed that the noise variance is $\sigma_{n}^{2}=0.01$, the variance of the primary user signal is $\sigma_{\omega}^{2}=0.5$, the step size $t$ is set to $0.001$ and tolerance $\epsilon$ is chosen as $10^{-7}$. Prior probabilities of the primary users being active and inactive in the channel are set to 0.4 and 0.6, respectively, i.e., $\Pr\{\mH_1\}=0.4$ and $\Pr\{\mH_0\}=0.6$. Unless mentioned explicitly, we also assume that the channel power gains $|h|^2$ and $|g|^2$ follow exponential distributions with unit mean, threshold for retransmission $Thr$ is chosen as $1.8$, the peak transmit power constraint is $P_{\pk}=10$ dB, the average transmit power constraint is $P_{\avg}=10$ dB, and the average interference power constraint is $Q_{\avg}=4$ dB. In order to present average simulation results in the presence of randomly-varying fading, results of 50 simulations are averaged. 

PSNR is chosen as the performance metric to measure the quality of the reconstructed data. PSNR is defined for an 8-bit-pixel image of size $m$ by $n$ pixels as
\begin{equation}
\text{PSNR}=10\log_{10}\Bigg(\frac{255^2}{\frac{1}{mn}\sum_{i=0}^{m-1}\sum_{j=0}^{n-1}(S_{m,n}-\hat{S}_{m,n})^2}\Bigg)
\end{equation}
where $S_{m,n}$ and $\hat{S}_{m,n}$ denote the pixel intensity values of the original image and the reconstructed image, respectively.

\begin{table}[h]
\begin{center}
\vspace{-.2cm}
\caption{Peak signal-to-noise ratio, PSNR vs. $\lambda$.} \label{table_PSNR_lambda}
\resizebox{0.4\textwidth}{!}{
    \begin{tabular}{| c | c | c | c | c | c |}
    \hline
    $\lambda$ & 0.1 & 0.3 & 0.5 & 0.7 & 0.9  \\ \hline
    PSNR & 40.1425 & 40.1425 & 40.1425 & 39.8876 & 39.8876  \\ \hline
$P_{0,opt}$ & 2.809 & 2.803 & 2.796 & 2.789 & 2.778 \\  \hline
$P_{1,opt}$ & 2.384 & 2.387 & 2.390 & 2.393 & 2.398 \\
    \hline
    \end{tabular}}
\end{center}
\vspace{-.3cm}
\end{table}

In Table \ref{table_PSNR_lambda}, we have listed the PSNR values of the test image and the optimal transmission power levels, $P_{0,opt}$ and $P_{1,opt}$, for different values of the weight factor $\lambda$, which determines the contributions of the BERs of HP bits and LP bits in the objective function. Packets are assumed to be modulated by 16-HQAM with $\alpha_0 = \alpha_1 =1$. The results in the table are obtained based on the statistical CSI subject to the peak transmit power constraint $P_{\pk}$, and average interference power constraint $Q_{\avg}$. It is seen that changing the value of $\lambda$ does not have significant impact on the PSNR of the reconstructed image. The reason is that giving more or less weight to the BER of HP data in the objective function does not result in much difference in the optimal transmission power levels $P_{0,opt}$ and $P_{1,opt}$ as shown in Table \ref{table_PSNR_lambda}, which leads to only slight changes in the image quality. A similar trend is also observed when optimal power control with instantaneous CSI is applied. Therefore, for the rest of the simulations, we set $\lambda=0.5$.

\subsection{The impact of channel sensing performance on multimedia quality}
In this subsection, we analyze the effects of the probabilities of detection and false alarm on the transmission of image and video data in CR systems. For instance, our main observations in Figs. \ref{fig:P0P1_Nre_PSNR_vsPd} and \ref{fig:P0P1_Nre_PSNR_vsPf}, which we discuss in detail below, are that as the sensing reliability improves (i.e., detection probability increases or false probability diminishes), the number of retransmissions decreases drastically and PSNR values tend to slightly grow or stay stable. Additionally, employing HQAM instead of conventional QAM and having instantaneous CSI rather than statistical CSI all improve the multimedia quality as evidenced by higher PSNR levels.

More specifically, in Fig. \ref{fig:P0P1_Nre_PSNR_vsPd}, we display the optimal power levels (only for the statistical CSI case, obtained either by solving (\ref{eq:Opt_Qavg_Ppeak_cons}) through exhaustive search on the boundary of constraints or solving a convex optimization problem using the aforementioned upper bound on BER expressions) and number of retransmissions and PSNR values as a function of the probability of detection, $P_{\nid}$. Cognitive users employ either 16-HQAM with $\alpha_0=\alpha_1=1$ or conventional QAM subject to peak transmit power constraint, $P_{\pk}$, and average interference constraint, $Q_{\avg}$. As $P_{\nid}$ increases while keeping $P_{\f}$ fixed to $0.1$, we have more reliable sensing performance. In this case, the cognitive users transmit at higher power, $P_0$, in an idle-sensed channel. In particular, $P_0$ takes its maximum value $P_{\pk}$ when $P_{\nid}=1$. Since more reliable sensing enables the cognitive user to transmit at higher power level, the number of retransmissions decreases with increasing $P_{\nid}$ for both scenarios where power control is performed based on either the statistical CSI or instantaneous CSI. On the other hand, it is seen that PSNR performance, while showing a slight tendency to improve with increasing $P_{\nid}$, is relatively robust to variations in $P_{\nid}$, mainly due to the presence of the retransmission mechanism. In particular, we notice that approximately the same PSNR value can be attained in the presence of increased sensing uncertainty (i.e., lower $P_{\nid}$) at the cost of higher number of retransmissions under both scenarios\footnote{Instead, if no retransmissions are allowed or a certain upper bound on the number of retransmissions is imposed, PNSR increases as $P_{\nid}$ increases. Hence, we will have better image quality as the sensing performance improves.}. In the figure, it is also observed that HQAM gives better PSNR performance compared to conventional QAM since HP data is protected better in HQAM signaling. Notice that this improved performance is achieved interestingly with similar number of retransmission requests and at similar power levels. It is also seen that the difference between the optimal transmission power levels obtained by solving (\ref{eq:Opt_Qavg_Ppeak_cons}) exactly or using an upper bound on the objective function obtained by eliminating the $Q$ functions with negative weights is very small. Hence, we can conclude that the upper bound on BER expressions can effectively be used to determine the transmission power levels $P_0$ and $P_1$ by using standard convex optimization tools.

In Fig. \ref{fig:P0P1_Nre_PSNR_vsPf}, we plot the optimal power levels (only for the statistical CSI case, obtained either by solving (\ref{eq:Opt_Qavg_Ppeak_cons}) through exhaustive search on the boundary of constraints or solving a convex optimization problem using the aforementioned upper bound on BER expressions) and number of retransmissions and PSNR values as a function of the probability of false alarm, $P_{\f}$.
As $P_{\f}$ increases while keeping $P_{\nid}$ fixed at $0.9$, the cognitive users experience false alarm events more frequently. We notice that unless the false alarm probability $P_{\f}$ is close to 1, $P_1$ is generally smaller than $P_0$ in order to protect the primary users by limiting the interference in a busy-sensed channel. We also note that initially as $P_{\f}$ increases, cognitive secondary users more often misperceive an idle channel as busy and consequently transmit unnecessarily at the lower power level of $P_1$ instead of $P_0$. In addition, the optimal value of $P_0$ diminishes with increasing $P_{\f}$. As a result, as seen in Fig. \ref{fig:P0P1_Nre_PSNR_vsPf}(b), the number of retransmissions increases due to these low transmission power levels when $P_{\f}$ increases. When $P_{\f}$ is close to $1$, the number of retransmissions levels off and even slightly decreases as $P_1$ exceeds $P_0$. Again, PNSR quality does not get affected much with changing $P_{\f}$ due to the same reasoning explained in the discussion of the impact of $P_{\nid}$.  Also, hierarchical QAM again outperforms conventional QAM in terms of PSNR. Another important remark is that when instantaneous CSI is used to determine the optimal power levels, the secondary users obtain better image quality with smaller number of retransmissions compared to that attained by optimal power levels based only on the statistical CSI. More specifically, up to $6$ dB improvement in PSNR is achieved and the number of retransmissions is reduced by nearly half. We note that similar results are observed when average transmit power and average interference power constraints are imposed. However, we have not included the corresponding simulation results for the sake of brevity.

\begin{figure*}[ht]
\centering
\begin{subfigure}[b]{0.32\textwidth}
\centering
\includegraphics[width=\textwidth]{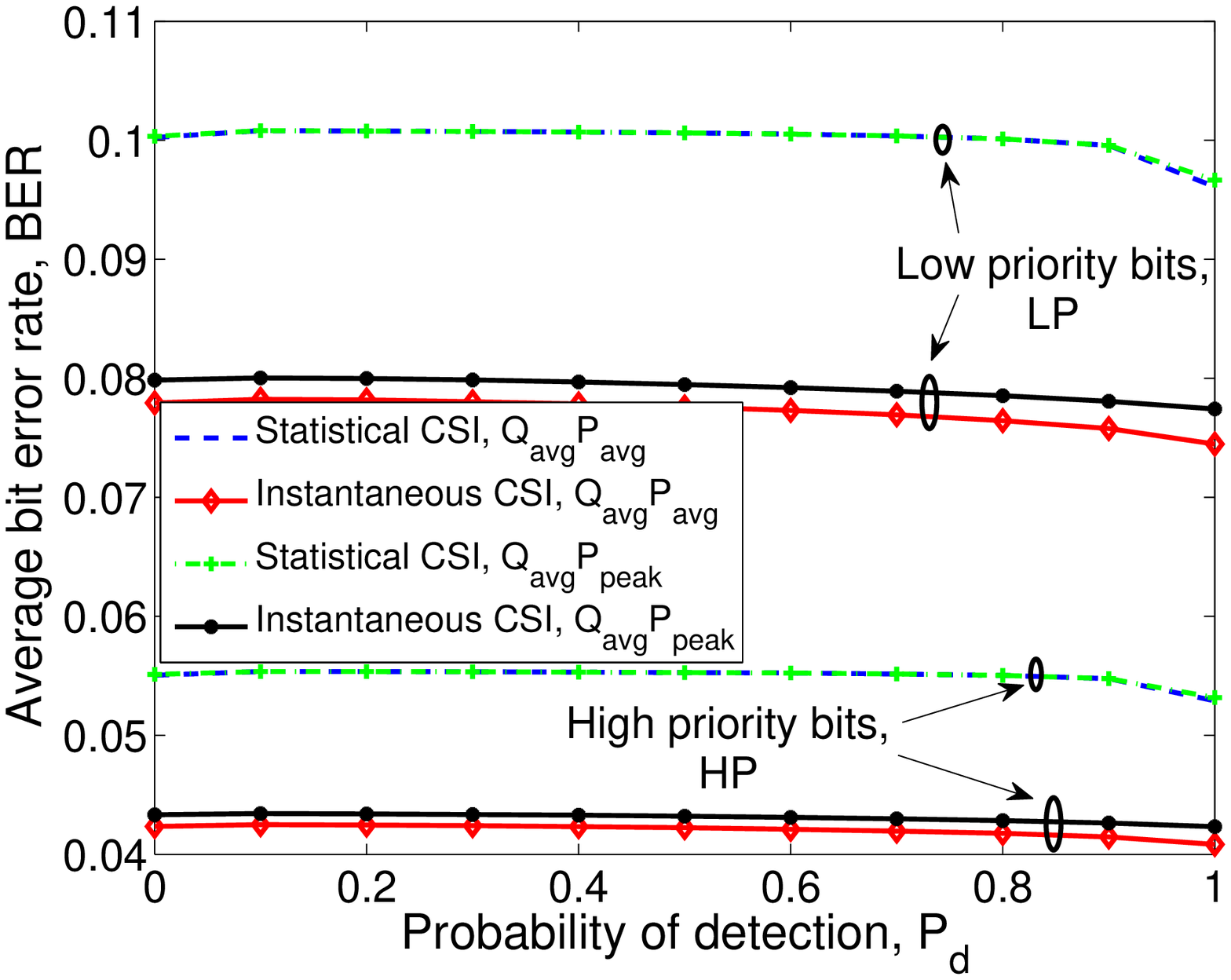}
\caption{Average BER vs. $P_{\nid}$}
\end{subfigure}
\begin{subfigure}[b]{0.32\textwidth}
\centering
\includegraphics[width=\textwidth]{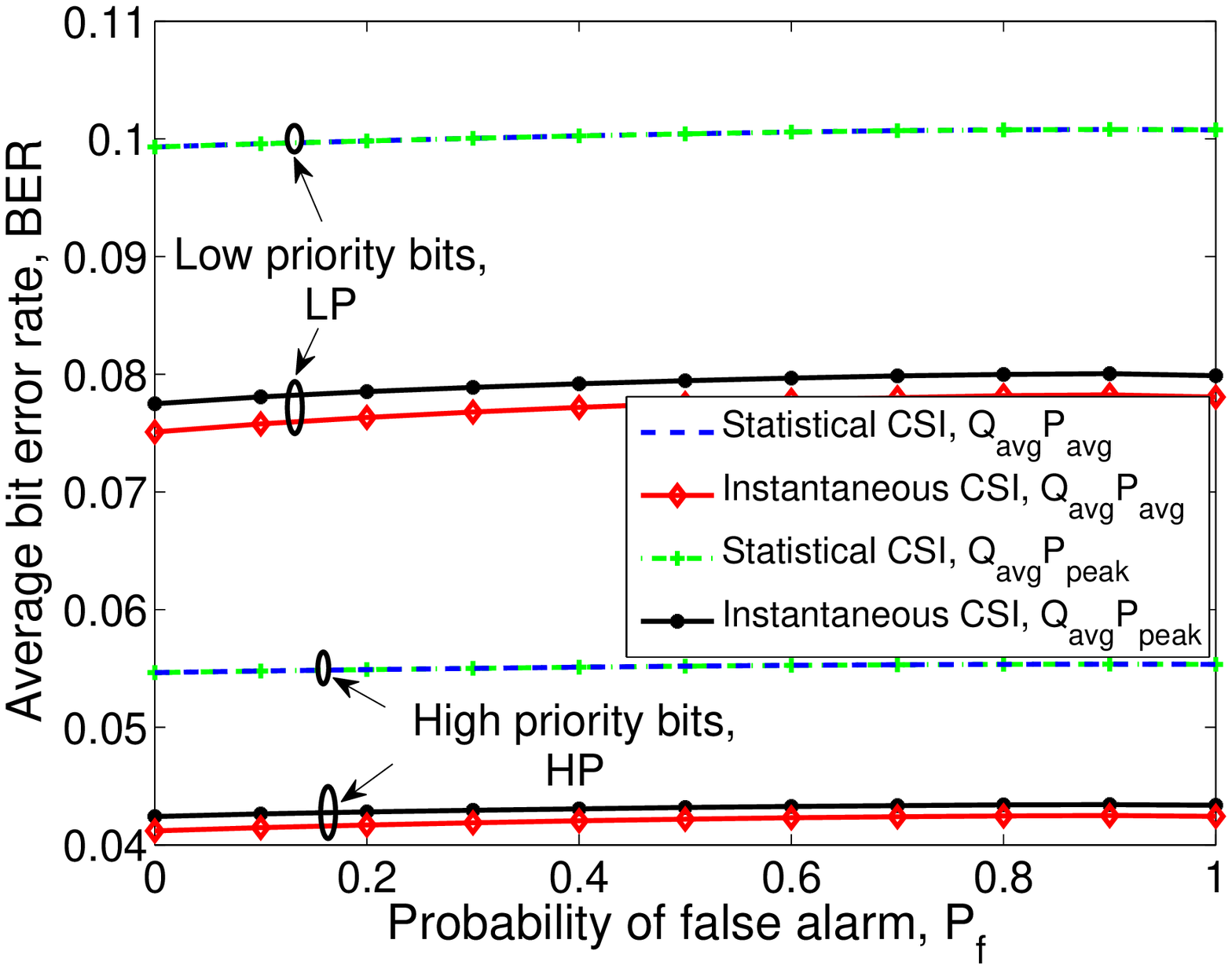}
\caption{Average BER vs. $P_{\f}$}
\end{subfigure}
\caption{\small{(a) Average BER vs. probability of detection, $P_{\nid}$; (b) Average BER vs. probability of false alarm, $P_{\f}$.}}\label{fig:BER_Pd_Pf}
\end{figure*}

In Fig. \ref{fig:BER_Pd_Pf}, we plot average BERs of HP bits and LP bits as a function of the detection probability, $P_{\nid}$, (left subfigure) and false alarm probability, $P_{\f}$ (right subfigure). We consider the cases in which either peak transmit power/average interference power constraints denoted by ($P_{\pk}, Q_{\avg}$) or average transmit power/average interference power constraints denoted by ($P_{\avg}, Q_{\avg}$) are imposed. Optimal power allocation is performed by assuming the availability of either instantaneous CSI or statistical CSI at the secondary transmitter.
In the left subfigure, as $P_{\nid}$ increases while keeping $P_{\f}$ fixed to $0.1$, average BERs of HP bits and LP bits decrease. In the right subfigure, where $P_{\nid}=0.9$, BER performance deteriorates with increasing $P_{\f}$ because of the same reasoning explained in the discussion of Fig. \ref{fig:P0P1_Nre_PSNR_vsPf}. It is also seen that power control with instantaneous CSI yields better BER performance than power allocation with statistical CSI. In addition, power control with instantaneous CSI under average transmit power constraint provides smaller BERs for both HP bits and LP bits compared to that attained under peak transmit power limitations since average transmit power constraint is more flexible than the peak transmit power constraint. In contrast, if power allocation based on statistical CSI is applied, BERs of HP bits are the same for all values of $P_{\f}$ and $P_{\nid}$ (except when $P_{\nid}=0$ or $1$) under both ($P_{\pk}, Q_{\avg}$) and ($P_{\avg}, Q_{\avg}$) constraints since optimal power levels are determined by only the average interference constraints rather than the peak/average transmit power constraints. For $P_{\nid}=0$ or $1$, the peak transmit power constraint limits the power levels and average transmit power constraint determines the optimal power levels along with the average interference constraint, which leads to different BERs for HP bits. As seen in Fig. \ref{fig:BER_Pd_Pf}(b), the same trend is also observed for BERs for LP bits.

\begin{figure}[h]
\centering
\begin{subfigure}[b]{0.15\textwidth}
\centering
\includegraphics[width=\textwidth]{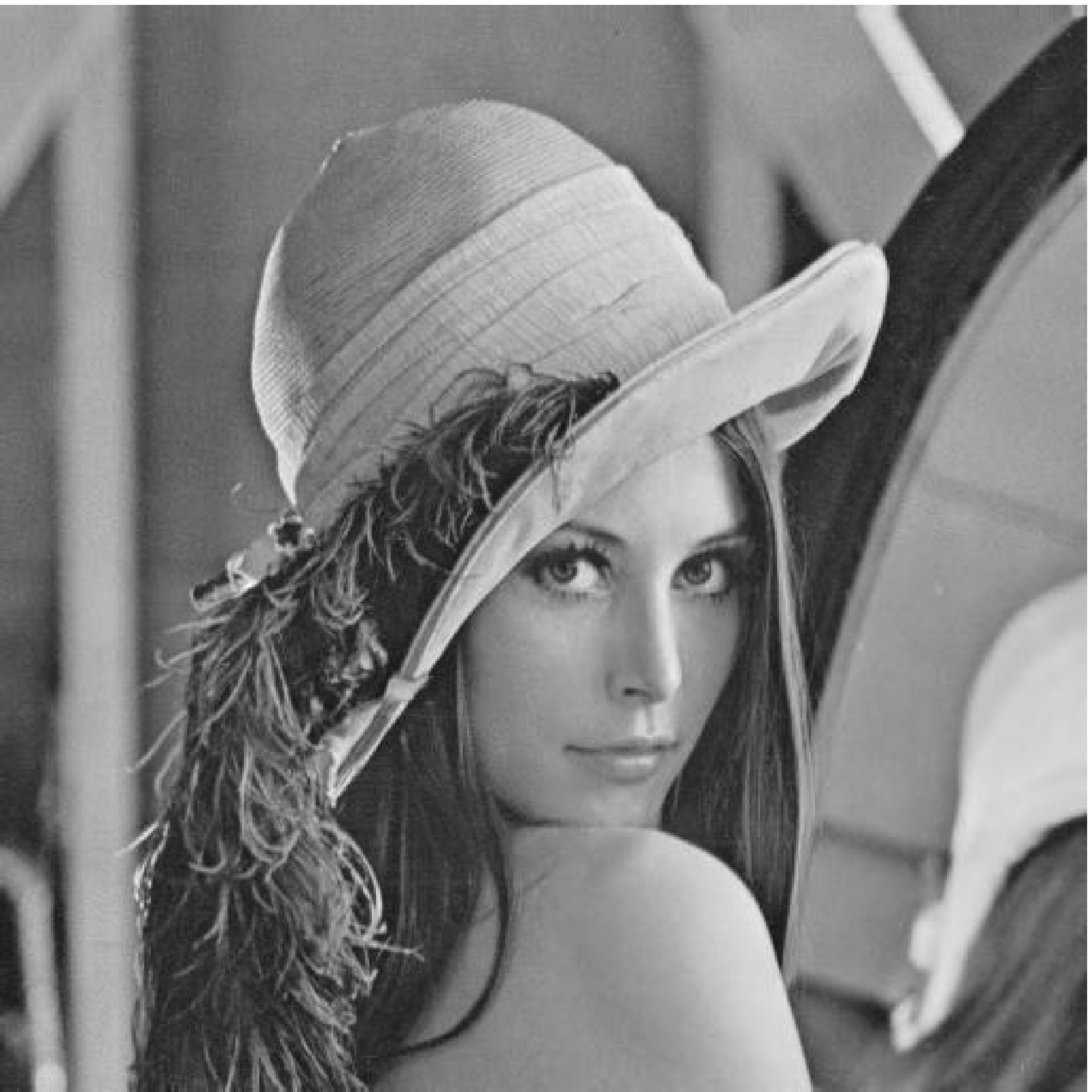}
\caption{}
\end{subfigure}
\begin{subfigure}[b]{0.15\textwidth}
\centering
\includegraphics[width=\textwidth]{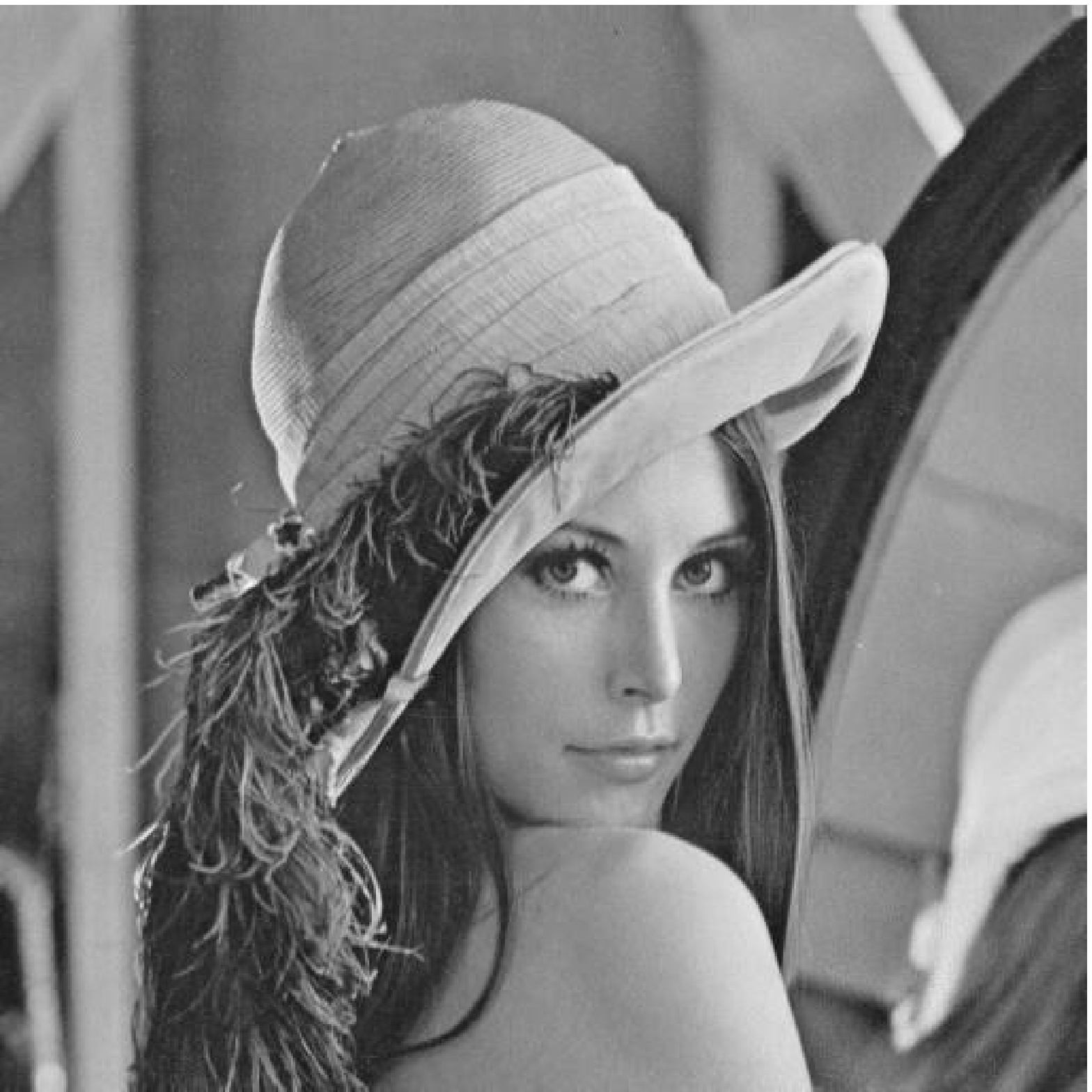}
\caption{}
\end{subfigure}
\begin{subfigure}[b]{0.15\textwidth}
\centering
\includegraphics[width=\textwidth]{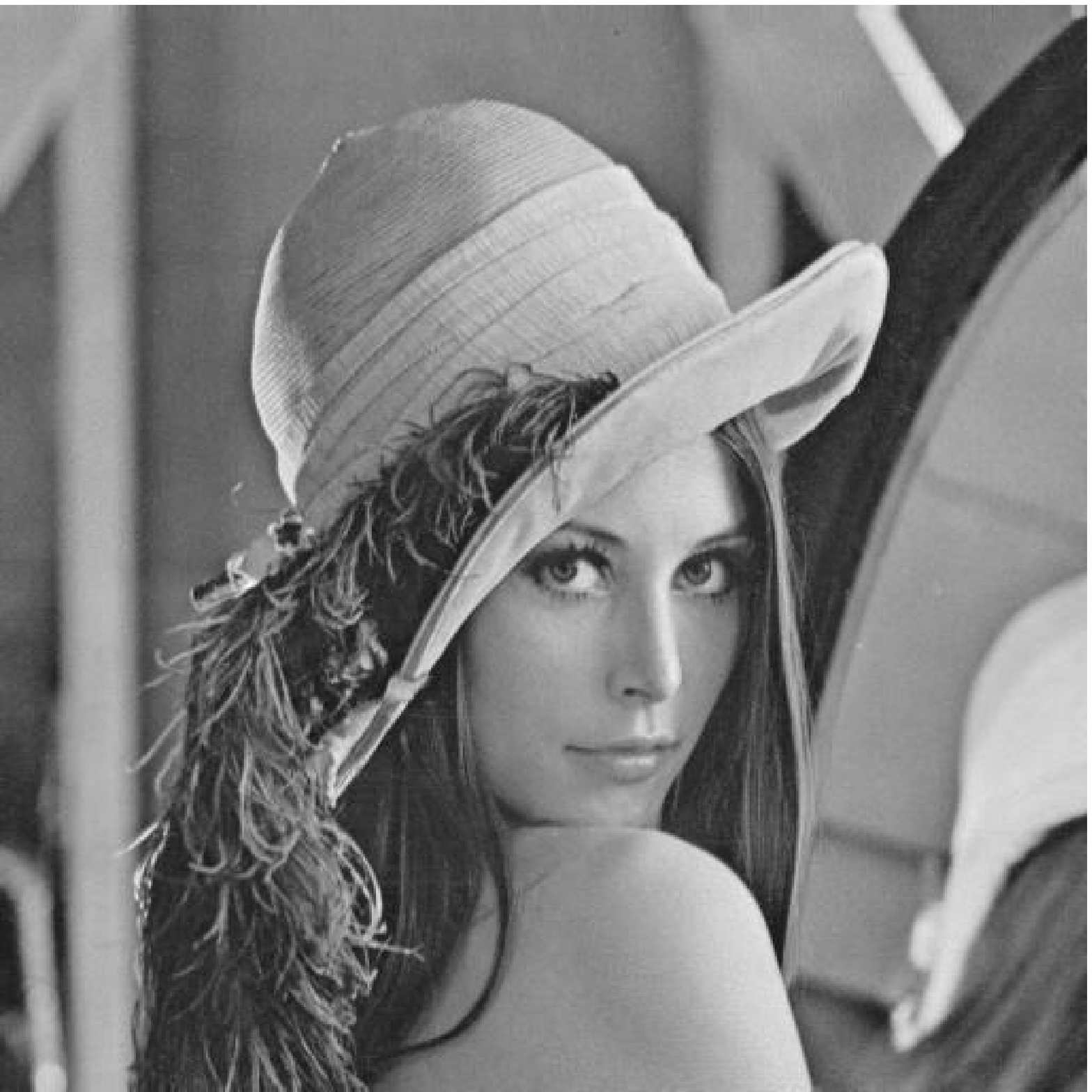}
\caption{}
\end{subfigure}
\caption{Reconstructed images with \small{(a) $P_{\nid}=0.5$, $P_{\f}=0$, $\PSNR=39.3902$ and $N_{\re}=1013$;  (b) $P_{\nid}=1$, $P_{\f}=0$, $\PSNR=40.5357$ and $N_{\re}=452$; (c) $P_{\nid}=1$, $P_{\f}=0.2$, $\PSNR=40.2706$ and $N_{\re}=541$.}}\label{fig:images_withoutboud}
\end{figure}

\begin{figure*}[htb]
\centering
\begin{subfigure}[b]{0.15\textwidth}
\centering
\includegraphics[width=\textwidth]{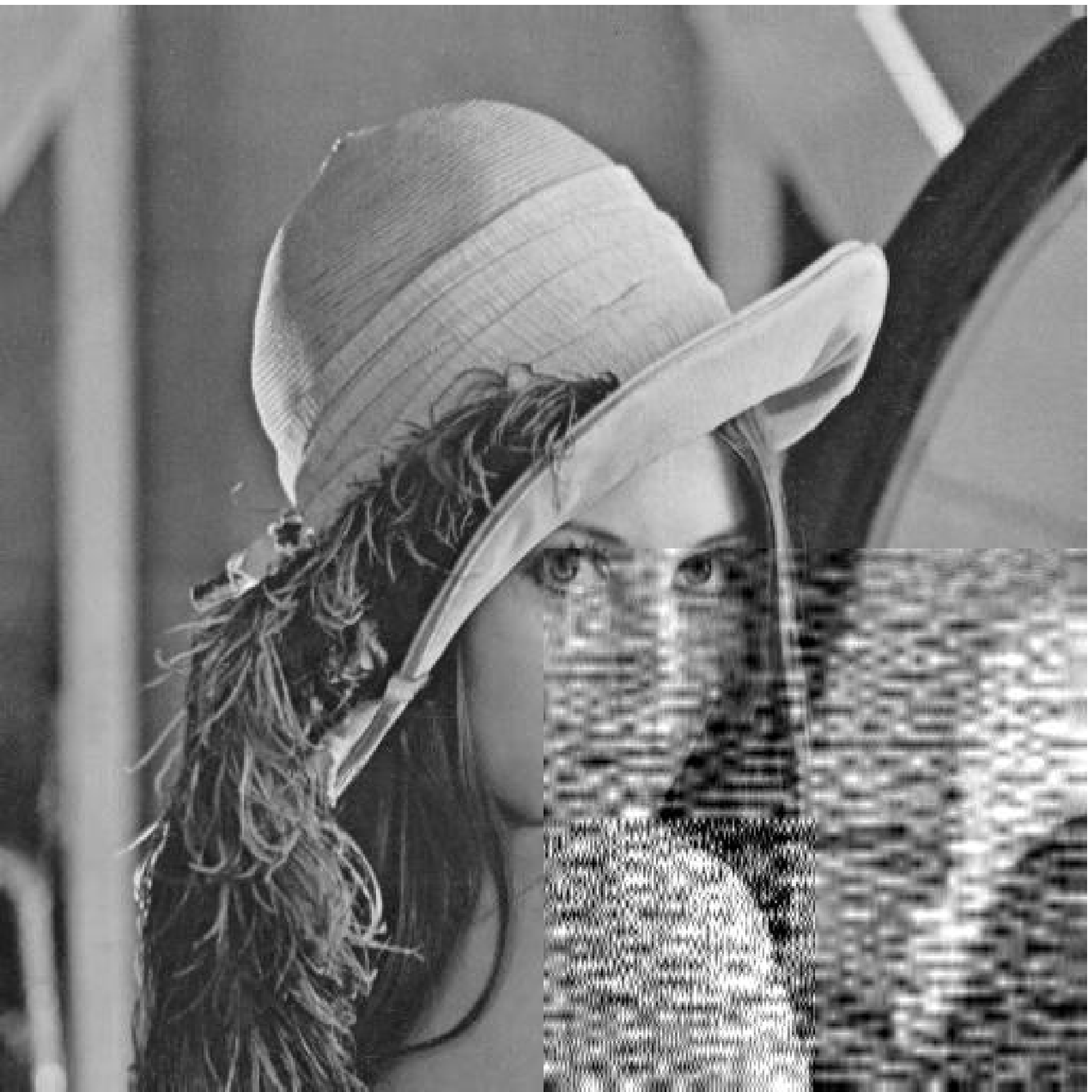}
\caption{}
\end{subfigure}
\begin{subfigure}[b]{0.15\textwidth}
\centering
\includegraphics[width=\textwidth]{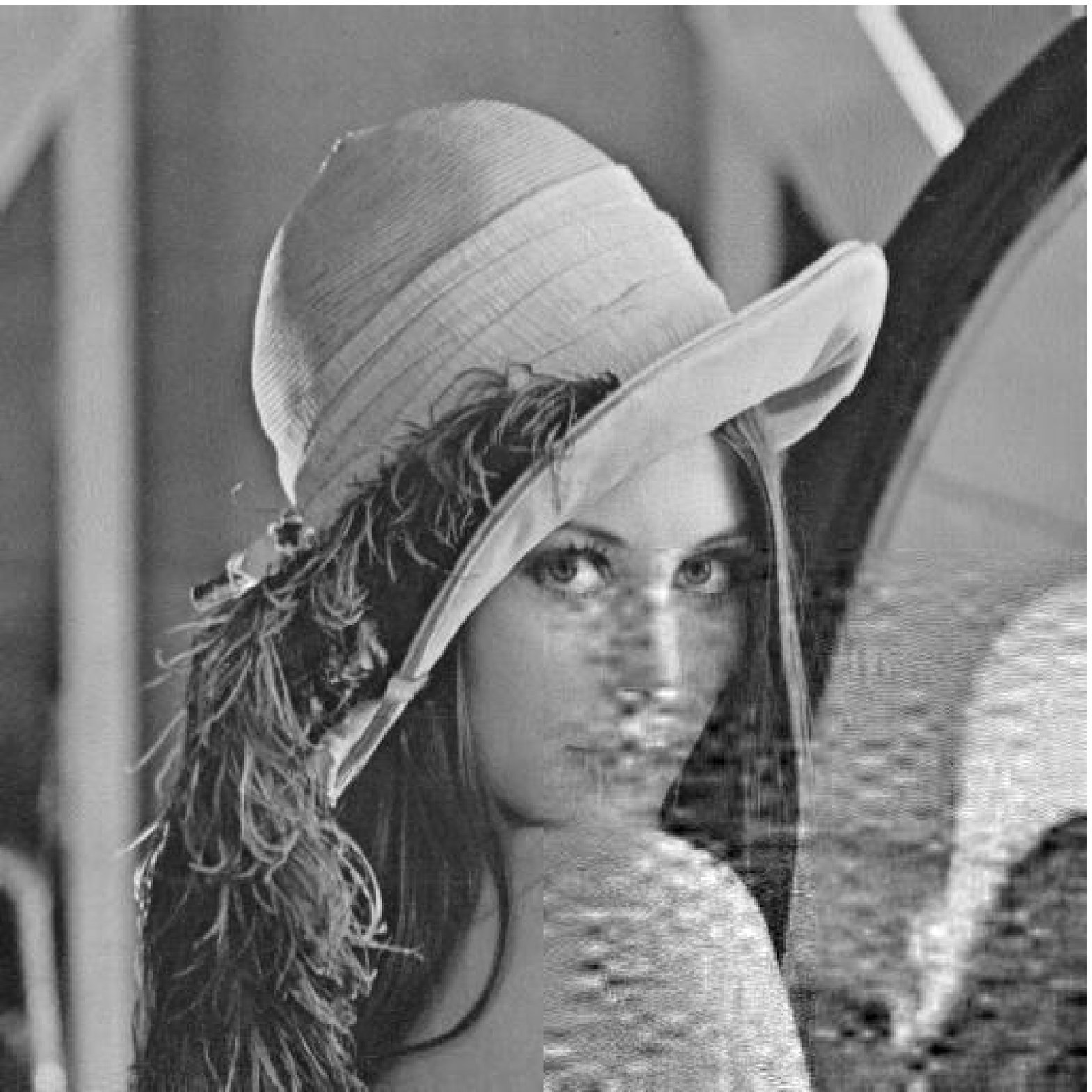}
\caption{}
\end{subfigure}
\begin{subfigure}[b]{0.15\textwidth}
\centering
\includegraphics[width=\textwidth]{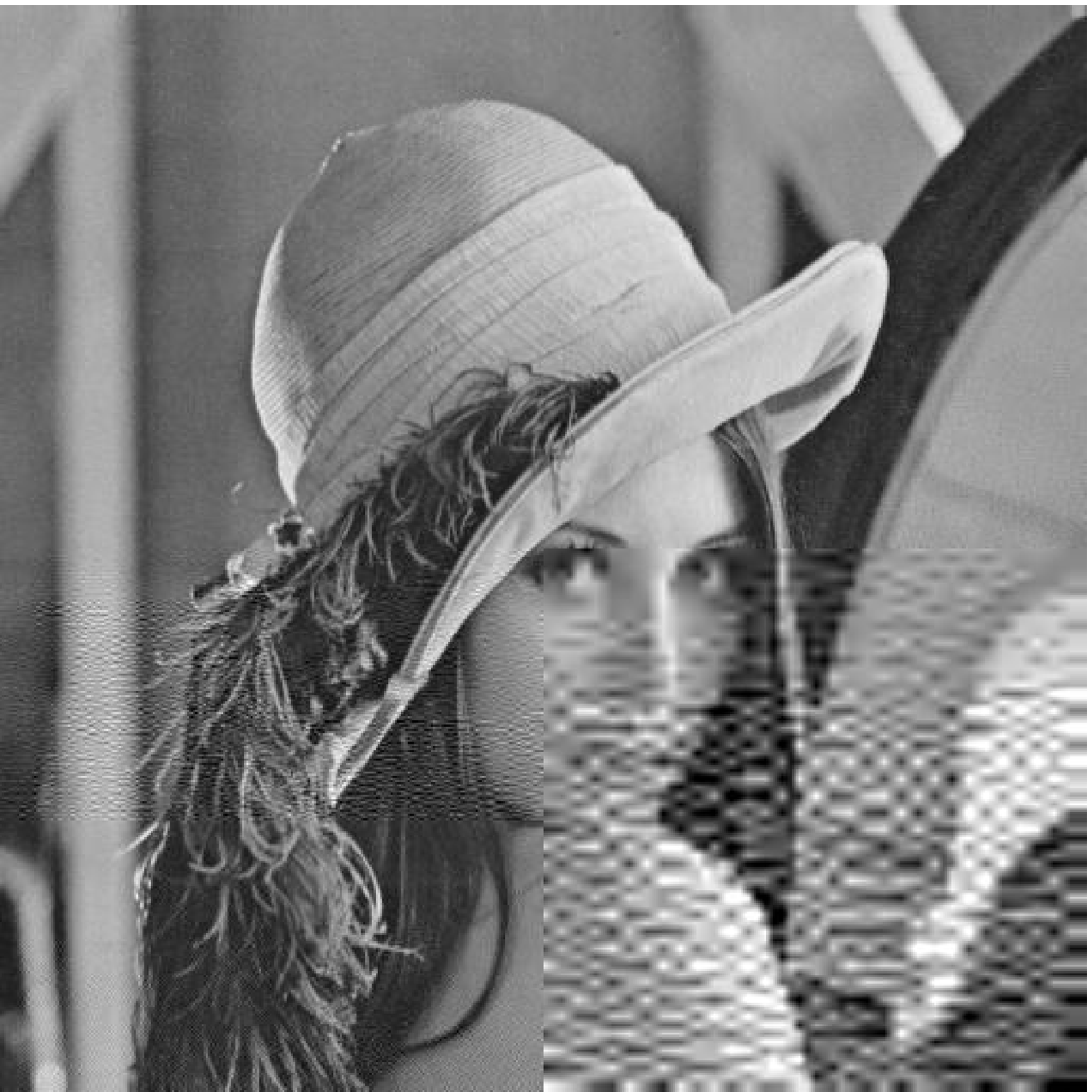}
\caption{}
\end{subfigure}
\begin{subfigure}[b]{0.15\textwidth}
\centering
\includegraphics[width=\textwidth]{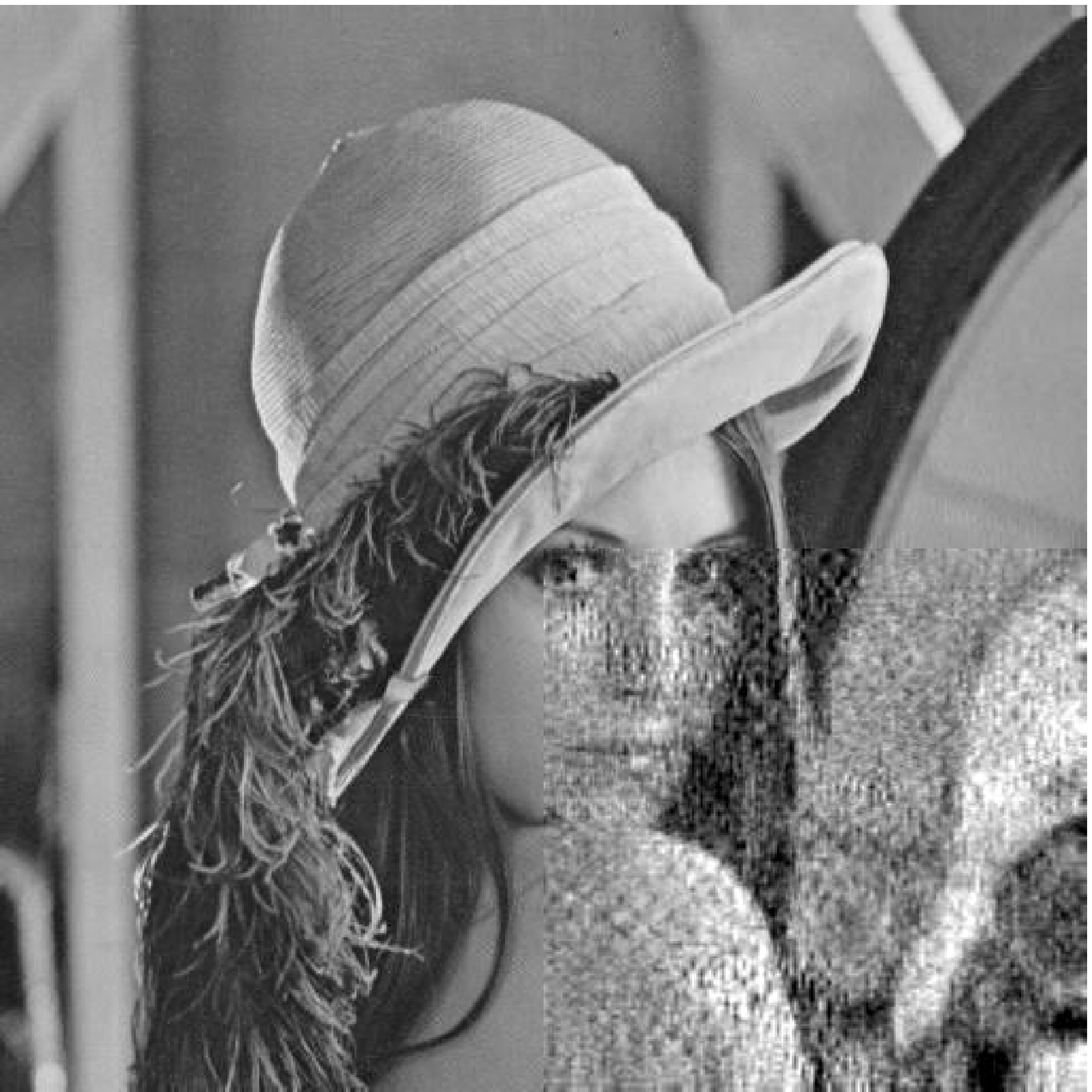}
\caption{}
\end{subfigure}
\begin{subfigure}[b]{0.15\textwidth}
\centering
\includegraphics[width=\textwidth]{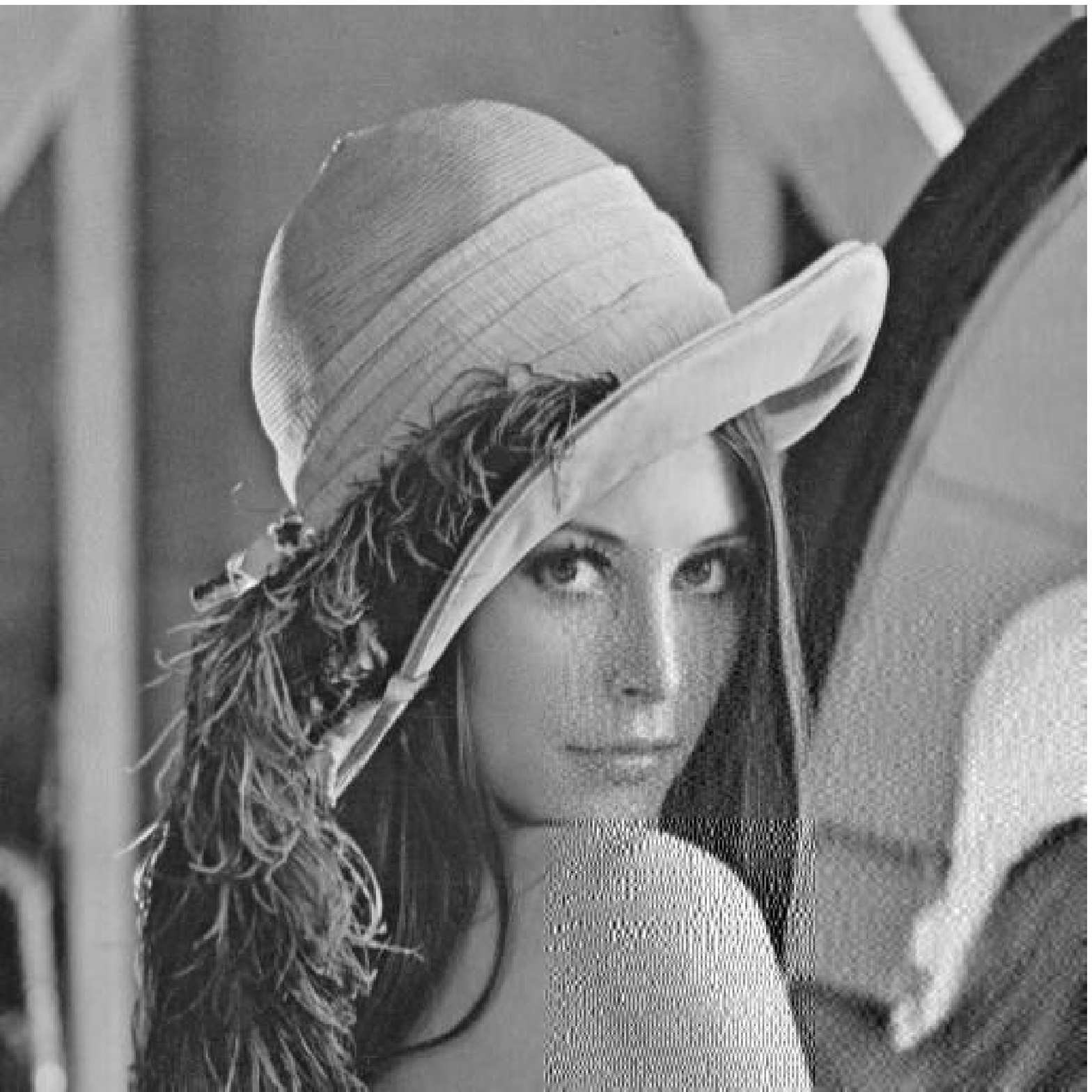}
\caption{}
\end{subfigure}
\begin{subfigure}[b]{0.15\textwidth}
\centering
\includegraphics[width=\textwidth]{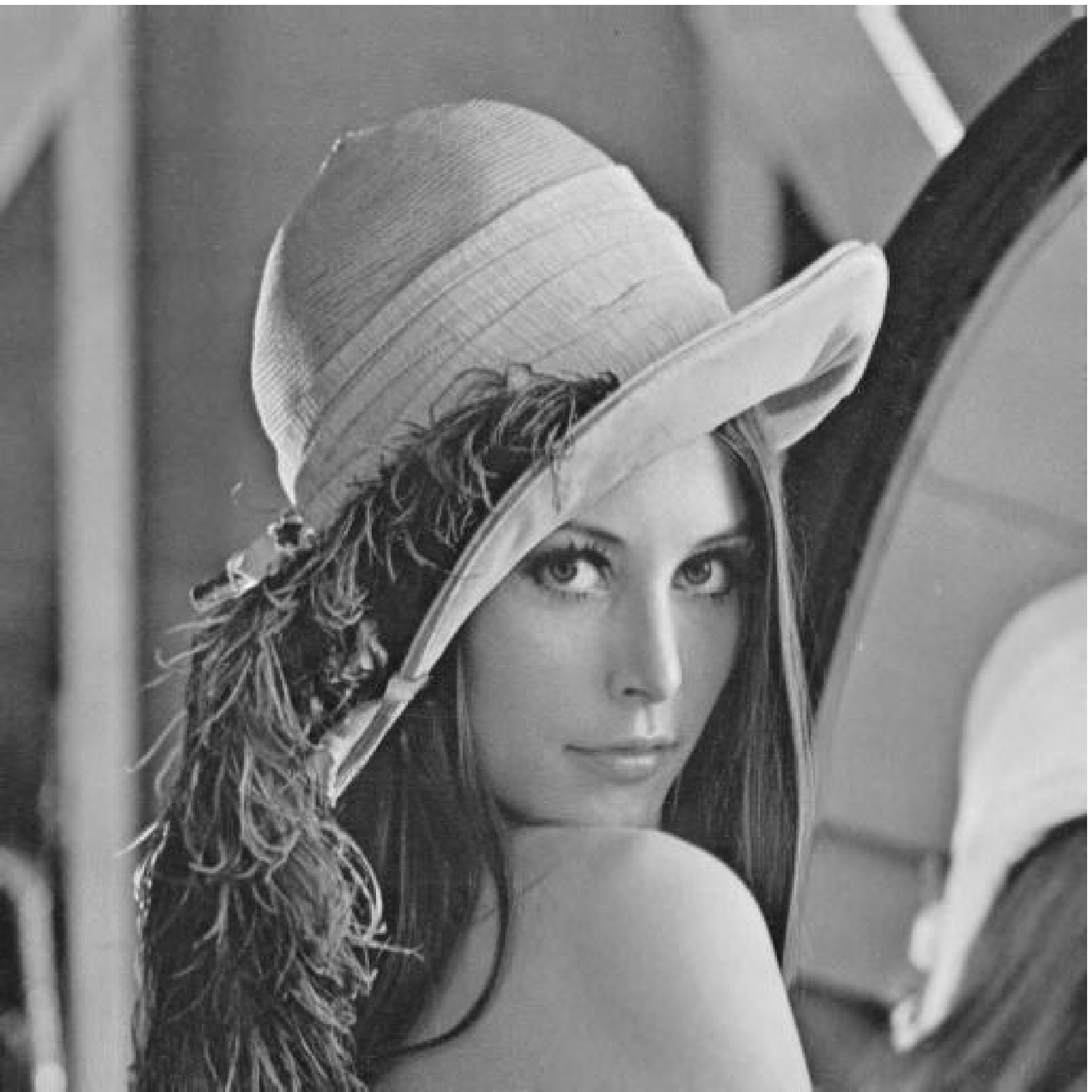}
\caption{}
\end{subfigure}
\caption{Reconstructed images with \small{(a) $P_{\f}=0.2$, $P_{\nid}=0.5$, $\PSNR=22.4303$ and $N_{\re}=800$;  (b) $P_{\f}=0.2$, $P_{\nid}=0.7$, $\PSNR=22.7172$ and $N_{\re}=800$; (c) $P_{\f}=0.2$, $P_{\nid}=1$, $\PSNR=40.3389$ and $N_{\re}=545$; (d) $P_{\f}=0.1$, $P_{\nid}=0.5$, $\PSNR=22.5194$ and $N_{\re}=800$;  (e) $P_{\f}=0.1$, $P_{\nid}=0.7$, $\PSNR=24.6407$ and $N_{\re}=800$; (f) $P_{\f}=0.1$, $P_{\nid}=1$, $\PSNR=40.349$ and $N_{\re}=498$.}}
\label{fig:images_withboud}
\end{figure*}

In Fig. \ref{fig:images_withoutboud}, the reconstructed images for different values of $P_{\f}$ and $P_{\nid}$ are displayed while power allocation is performed based on the statistical CSI subject to peak transmit power and average interference constraints. It is assumed that there is no upper bound on the number of retransmissions and 16-HQAM with $\alpha_0=\alpha_1=1$ is employed. It is seen that the received image quality for each scenario is nearly the same. Indeed, their PSNRs are around $40$ dB. However, the number of retransmissions is different in each scenario. In perfect sensing, i.e., when $P_{\nid}=1$ and $P_{\f}=0$, we have the least number of retransmissions with $N_{\re}=452$. On the other hand, in the case of $P_{\nid}=0.5$ and $P_{\f}=0$, a similar received image quality is attained with $N_{\re}=1013$ retransmissions. Note that this significant increase in $N_{\re}$ implies higher delays and higher energy expenditure. Under the same setting, we have performed simulations for other cases where power control with instantaneous CSI rather than statistical CSI is applied or average transmit power/interference power constraints are imposed instead of peak transmit power/average interference power constraints. Due to the sake of brevity, the corresponding results are not displayed but we have the following important observations:
\begin{itemize}
\item When power control with instantaneous CSI is applied under the same power constraints, PSNR performance is improved by around $1$ dB with up to $49 \%$ reduction in the number of retransmissions, yielding lower retransmission delay compared to power allocation with statistical CSI.
\item When optimal power allocation with statistical CSI is performed, imposing either peak transmit power constraint or average transmit power constraint provides nearly the same PSNR performance. However, the impact on the number of retransmissions is profound especially when instantaneous CSI is employed
and sensing result is reliable, e.g., the number of retransmissions is reduced by up to $47 \%$.
\end{itemize}

In Fig. \ref{fig:images_withboud}, the reconstructed images for different values of $P_{\f}$ and $P_{\nid}$ are shown. The statistical CSI is used to determine the optimal power levels. Different from the previous figure, we now set an upper bound on the number of retransmissions, i.e., $N_{\upper}=800$. Cognitive transmission is again subject to peak power and average interference constraints. In contrast to the previous reconstructed images, for which PSNR is nearly the same, the received quality is now affected by the channel sensing performance. More specifically, as $P_{\nid}$ increases and hence sensing reliability improves, PSNR increases and the received image quality becomes better. On the other hand, increasing $P_{\f}$ results in lower PSNR values. In Fig. \ref{fig:images_withboud}, we also observe that the degradation in the image quality is generally in the lower right portion of the images. This is due to the fact that this part of the image is transmitted the last by which time the number of retransmissions has  generally reached the upper bound and no more retransmissions are allowed.

\begin{figure}
\centering
\begin{subfigure}[b]{0.17\textwidth}
\centering
\includegraphics[width=\textwidth]{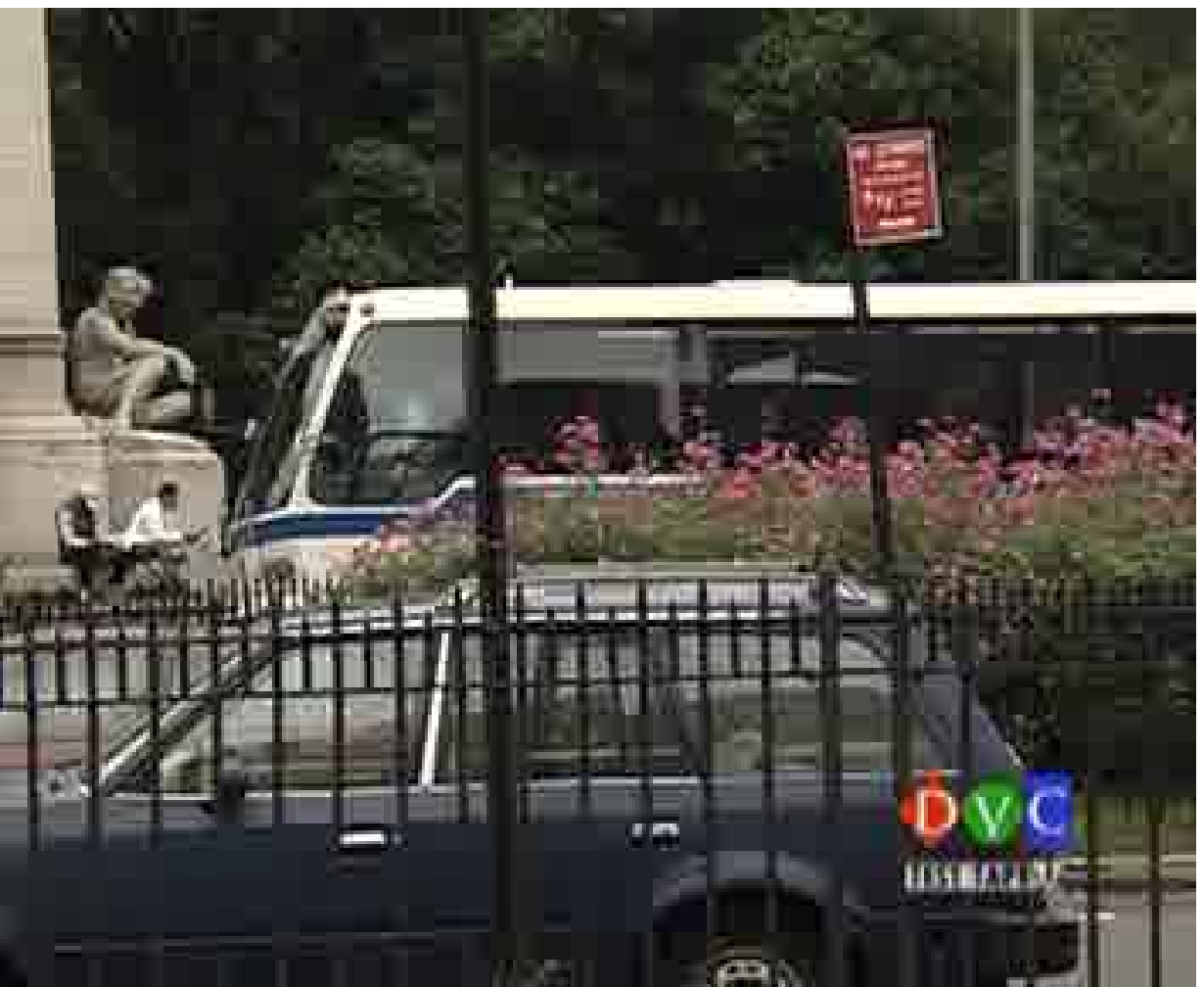}
\caption{Imperfect sensing}
\end{subfigure}
\begin{subfigure}[b]{0.17\textwidth}
\centering
\includegraphics[width=\textwidth]{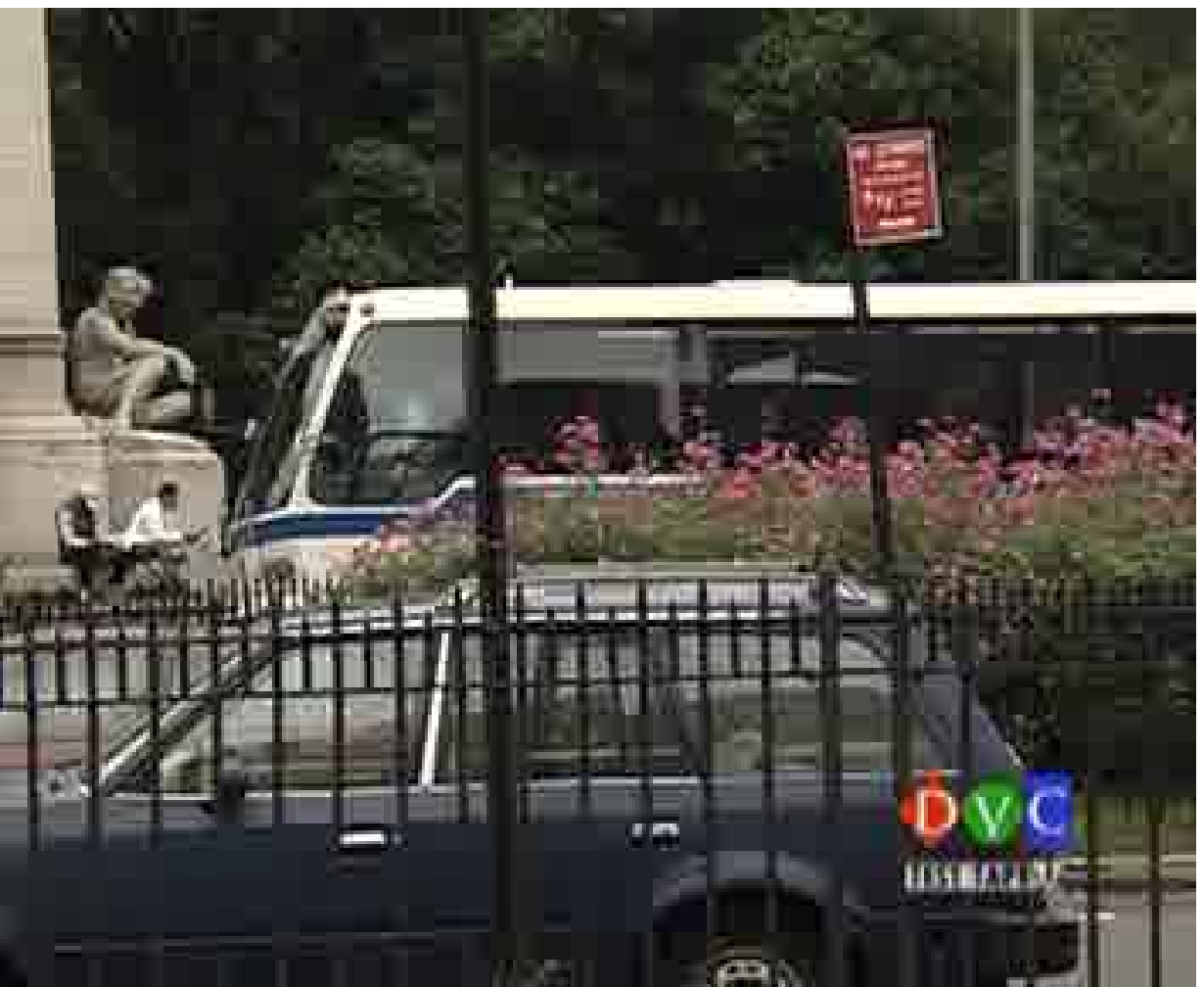}
\caption{Perfect sensing}
\end{subfigure}
\caption{(a) Imperfect channel sensing with $P_{\nid} = 0.8$, $P_{\f} = 0.2$, $\PSNR=15.3170$ dB and $N_{\re}=3059$ (b) Perfect channel sensing with $P_{\nid} = 1$, $P_{\f} = 0$, $\PSNR=15.6711$ dB and $N_{\re}=889$.}\label{fig:video_sensing}
\vspace{-.6cm}
\end{figure}

\begin{figure*}
\centering
\begin{subfigure}[b]{0.32\textwidth}
\centering
\includegraphics[width=\textwidth]{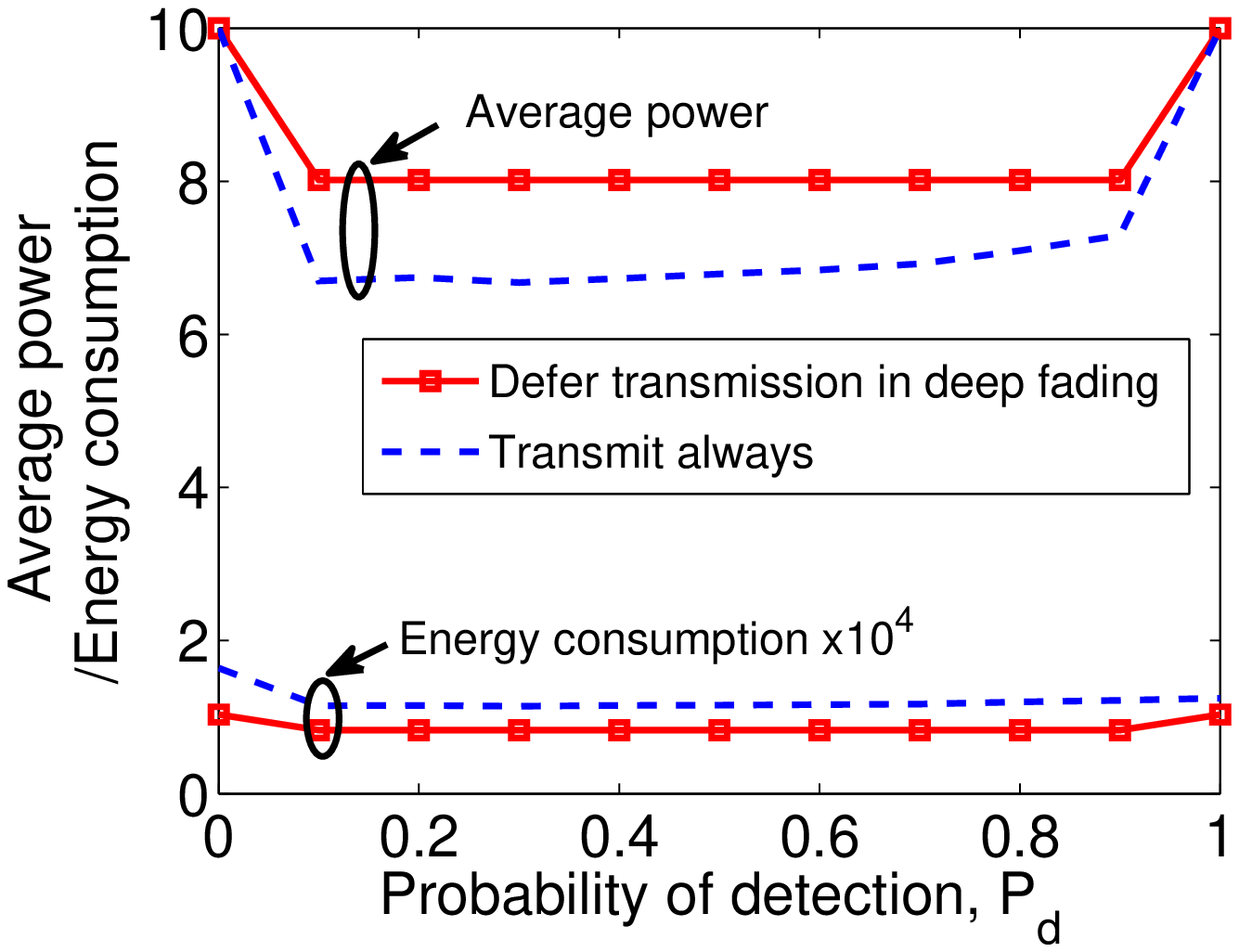}
\caption{Power and energy consumption vs. $P_{\nid}$}
\end{subfigure}
\begin{subfigure}[b]{0.34\textwidth}
\centering
\includegraphics[width=\textwidth]{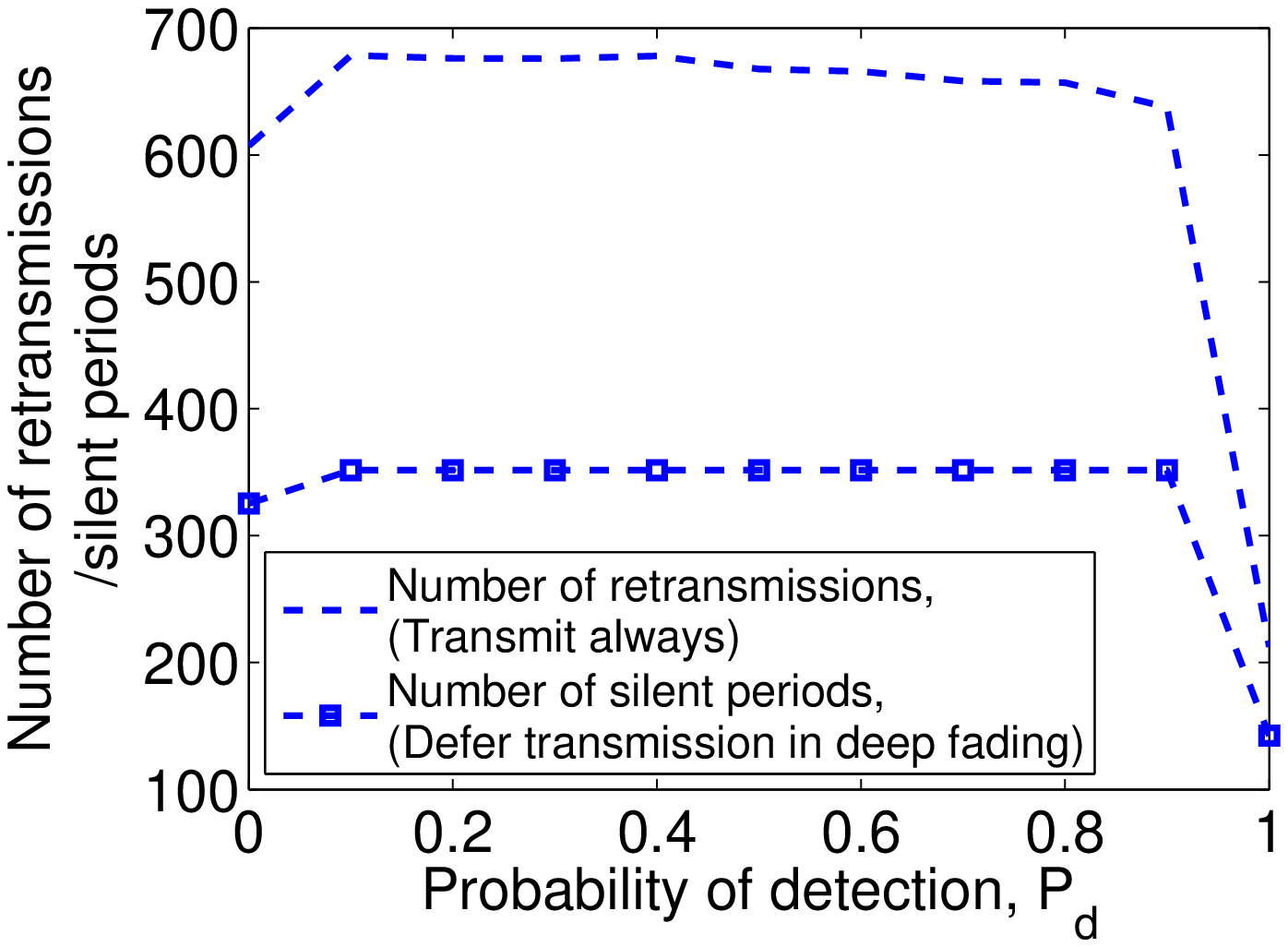}
\caption{$N_{re}$ and number of silent periods vs. $P_{\nid}$}
\end{subfigure}
\begin{subfigure}[b]{0.32\textwidth}
\centering
\includegraphics[width=\textwidth]{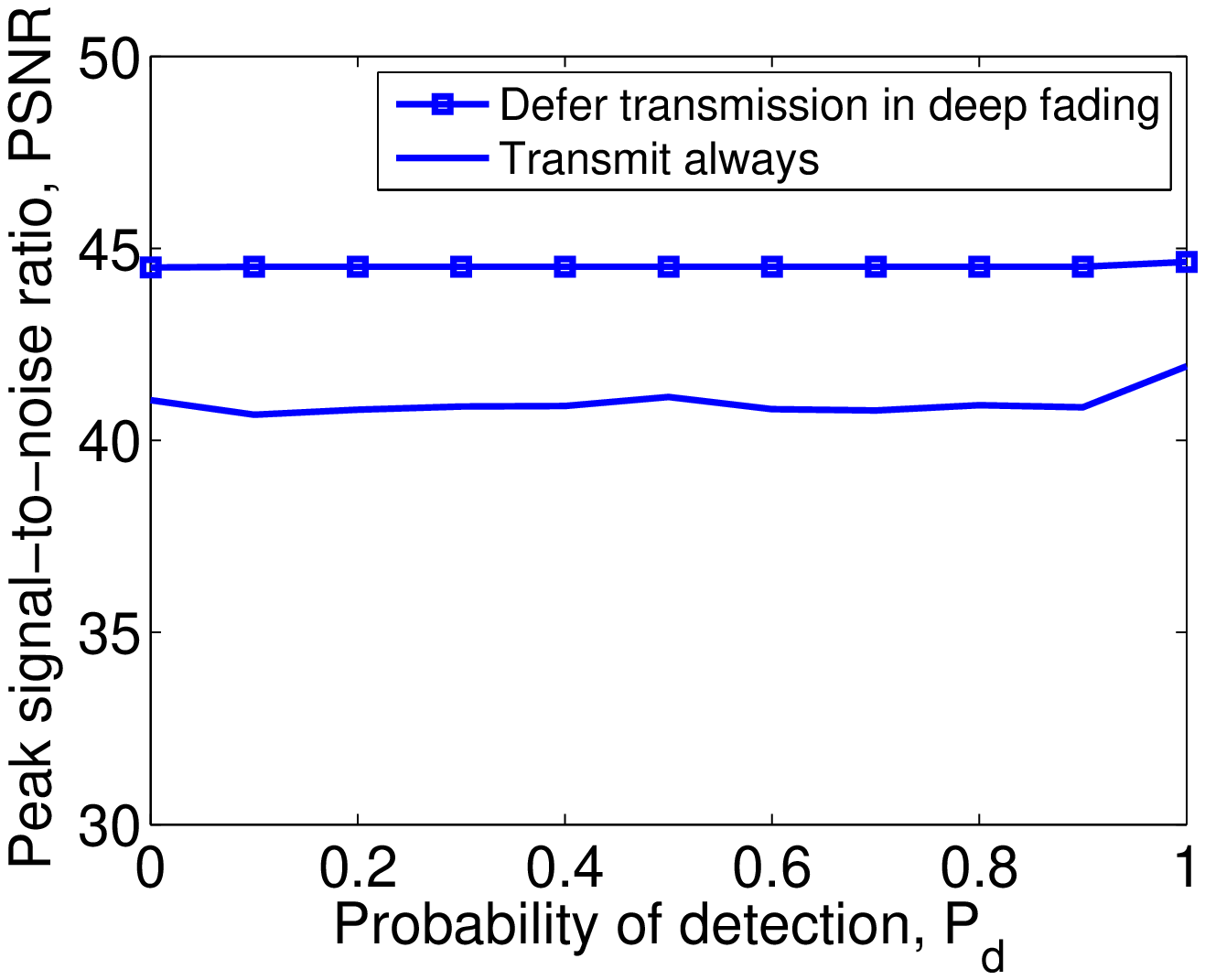}
\caption{PSNR vs. $P_{\nid}$}
\end{subfigure}
\caption{\small{(a) Average energy and power consumption vs. probability of detection, $P_d$; (b) Number of retransmissions/silent periods vs. $P_d$; (c) Peak signal-to-noise ratio, PSNR vs. $P_d$.}}\label{fig:threshold}
\end{figure*}

\begin{figure*}[ht]
\centering
\begin{subfigure}[b]{0.32\textwidth}
\centering
\includegraphics[width=\textwidth]{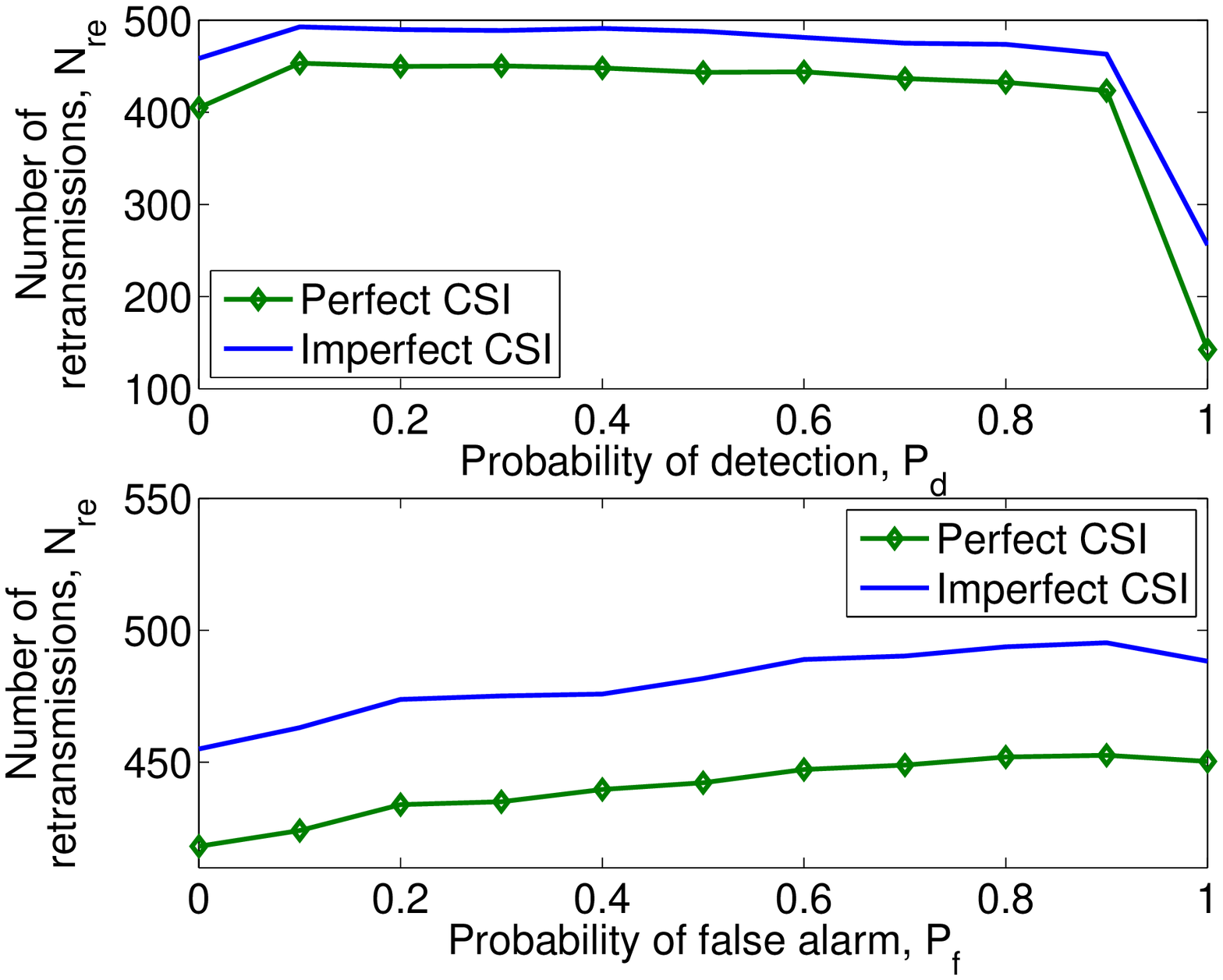}
\caption{$N_{re}$ vs. $P_{\nid}$ and $P_{\f}$}
\end{subfigure}
\begin{subfigure}[b]{0.32\textwidth}
\centering
\includegraphics[width=\textwidth]{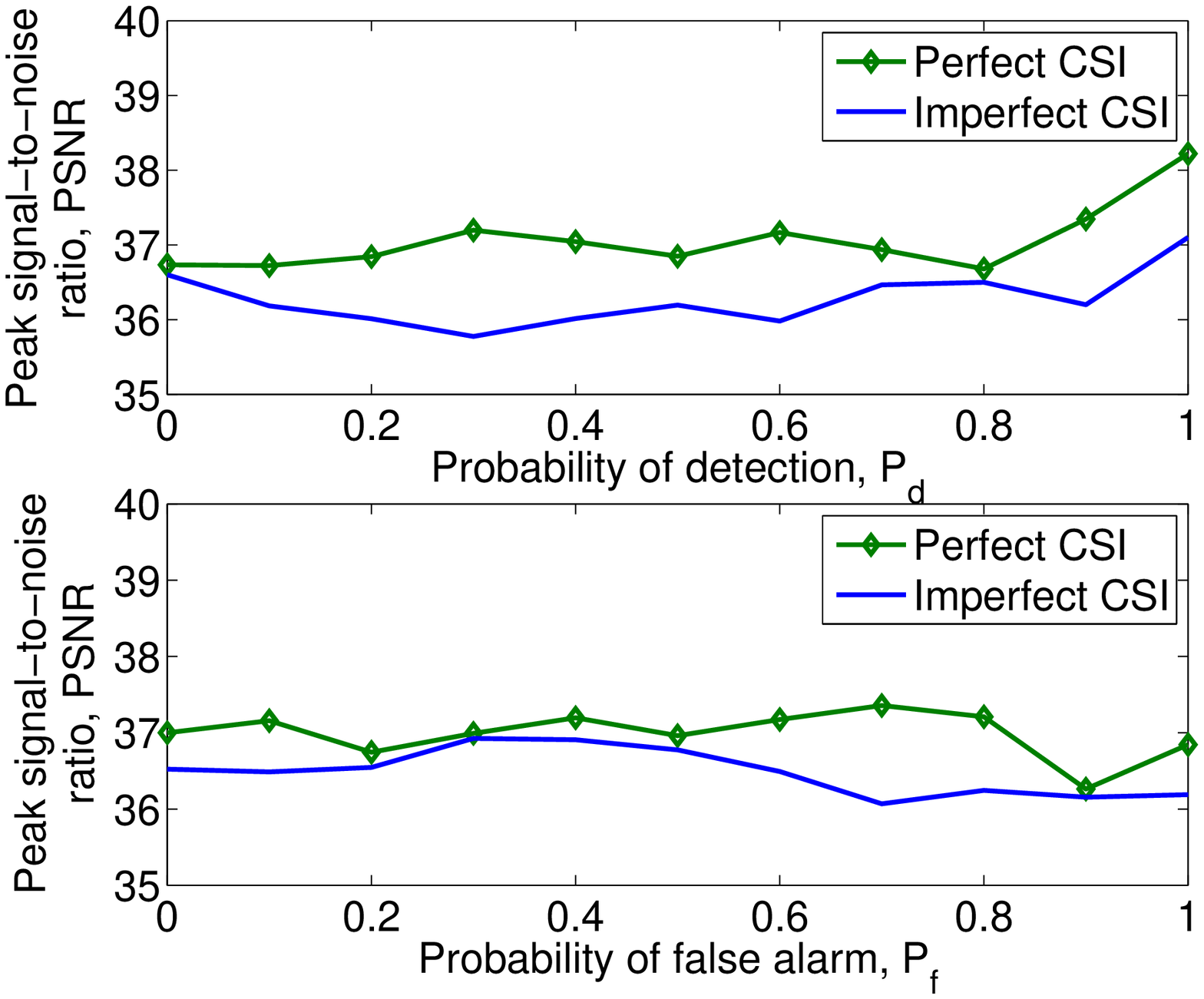}
\caption{PSNR vs. $P_{\nid}$ and $P_{\f}$}
\end{subfigure}
\caption{\small{(a) Number of retransmissions, $N_{re}$ vs. $P_{\nid}$ and $P_{\f}$; (b) Peak signal-to-noise ratio, PSNR vs. $P_{\nid}$ and $P_{\f}$.}}\label{fig:Imperfect_Pd_Pf}
\end{figure*}

In Fig. \ref{fig:video_sensing}, we display a single frame from the received video with both imperfect sensing (i.e., $P_{\nid} = 0.8$, $P_{\f} = 0.2$) and perfect sensing ($P_{\nid} = 1$, $P_{\f} = 0$) subject to average transmit power and average interference constraints. Power control based on instantaneous CSI is applied. In the simulation of our video transmissions, cognitive users again employ $16$-HQAM with $\alpha_0=\alpha_1 = 1$ for transmission. Threshold for transmission, $Thr$, is set to $2.1$. More retransmissions are required when sensing is imperfect. The averages of $\PSNR$ and $N_{\re}$ values are obtained by simulating the wireless transmission of the same video sequence 60 times. In Fig. \ref{fig:video_sensing}, the 11th frame out of 60 frames in the video sequence is displayed in both cases. We observe that while image quality is similar under imperfect and perfect sensing decisions, imperfect sensing can have substantial impact on the number of retransmissions. We also analyze power allocation with statistical CSI, which gives almost the same PSNR value at the cost of higher number of retransmissions, e.g., around $49\%$ higher under imperfect sensing and around $100\%$ higher under perfect sensing.

In Fig. \ref{fig:threshold}, we plot the average power and energy consumption, number of retransmissions/silent periods, and PSNR as a function of detection probability $P_d$. It is assumed that $P_f = 0.1$. We consider two cases: either the packets are always transmitted or there is no packet transmission during deep fades. It is seen that we have less energy consumption, smaller number of retransmissions or silent periods, and better PSNR performance when there is no transmission in deep fading since the transmitter does not unnecessarily use power budget in case of unfavorable channel conditions and allocates more power to better channels.

\subsection{The impact of imperfect CSI of interference link on multimedia transmission}
In this section, we analyze the performance of multimedia transmission in the presence of imperfect CSI of the interference link subject to average transmit power and average interference power constraints. It is assumed that the variance of the estimation error is $\sigma_e^2 = 0.0124$. In Fig. \ref{fig:Imperfect_Pd_Pf}, we plot the number of retransmissions and PSNR as a function of the probability of detection and probability of false alarm. It is seen that having perfect CSI of the interference link results in a smaller number of retransmissions and higher PSNR as compared to having only imperfect CSI of this link, as expected. It is also observed that the number of retransmissions increases with increasing $P_{\f}$ or decreasing $P_{\nid}$ due to the same reasoning explained in the discussions of Fig. \ref{fig:P0P1_Nre_PSNR_vsPd} and Fig. \ref{fig:P0P1_Nre_PSNR_vsPf}.

\subsection{The impact of unequal error protection (HQAM) vs. equal error protection (conventional QAM) on multimedia quality}

\begin{figure*}
\centering
\begin{subfigure}[b]{0.15\textwidth}
\centering
\includegraphics[width=\textwidth]{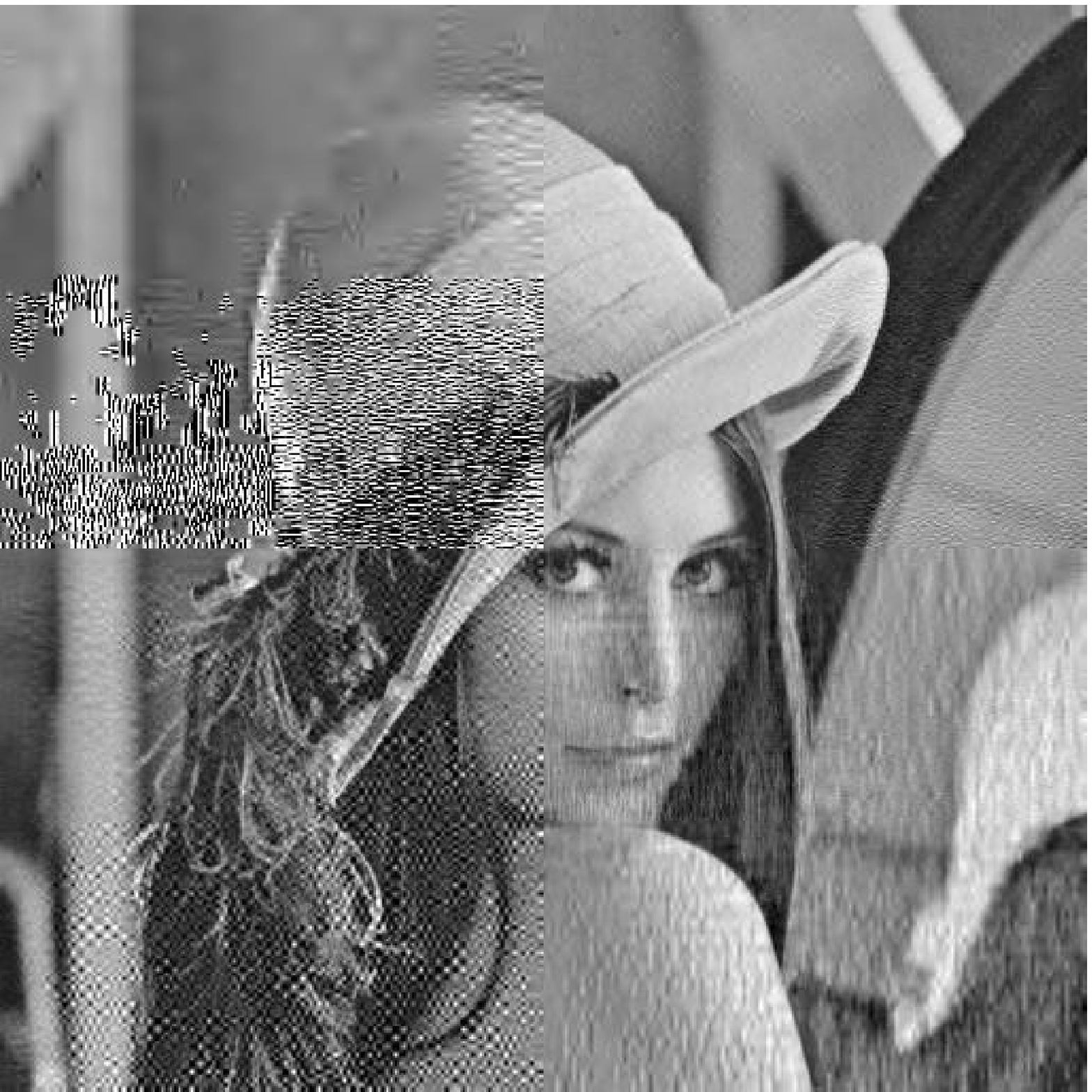}
\caption{}
\end{subfigure}
\begin{subfigure}[b]{0.15\textwidth}
\centering
\includegraphics[width=\textwidth]{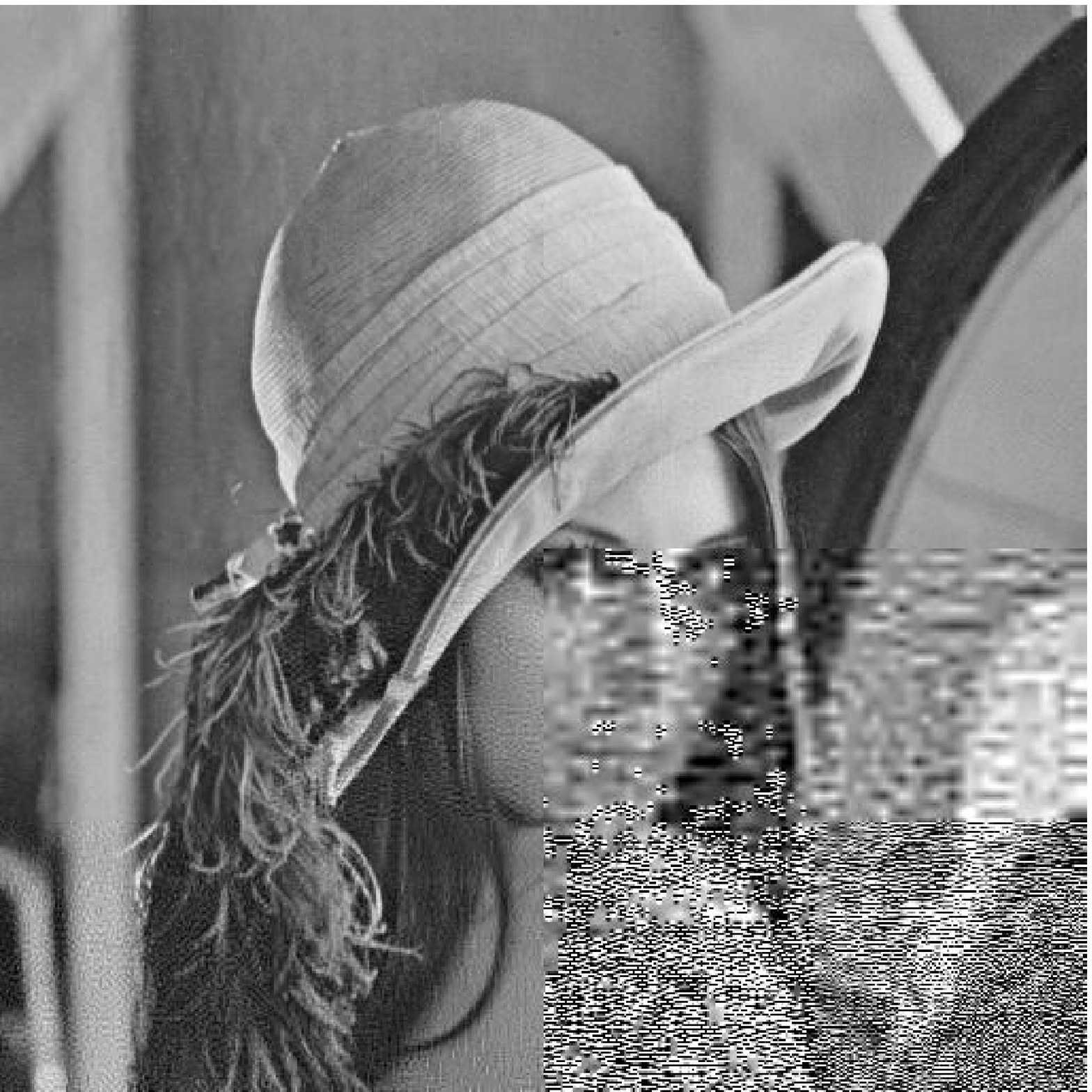}
\caption{}
\end{subfigure}
\begin{subfigure}[b]{0.15\textwidth}
\centering
\includegraphics[width=\textwidth]{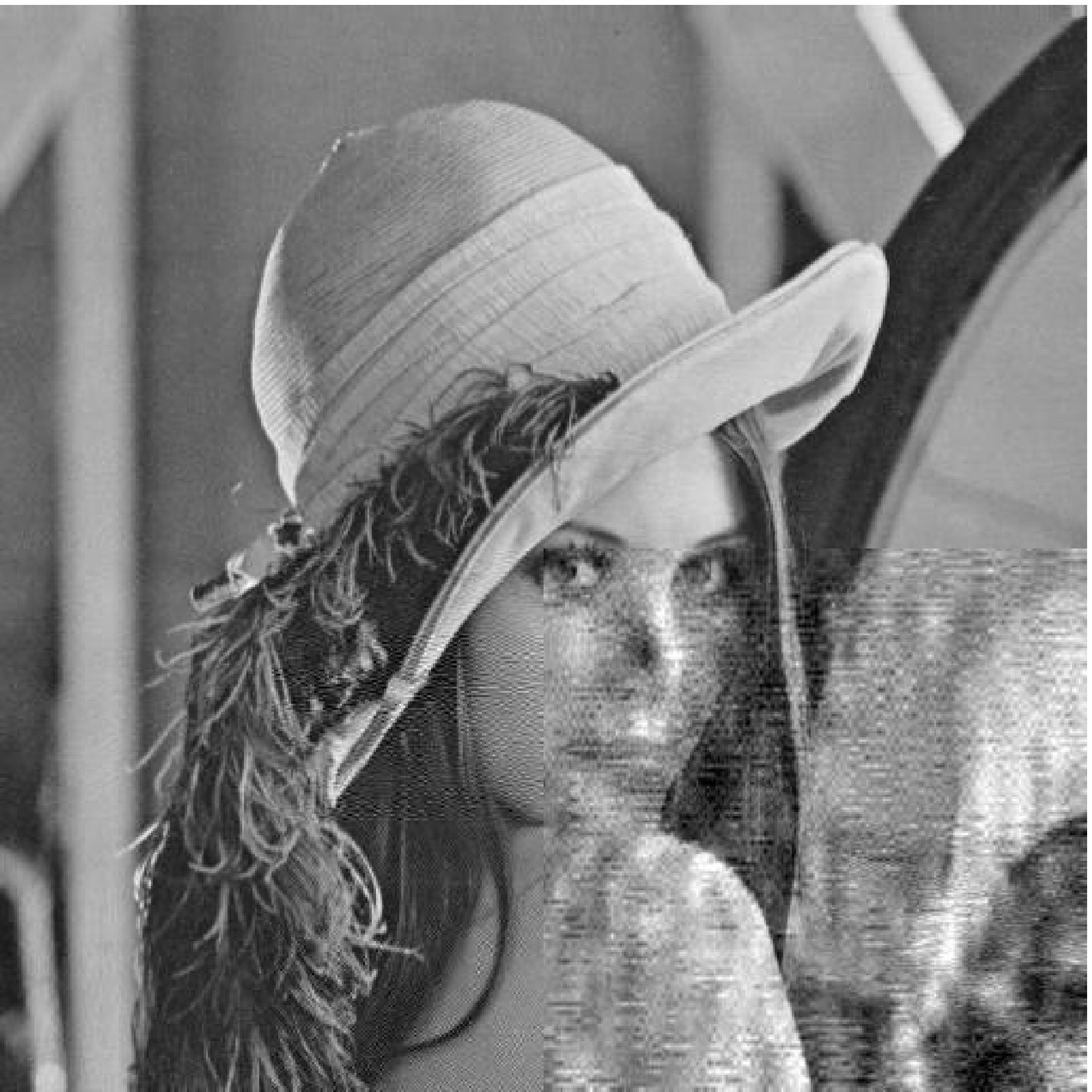}
\caption{}
\end{subfigure}
\begin{subfigure}[b]{0.15\textwidth}
\centering
\includegraphics[width=\textwidth]{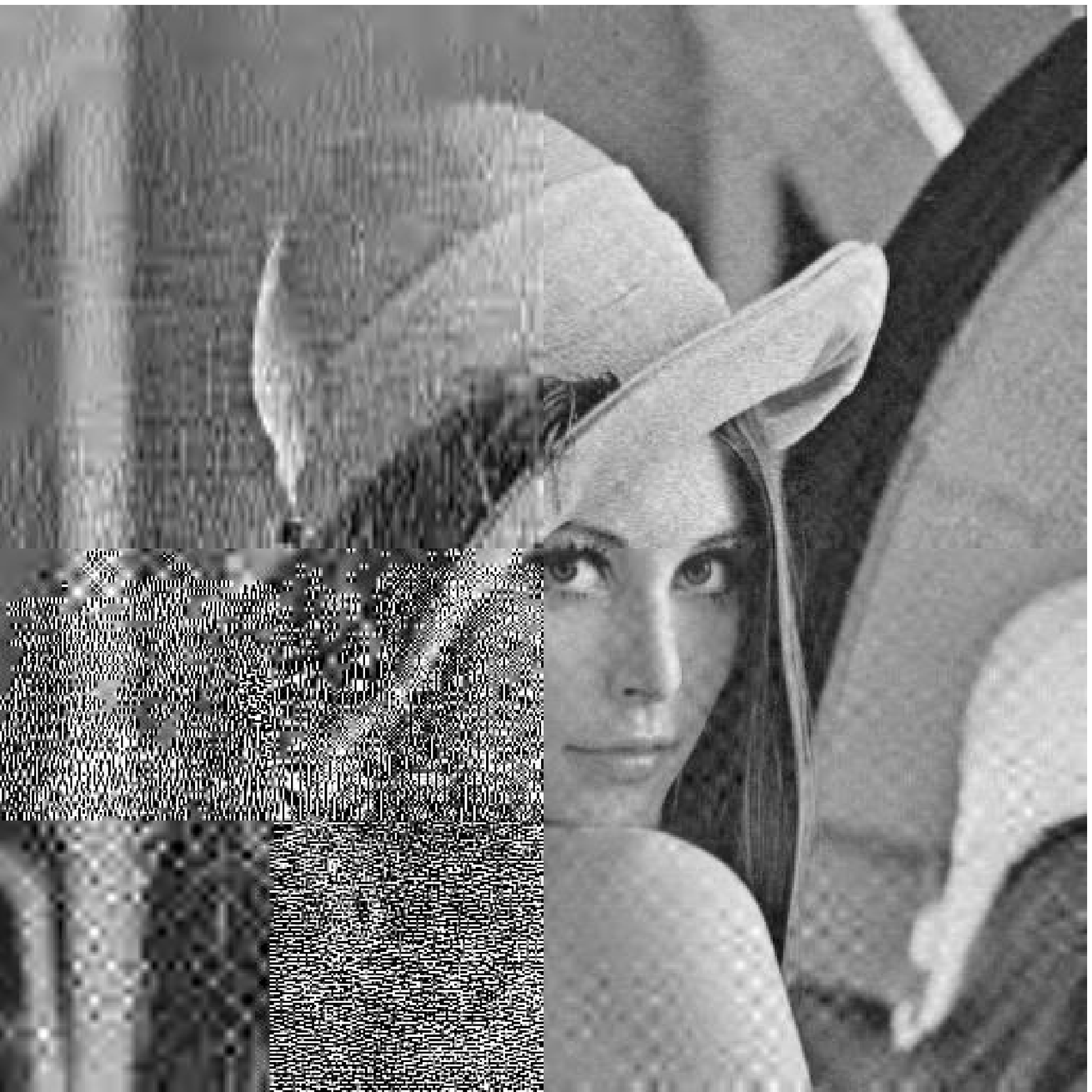}
\caption{}
\end{subfigure}
\begin{subfigure}[b]{0.15\textwidth}
\centering
\includegraphics[width=\textwidth]{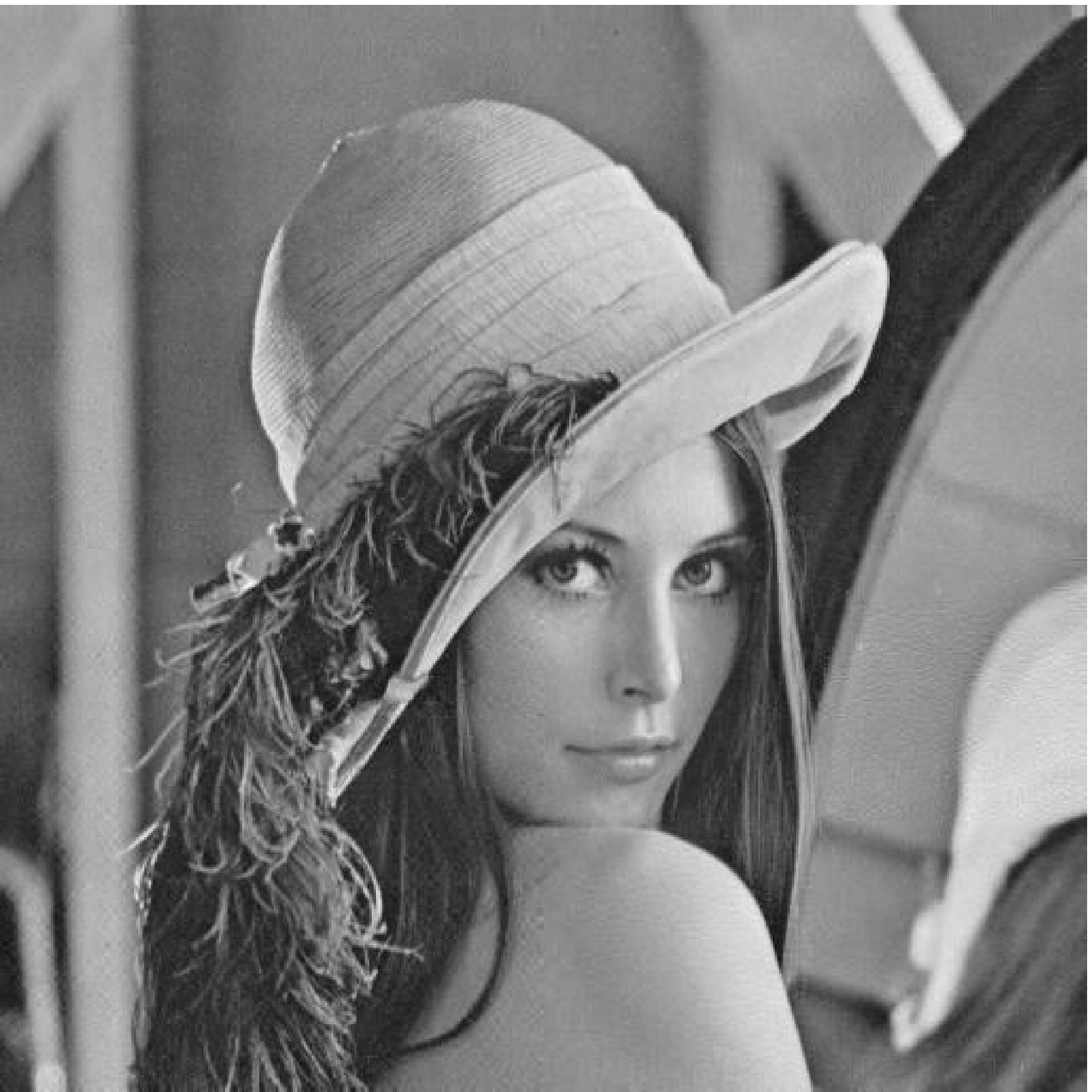}
\caption{}
\end{subfigure}
\begin{subfigure}[b]{0.15\textwidth}
\centering
\includegraphics[width=\textwidth]{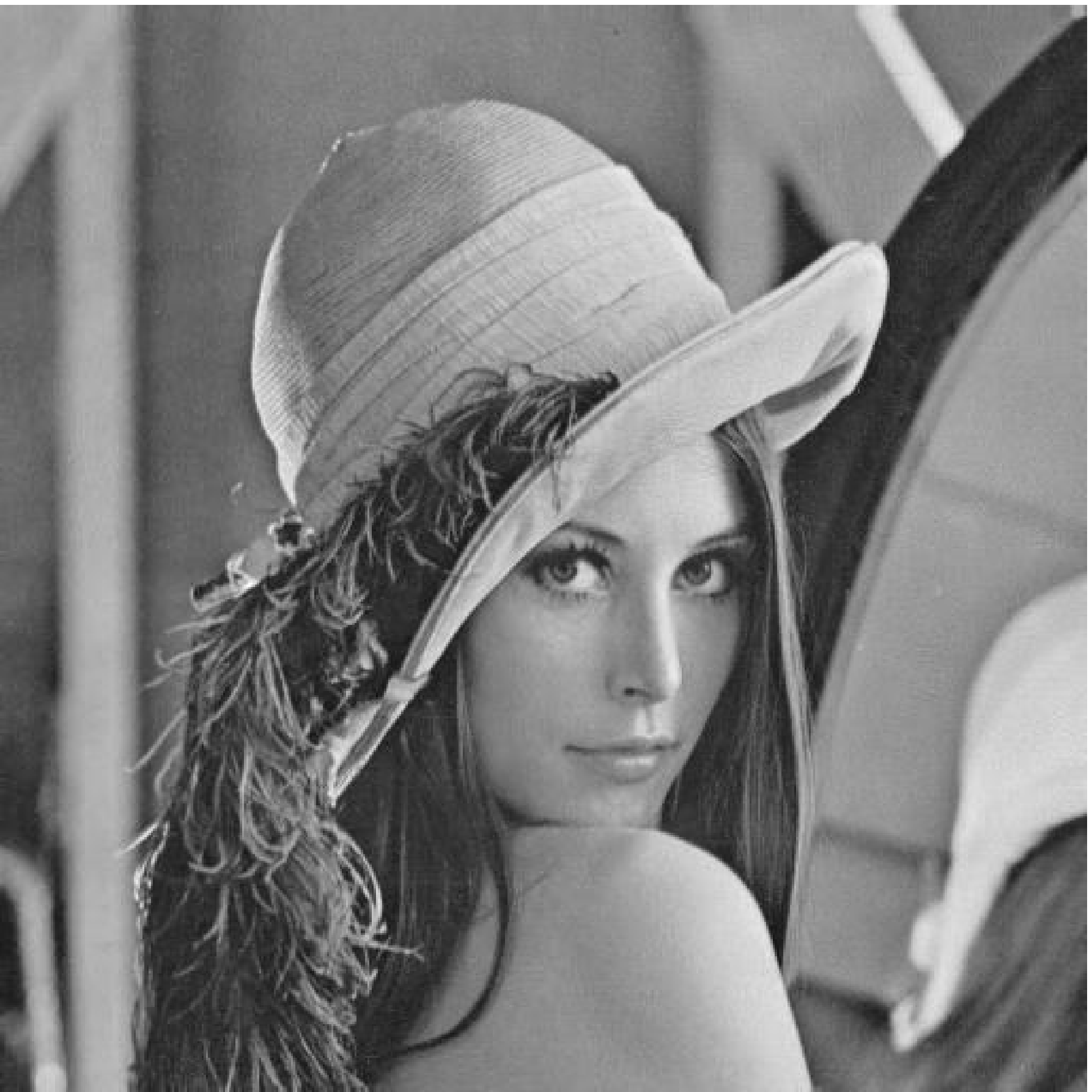}
\caption{}
\end{subfigure}
\caption{Reconstructed images with \small{(a) QAM, $m=1$, $\PSNR=18.8928$ and $N_{\re}=410$;  (b) HQAM, $m=1$, $\alpha_0=\alpha_1=1$, $\PSNR=21.3205$ and $N_{\re}=410$; (c) HQAM, $m=1$, $\alpha_0=\alpha_1=2$, $\PSNR=22.4221$ and $N_{\re}=410$; (d) QAM, $m=2$, $\PSNR=19.1760$ and $N_{\re}=396$;  (e) HQAM, $m=2$, $\alpha_0=\alpha_1=1$, $\PSNR=35.6625$ and $N_{\re}=400$; (f) HQAM, $m=2$, $\alpha_0=\alpha_1=2$, $\PSNR=37.8730$ and $N_{\re}=366$.}}
\label{fig:images_QAM_HQAM}
\end{figure*}

While the improvements with the use of HQAM rather than QAM have already been pointed out, we in this subsection further compare the performances of image and video transmissions with conventional QAM and HQAM. In Fig. \ref{fig:images_QAM_HQAM}, we display the reconstructed images for different values of the fading parameter $m$ when conventional QAM and HQAM with different values of $\alpha$ are employed, in which we consider the same modulation parameter in both sensing decisions, i.e., $\alpha_0=\alpha_1 =\alpha$. It is assumed that power allocation with statistical CSI is applied. All image data is protected equally with conventional QAM. On the other hand, critical bits, i.e., HP bits receive higher protection with HQAM. With this, we see in the figure that HQAM generally provides better image quality when compared to conventional QAM signaling. This is further confirmed with the higher PSNR values for HQAM. We also observe that increasing $\alpha_i$ from $1$ to $2$ (i.e., increasing the protection level of HP bits) results in even higher PSNR values. Finally, we see that the received image quality expectedly improves as the fading parameter $m$ is increased from $1$ to $2$ for which we have more favorable fading conditions. In our additional simulations, we have observed that as fading becomes more severe, employing power control with instantaneous CSI substantially affects the PSNR performance, e.g., we see around 9 dB of improvement over power allocation with statistical CSI. On the other hand, when fading is less severe, there is only a slight change in image quality when power control based on either instantaneous CSI or statistical CSI is performed.

\begin{figure}
\centering
\begin{subfigure}[b]{0.17\textwidth}
\centering
\includegraphics[width=\textwidth]{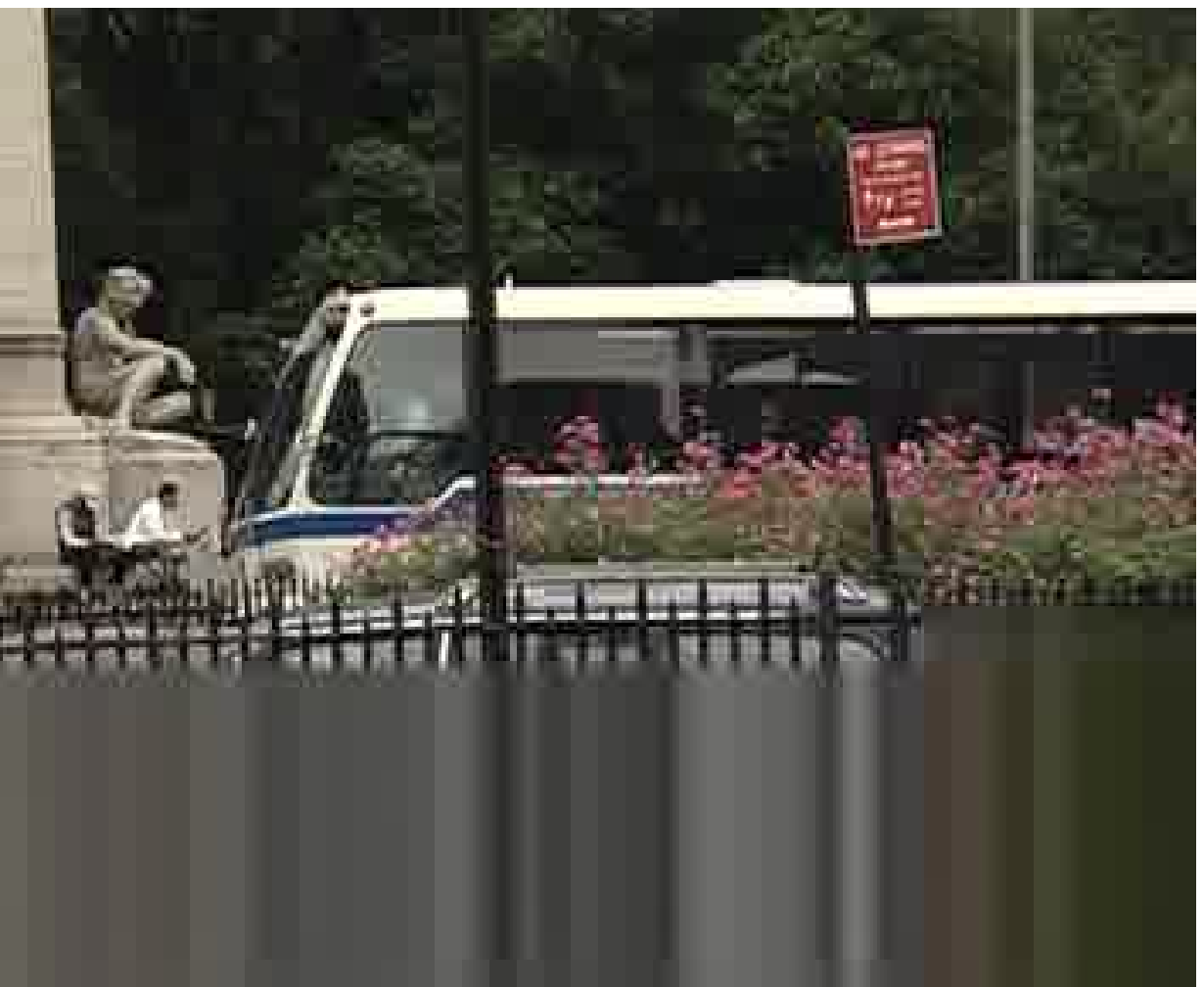}
\caption{}
\end{subfigure}
\begin{subfigure}[b]{0.17\textwidth}
\centering
\includegraphics[width=\textwidth]{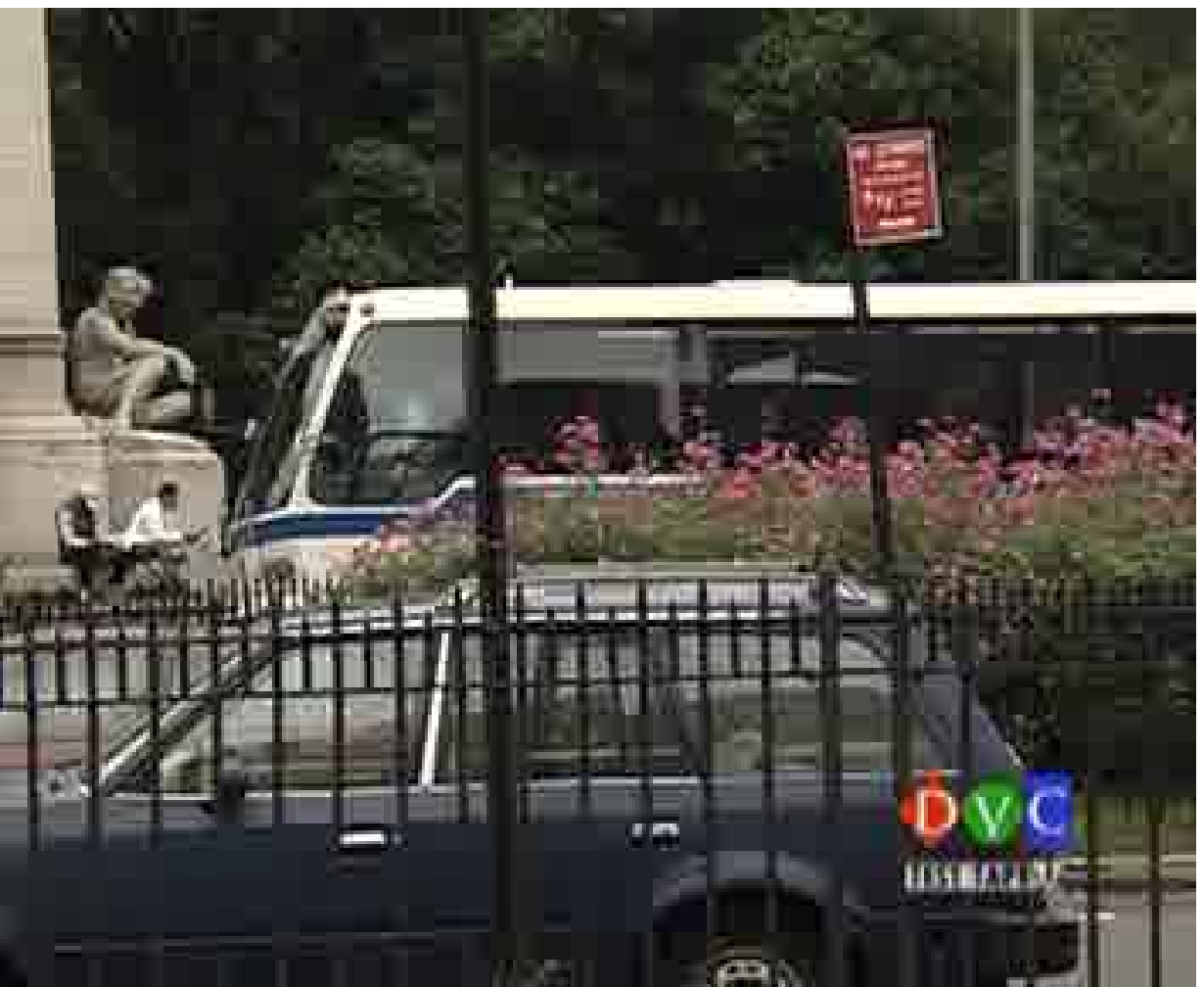}
\caption{}
\end{subfigure}
\begin{subfigure}[b]{0.17\textwidth}
\centering
\includegraphics[width=\textwidth]{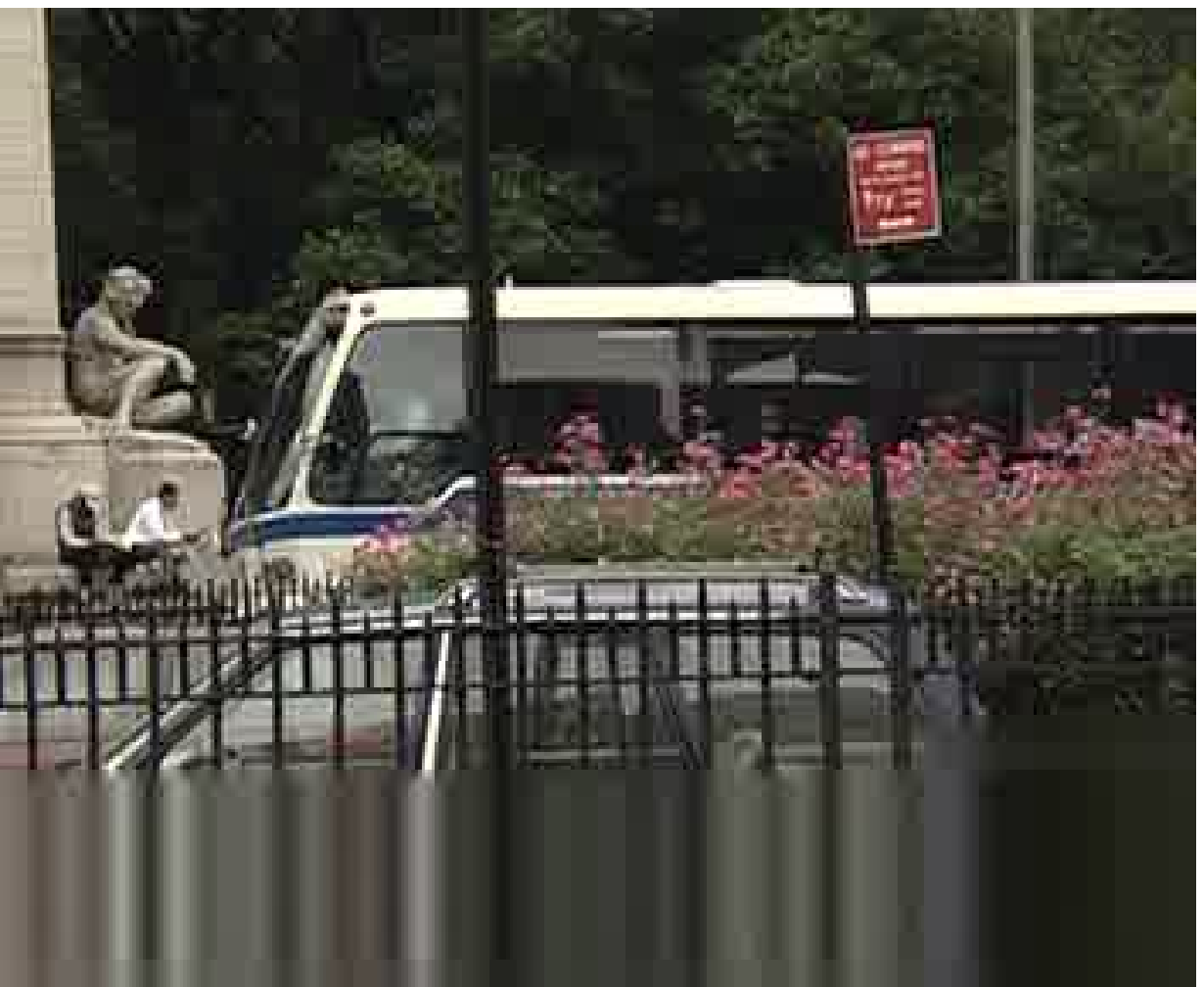}
\caption{}
\end{subfigure}
\begin{subfigure}[b]{0.17\textwidth}
\centering
\includegraphics[width=\textwidth]{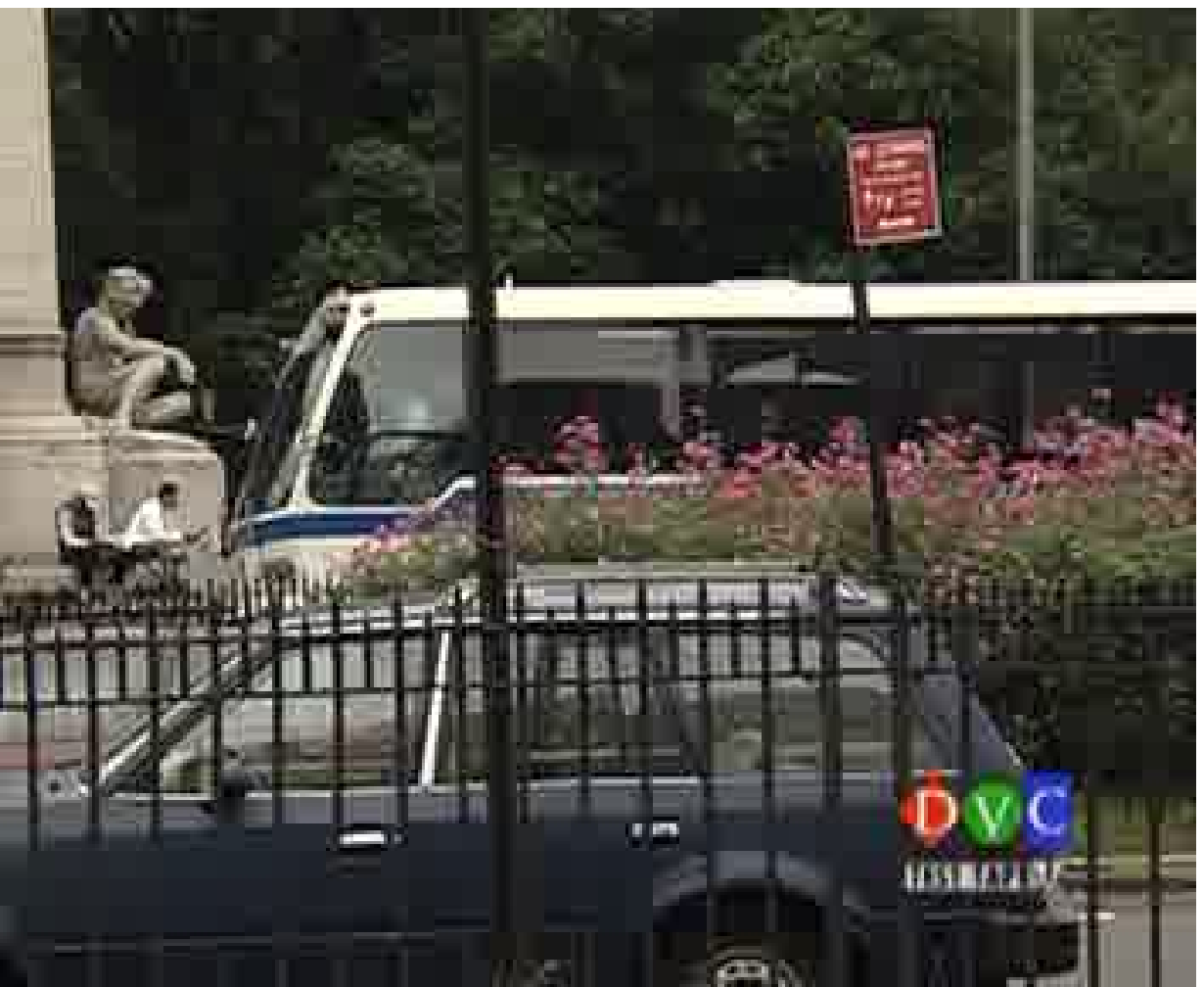}
\caption{}
\end{subfigure}
\caption{Video transmission based on power control with (a) statistical CSI and conventional QAM, $\PSNR=12.4644$ and $N_{\re}=3694$; (b) statistical CSI and hierarchical QAM, $\PSNR=15.1463$ and $N_{\re}=3855$; (c) instantaneous CSI and conventional QAM, $\PSNR=14.4713$ and $N_{\re}=2492$; (d) instantaneous CSI and hierarchical QAM, $\PSNR=15.4607$ and $N_{\re}=2976$.}\label{fig:video_QAM_HQAM}
\end{figure}
In Fig. \ref{fig:video_QAM_HQAM}, we display a single frame from the reconstructed video sequences which are transmitted by using conventional QAM and HQAM with power control applied based on either statistical CSI or instantaneous CSI. Imperfect sensing with $P_{\nid}=0.9$, and $P_{\f}=0.1$ is considered. It is also assumed that $P_{\avg} = 10$ dB, $Q_{\avg} = 4$ dB, and $Thr = 2.1$. The 11th frame of the video sequence is shown. While the average numbers of retransmissions for both modulation schemes are close to each other, it is seen that HQAM can lead to significant improvements in video quality compared to conventional QAM. Also, it is observed that applying optimal power control with instantaneous CSI reduces the number of retransmissions and improves the PSNR performance. In addition, when average transmit power and average interference power constraints are imposed, nearly the same PSNR values are obtained with smaller number of retransmissions.

\begin{figure}
\centering
{\includegraphics[width = 0.7\linewidth]{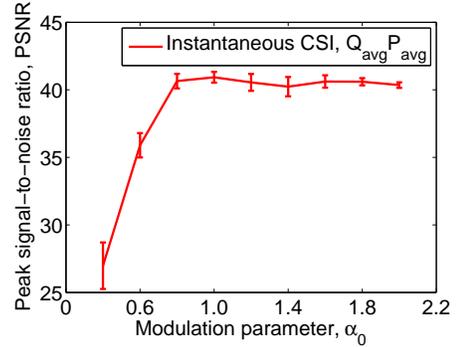}}
\caption{Peak signal-to-noise ratio, PSNR as a function of $\alpha_0$.}
\label{fig:PSNR_alpha}
\end{figure}

In Fig. \ref{fig:PSNR_alpha}, we display PSNR values as a function of $\alpha_0$ when $P_{\nid}=0.9$ and $P_{\f}=0.1$. We set $\alpha_1=1$ for busy sensing decision and change the values of $\alpha_0$ for idle sensing decision. In the figure, we consider confidence intervals in which the confidence level is set to $95\%$. Average transmit power and average interference power constraints are imposed, it is assumed that $Thr=1.8$, and instantaneous CSI is utilized in power control. It is observed that PSNR performance first improves with increasing $\alpha_0$ since the distance between quadrants increases, which leads to higher protection for HP data and hence lower BERs for HP bits. By further increasing $\alpha_0$, the image quality does not significantly change. This is because HP data is already protected well and  BER of HP bits is much smaller than the BER of LP data bits. Hence, allocating more power to the HP data bits does not substantially affect the BER of HP data bits, which leads to almost constant PSNR values. Similar trends are also observed under peak transmit power and average interference power constraints.

\begin{table}[h]
\begin{center}
\caption{Performance comparison of optimal power controls (exact and approximate)} \label{table_app_exact}
\hspace{-0.2cm}
\resizebox{0.4\textwidth}{!}{
\begin{tabular}{|cc|c|c|c|c|}
\hline
& & \multicolumn{2}{ c| }{($P_{\avg}, Q_{\avg}$) dB} & \multicolumn{2}{ c| }{($P_{\pk}, Q_{\avg}$) dB} \\
\cline{3-6}
& & ($10, 4$) & ($15, 9$) &($10, 4$) & ($15, 9$) \\ \hline
\multicolumn{1}{ |c|  }{BER} &  App. & 0.0428 & 0.0182 & 0.044 & 0.0186  \\  \cline{2-6}
\multicolumn{1}{ |c|  }{of HP bits} & Exact & 0.0404 & 0.0174 & 0.042 & 0.0178 \\  \cline{1-6}
\multicolumn{1}{ |c|  }{BER} & App.  & 0.0752& 0.0334 & 0.0782 & 0.0344 \\  \cline{2-6}
\multicolumn{1}{ |c|  }{of LP bits} & Exact & 0.074 & 0.033 & 0.0772 & 0.034\\ \cline{1-6}
\multicolumn{1}{ |c|  }{\multirow{2}{*}{PSNR} } &  App. & 41.9002 & 44.7711 & 41.1324& 44.5025  \\  \cline{2-6}
\multicolumn{1}{ |c|  }{} & Exact & 42.8420 & 46.3720 & 42.2010 & 45.3714 \\ \cline{1-6}
\end{tabular}}
\end{center}
\end{table}

In Table \ref{table_app_exact}, we have listed BERs of HP bits and LP bits, and PSNR values when exact optimal power control and approximate power control given in Propositions \ref{prop_2} and \ref{prop_4} at high SNR levels are employed under perfect sensing subject to different peak transmit power/average interference power and average transmit power/average interference power constraints. It is seen that exact and approximate power levels result in very similar error rates and PSNR performances at moderate and high SNRs, which is in agreement with Propositions \ref{prop_2} and \ref{prop_4}. Hence, instead of solving the exact optimal power control by bisection search, we can employ the approximate power control given in terms of the Lambert-W function, which is easier to evaluate.

\section{Conclusion}\label{sec:conclusion}
We have studied the performance of multimedia transmissions with HQAM in CR systems in the presence of imperfect sensing results and constraints on both the transmit and interference power levels. By exploiting the unequal importance of the compressed data bits, we have provided more protection to high priority bits of JPEG2000 coded image and H.264/MPEG-4 coded video by employing 16-HQAM. We have obtained closed-form expressions for the error probabilities of HP and LP bits in HQAM over Nakagami-$m$ fading channels under sensing uncertainty. We have determined the optimal power levels that minimize the total average error probability or its upper bound under peak power and average interference constraints by assuming the availability of instantaneous CSI or statistical CSI. Via simulations, we have analyzed the impact of channel sensing performance, modulation parameter $\alpha_i$, and severity of the fading on the received data quality. Simulation results demonstrate that HQAM performs better than conventional QAM in terms of average PSNR. In addition, power control with instantaneous CSI outperforms power allocation with statistical CSI. We have shown that received data quality is robust to imperfect channel sensing results if there is no upper bound on the number of retransmissions. In these cases, the number of retransmissions increases with decreasing $P_{\nid}$ or increasing $P_{\f}$, resulting in larger delays and energy consumption.  If there is a constraint on the number of retransmissions, PSNR performance of multimedia transmission is affected by sensing errors. We have observed that improved sensing performance leads to better quality at reception. Less severe fading (i.e., larger $m$) is also shown to improve the received multimedia data quality.

\appendix

\subsection{Derivation of Equations (\ref{eq:BER_HP_averaged_fading}) and (\ref{eq:BER_LP_averaged_fading})} \label{app:derivation-eqs}

In order to find the averaged BER of HP bits and LP bits over Nakagami-$m$ fading distribution, we evaluate the expectations below with respect to channel power gain $z=|h|^2$:
\begin{align} \nonumber
\hspace{-0.3cm}P_{\text{HP}}(\bP) &=\frac{m^m}{\Omega^m\Gamma(m)}\int_0^{\infty} P_{\text{HP}}(\bP,z)z^{m-1}\rme^{-\frac{m}{\Omega}z}dz\\
P_{\text{LP}}(\bP) &= \frac{m^m}{\Omega^m\Gamma(m)}\int_0^{\infty} P_{\text{LP}}(\bP,z)z^{m-1}\rme^{-\frac{m}{\Omega}z}dz
\end{align}
where $\Gamma(.)$ is the gamma function \cite[eq. 6.1.1]{abramowitz}, $m$ is the fading parameter that controls the severity of the amplitude fading, $m\geq 0.5$, and $P_{\text{HP}}(\bP,h)$ and $P_{\text{LP}}(\bP,h)$  are given in (\ref{eq:BER_HP}) and (\ref{eq:BER_LP}), respectively. In order to evaluate the above integrals, the following alternative representation of the Gaussian $Q$ function is employed:
\begin{align}
Q(x)=\frac{1}{2\sqrt{\pi}}\Gamma\Bigg(\frac{1}{2},\frac{x^2}{2}\Bigg).
\end{align}
Inserting the above $Q$ function expression into (\ref{eq:BER_HP}) and (\ref{eq:BER_LP}), and using the identity \cite[eq. 6.455.1]{gradshteyn}, we obtain the closed-form BER expressions for HP and LP bits, respectively in (\ref{eq:BER_HP_averaged_fading}) and (\ref{eq:BER_LP_averaged_fading}).

\subsection{Proof of Proposition \ref{prop_1}} \label{app:proof-prop1}
By removing the $Q$ functions with negative weight in (\ref{eq:BER_LP}), the objective function becomes convex subject to affine inequality constraints given in (\ref{eq:peak_P0_constraint}), (\ref{eq:peak_P1_constraint}) and (\ref{eq:interference_power_constraint_inst}). Hence, the optimal power can be obtained by using the Lagrangian optimization approach as follows:
\begin{equation}
\small
\begin{split} \label{eq:Lagrange_function}
&L(P_0(h,g),P_1(h,g),\mu_1)=\E\big\{\lambda P_{\text{HP}}(\bP,h) + (1-\lambda)P_{\text{LP}}^{u}(\bP,h)\big\}\\&+\mu_1(\E\{(1-P_{\nid})\,P_0(h,g) \,|g|^2 + P_{\nid} \,P_1(h,g) \, |g|^2\} -Q_{\avg}).
\end{split}
\normalsize
\end{equation}
Above, the superscript $u$ in $P_{\text{LP}}^{u}(\bP,h)$ indicates that this is the upper bound on $P_{\text{LP}}(\bP,h)$ and $\mu_1$ is the nonnegative Lagrange multiplier. The Lagrange dual problem is defined as
\begin{equation}
\small
\begin{split} \label{eq:Lagrange_dual}
\max_{\mu_1 \ge 0} \min_{
\substack{0 \le P_0(h,g) \le P_{\pk} \\ 0 \le P_1(h,g) \le P_{\pk}}} L(P_0(h,g),P_1(h,g),\mu_1),\mu_1).
\end{split}
\normalsize
\end{equation}
For fixed $\mu_1$ and fading coefficients, the subproblem is formulated, by applying the Lagrange dual decomposition method, as follows:
\begin{align}
\small
\begin{split}
 &\min_{
\substack{0 \le P_0(h,g) \le P_{\pk} \\ 0 \le P_1(h,g) \le P_{\pk}}} \lambda P_{\text{HP}}(\bP,h) + (1-\lambda)P_{\text{LP}}^{u}(\bP,h)\\&\hspace{2cm}+ \mu_1 \big((1-P_{\nid})\,P_0(h,g) \,|g|^2 + P_{\nid} \,P_1(h,g) \, |g|^2\big).
\end{split}
\normalsize
\end{align}
According to the Karush-Kuhn-Tucker (KKT) conditions, the optimal power levels $P^{(0)}_{\text{opt}}(h,g)$ and $P^{(1)}_{\text{opt}}(h,g)$ must satisfy the following:
\small
\begin{align} \nonumber
&\hspace{-0.1cm}\sum_{j,l=0}^{1} \!\!\!\frac{P(\mH_j,\!\hH_0)}{4\sqrt{2\pi}}\!\Bigg\{\lambda\frac{\rme^{\frac{-c_{l,0}P_0(h,g)|h|^2}{2\sigma_j^2}}}{\sqrt{\frac{\sigma_j^2 P_0(h,g)}{c_{l,0}|h|^2}}}\!+\!(1\!-\!\lambda)\rho_l \frac{\rme^{\frac{-\beta_{l,0}P_0(h,g)|h|^2}{2\sigma_j^2}}}{\sqrt{\frac{ \sigma_j^2 P_0(h,g)}{\beta_{l,0}|h|^2}}}\!\Bigg\}\\ \label{eq:P0_sol_1}&\hspace{5cm}-\mu_1(1-P_{\nid})|g|^2=0,\\ \nonumber
&\hspace{-0.1cm}\sum_{j,l=0}^{1}\frac{P(\mH_j,\!\hH_1)}{4\sqrt{2\pi}}\Bigg\{\!\lambda\frac{\rme^{\frac{-c_{l,1}P_1(h,g)|h|^2}{2\sigma_j^2}}}{\sqrt{\frac{\sigma_j^2 P_1(h,g)}{c_{l,1}|h|^2}}}\!+\!(1\!-\!\lambda)\rho_l \frac{\rme^{\frac{-\beta_{l,1}P_1(h,g)|h|^2}{2\sigma_j^2}}}{\sqrt{\frac{\sigma_j^2 P_1(h,g)}{\beta_{l,1}|h|^2}}}\Bigg\}\\ \label{eq:P1_sol_1}&\hspace{5.7cm}-\mu_1P_{\nid} |g|^2=0,\\
\label{eq:peak_power_average_inter_cond2}
&\mu_1(\E\{(1-P_{\nid})\,P_0(h,g) \,|g|^2 + P_{\nid} \,P_1(h,g) \, |g|^2\} -Q_{\avg})=0, \\ \label{eq:peak_power_average_inter_cond3}
&\mu_1 \ge 0,\\ \label{eq:peak_power_average_inter_cond4}
&\E\{(1-P_{\nid})\,P_0(h,g) \,|g|^2 + P_{\nid} \,P_1(h,g) \, |g|^2\} -Q_{\avg}\le 0.
\normalsize
\end{align}
\normalsize
Solving the above equations (\ref{eq:P0_sol_1}) and (\ref{eq:P1_sol_1}), and combining the solutions denoted by $P_0^*$ and $P_1^*$ with peak power constraints (\ref{eq:peak_P0_constraint}) and (\ref{eq:peak_P1_constraint}), respectively, yield the desired result in (\ref{eq:P0_opt_QavgPpeak_ins}) and (\ref{eq:P1_opt_QavgPpeak_ins}). \hfill $\square$
\subsection{Proof of Proposition \ref{prop_2}} \label{app:proof-prop2}
When the sensing is perfect (i.e., $P_{\nid}=1$ and $P_{\f}=0$), the optimal power levels that minimize the BER of HP bits can be found by solving the following optimization problem:
\begin{align}
\label{eq:Opt_Qavg_Ppeak_perfect}
&\hspace{0.5cm}\min_{
\substack{P_0(h,g), P_1(h,g)}} \E\big\{P_{\text{HP}}(\bP,h)\big\} \\ \nonumber &\hspace{-2cm}\text{subject to} \\ &\label{eq:peak_P0_constraint_perfect}\hspace{-2cm}P_0(h,g) \le P_{\pk}, \hspace{0.1cm}P_1(h,g) \le P_{\pk} \\ \label{eq:interference_power_constraint_inst_perfect}
&\hspace{-2cm}\E\{P_{\nid} \,P_1(h,g) \, |g|^2\} \le Q_{\avg}
\end{align}
Since $Q$ function decreases rapidly in its argument, BER in (\ref{eq:Opt_Qavg_Ppeak_perfect}) is dominated by the $Q$ function with the smaller argument at high SNRs. Therefore, the objective function becomes
\begin{align}
\small
\begin{split}\label{eq:BER_MIN}
&\frac{1}{2}\Pr\{\mH_0\}\E\Bigg\{Q\Bigg(\sqrt{\frac{c_{1,0} P_0(h,g)|h|^2}{\sigma_n^2}}\Bigg)\Bigg\}\\&\hspace{2.5cm}+\frac{1}{2}\Pr\{\mH_1\}\E\Bigg\{Q\Bigg(\sqrt{\frac{c_{1,1} P_1(h,g)|h|^2}{\sigma_n^2+\sigma_w^2}}\Bigg)\Bigg\}.
\end{split}
\normalsize
\end{align}
\normalsize
It is seen that the only constraint related to $P_0$ is the peak transmit power constraint in (\ref{eq:peak_P0_constraint_perfect}), and hence the minimum BER is achieved when the secondary user transmits at the maximum available instantaneous power. Therefore, $P^{(0)}(h,g)=P_{\pk}$. In order to find the optimal $P_1$, we first express the Lagrangian function and take its derivative with respect to $P_1$ and set it to zero, which results in
\begin{align}
\small
\begin{split}
P_1 \rme^{\frac{c_{1,1} |h|^2P_1}{\sigma_n^2+\sigma_w^2}}=\frac{c_{1,1} |h|^2P(\mH_1)^2}{32 \pi (\mu_1|g|^2)^2 (\sigma_n^2+\sigma_w^2)}.
\end{split}
\normalsize
\end{align}
Solving for $P_1$ in the above equation and combining the result with peak transmit power constraint in (\ref{eq:peak_P0_constraint_perfect}) provide the optimal power policy in (\ref{eq:P1_opt_perfect}).
\hfill $\square$


\end{document}